\documentclass[final,5p,times,twocolumn,number,sort&compress]{elsarticle}
\usepackage[T1]{fontenc}
\usepackage[utf8]{inputenc}
\usepackage{textcomp}
\usepackage{amsmath,amssymb}
\usepackage{graphicx}
\usepackage{booktabs,longtable,array,calc,rotating}
\usepackage{lineno}
\usepackage[hidelinks]{hyperref}
\usepackage{orcidlink}
\makeatletter\newcommand{\authororcid}[1]{\ifx\corref\@gobble\else\,\orcidlink{#1}\fi}\makeatother
\graphicspath{{figures/}}

\journal{Composite Structures}

\begin{document}
\begin{frontmatter}

\title{Process-zone condensation: a discretization-exact reduced cell model for FE\textsuperscript{2} analysis of composites with fibre--matrix debonding and matrix plasticity}

\author[a]{Leonilson Toffoli Pedrini\authororcid{0000-0002-9482-1007}\corref{cor1}}
\ead{leonilson.pedrini01@estudante.ifes.edu.br}
\author[b]{Maycoln Conci\authororcid{0000-0002-3985-0307}}
\author[a,b]{Rafael Marin Ferro\authororcid{0000-0002-3229-2170}}
\cortext[cor1]{Corresponding author. ORCID: 0000-0002-9482-1007.}
\affiliation[a]{organization={Programa de Pós-Graduação em Engenharia Metalúrgica e de Materiais (PROPEMM), Instituto Federal do Espírito Santo (IFES), Campus Vitória},
  addressline={Avenida Vitória, 1729, Jucutuquara}, postcode={29040-780},
  city={Vitória}, state={ES}, country={Brazil}}
\affiliation[b]{organization={Instituto Federal do Espírito Santo (IFES), Campus Aracruz},
  addressline={Avenida Morobá, 248, Morobá}, postcode={29192-733},
  city={Aracruz}, state={ES}, country={Brazil}}

\begin{abstract}
Periodic composites with cohesive interfaces and matrix plasticity require nonlinear cell solutions in multiscale analysis. Process-zone condensation eliminates the linear elastic remainder of the cell by a Schur complement, retaining all interfacial and process-zone internal variables. Influence matrices recover macroscopic stress and consistent tangent. One back-substitution per increment checks that the matrix outside the retained zone remains elastic. While this holds, the reduction is exact relative to the full discretisation, also with a plastic matrix, and requires no training. In an E-glass/epoxy cell, eight of nine two-dimensional paths agree with the full cell to 1e-11 in stress and tangent; the check flags the ninth. Three paths on each of two three-dimensional meshes need zone expansion. Excluding setup and checks, two-dimensional path timings show an aggregate speed-up of 3.3; certified three-dimensional zones show none. In adaptive finite element squared (FE\textsuperscript{2}) simulations of a perforated plate, the reduced solver reproduces the full-cell response within solver tolerances, 1.51 times faster for glass/polypropylene and slower for glass/epoxy. Independent Ansys models agree to below 1e-3 in cell stress and about 0.6 \% in plate reactions, although local damage and plasticity onsets differ. The advantage depends on zone size, active set and adaptation costs.
\end{abstract}


\end{frontmatter}

\hypertarget{introduction}{%
\section{Introduction}\label{introduction}}

The structural response of two-phase composites depends on how the matrix and the interfaces redistribute load as inelastic deformation develops. Matrix plasticity and progressive interface debonding can interact within a small part of a periodic cell, while the surrounding material remains elastic. Resolving this interaction is relevant to inclusion-reinforced polymers and to architected composites with stiff and compliant phases. Multi-material additive manufacturing provides an additional motivation: the manufacturing process can alter interface stiffness, strength and fracture resistance, making interface characterisation part of the constitutive description \cite{ref1,ref2,ref3}.

Computational homogenisation connects these local mechanisms to the response of a structural component. In the FE² scheme, a nonlinear cell problem supplies the stress and consistent tangent at each macroscopic integration point, and repeated cell solutions can dominate the cost of the structural analysis. Reduced bases, hyper-reduction and non-intrusive constitutive approximations address this cost in different ways \cite{ref4,ref5,ref6}; their accuracy depends on the approximation and its assessment. For a cell whose interface can debond, the reliability of the reduced response matters as much as its cost. Nonlinear homogenisation of architected cells has been developed for elastic truss, beam-lattice and planar microstructures \cite{ref7,ref8,ref9}; interface debonding in periodic composites has been studied in hyperelastic settings without plasticity \cite{ref10,ref11,ref12}; and interface behaviour has also been considered in multi-material structural topology optimisation \cite{ref13,ref14}. For a cell with both matrix plasticity and cohesive interfaces, the practical question is how much of the repeated computation can be removed while preserving its response along the prescribed loading history.

Transformation field analysis (TFA) represents inelastic strains through transformation fields and elastic influence functions, with piecewise uniform fields in its classical form \cite{ref15,ref16}. Nonuniform TFA (NTFA) introduces spatially varying inelastic modes \cite{ref17,ref18}, including in three-dimensional finite element implementations \cite{ref19}. Its accuracy and cost depend on the retained representation and the loading \cite{ref20}. Cohesive interfaces have also been incorporated. Fritzen and Leuschner reduce interface jumps on a trained basis for an elastic bulk \cite{ref21} and extend the approach to viscoplastic constituents with dissipative interfaces \cite{ref22}. Marfia and Sacco combine constituent plasticity and cracking with interfacial debonding using analytical modes over subdomains and subinterfaces \cite{ref23}. The enhanced TFA (E-TFA) of Mishra et al.~provides a consistent tangent \cite{ref24} and a multiscale FE² formulation \cite{ref25}. With an elastic bulk, one subinterface per cohesive element and matching interpolation modes and finite element shape functions, E-TFA reproduces the finite element solution \cite{ref24,ref25}. With an inelastic bulk, its variables are defined at the nodes of a support mesh: agreement improves with subdivision, while the computational advantage decreases as the plastic region grows \cite{ref24}. Partitioned model order reduction of fracture problems retains the full model where damage grows and projects the remainder onto a reduced basis \cite{ref26}. That remainder is nonlinear and is approximated. Here, the remainder is linear and can be eliminated without approximation. The aim is to preserve the full discrete cell response with matrix plasticity confined to a retained zone and to check during loading whether that zone remains adequate.

We retain the inelastic variables individually and eliminate the linear elastic remainder of the cell by static condensation, using the classical Schur complement \cite{ref27}. The resulting influence matrices recover the macroscopic stress and consistent tangent from the retained solution. We call this formulation process-zone condensation.

The contribution combines four elements. First, the elastic remainder is eliminated exactly, without subdivision or modes. Second, the cohesive variables and the plastic variables of every integration point in a prescribed matrix zone are retained individually, preserving the discrete response also with matrix plasticity. Third, an incremental check tests whether the zone contains all matrix plasticity. Fourth, the FE² implementation enlarges the zone at the macroscopic integration points where this check fails. E-TFA already provides an exact interface representation under the matching-discretisation conditions described above \cite{ref24,ref25}. The present formulation combines exact retention of matrix plasticity within the zone with a check of its adequacy during loading.

The reduction is algebraically exact for the chosen finite element discretisation provided that the matrix outside the retained zone remains elastic throughout the loading history. At each converged increment, one back-substitution recovers the eliminated displacements, and a yield check tests that condition at the matrix integration points outside the zone. Failure of the check identifies an inadequate zone and triggers its expansion, in the cell and per integration point in the adaptive FE² implementation. We call this check a certificate in the sense of an admissibility test: it states whether the reduced and full solutions coincide, and it does not bound the error when they do not (Section 2.6). The certificate establishes zone adequacy; the computational benefit must be measured separately, because it also depends on the retained dimension, the size of the active set, the reference solver and the cost of adaptation.

The numerical study uses periodic inclusion cells with material parameters drawn from the glass/epoxy and glass/polypropylene literature \cite{ref28,ref29,ref30,ref31,ref32,ref33}. These systems provide experimental information for the constituent and interface laws, supplemented by the modelling assumptions stated in Section 4. They are benchmarks for the formulation; experimental validation of an additively manufactured material remains outside the present scope. Reduced and full cells are compared on identical meshes in two and three dimensions, and reduced FE² is compared with the corresponding full-cell structural calculation. Effective debonding and yield-onset surfaces and finite-difference checks of the macroscopic tangent provide additional tests. Independent Ansys models provide further comparisons at cell level and for perforated plates with the microstructure resolved. POD--Galerkin and NTFA models trained on snapshots examine the respective roles of training, approximation error and computational cost. The principal performance comparisons use the same hardware and thread counts, with PARDISO for the full-cell reference and dense linear algebra for the reduced calculations, with setup and certification costs identified separately.

Section 2 derives the condensation, macroscopic recovery and zone certificate, and relates them to TFA and reduced bases. Section 3 describes the implementation and reference models. Section 4 presents the cell and structural comparisons, the effects of zone growth and geometry, and the comparison with trained models. Section 5 summarizes the findings and their limits.

\hypertarget{cell-problem-and-process-zone-condensation}{%
\section{Cell problem and process-zone condensation}\label{cell-problem-and-process-zone-condensation}}

\hypertarget{unit-cell-and-macroscopic-loading}{%
\subsection{Unit cell and macroscopic loading}\label{unit-cell-and-macroscopic-loading}}

The material is a periodic two-phase composite. Its unit cell \(Y\), of area \(|Y|\), is analysed in plane strain with unit thickness (Fig. 1). A stiff inclusion \(\Omega_s\), linear elastic, is embedded in a ductile matrix \(\Omega_d\), elastoplastic; their common boundary \(\Gamma_i\) is a cohesive interface of zero thickness. Strains are small, loading is quasi-static and the response is rate independent. The macroscopic strain \(\boldsymbol{E}\) is prescribed and the displacement in the cell is split into an affine part and a periodic fluctuation \(\tilde{\boldsymbol{u}}\),

\begin{figure}[tbp]
\centering
\includegraphics[width=\linewidth]{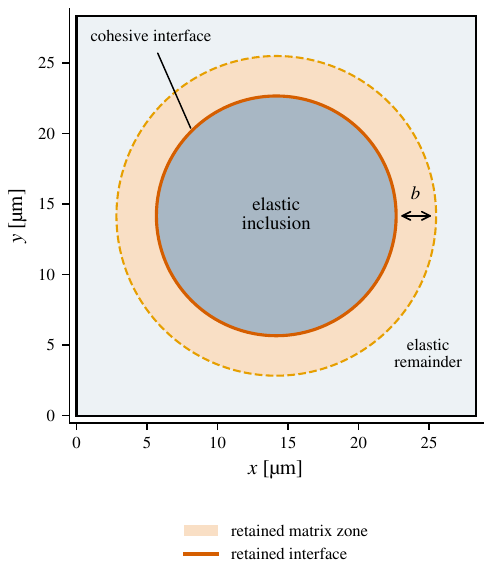}
\caption{Unit cell: stiff inclusion, ductile matrix, cohesive interface and process zone of width \(b\); schematic distinction between the retained matrix zone and cohesive interface and the condensed elastic phases; opposite faces satisfy periodic fluctuation conditions.}
\end{figure}

\begin{equation}
\begin{gathered}
\boldsymbol{u}(\boldsymbol{x}) = \boldsymbol{E}\,\boldsymbol{x} + \tilde{\boldsymbol{u}}(\boldsymbol{x}), \\
\tilde{\boldsymbol{u}} \ \text{periodic on } \partial Y.
\end{gathered}
\end{equation}
The macroscopic stress is the volume average of the cell stress,

\begin{equation}
\boldsymbol{\Sigma} = \frac{1}{|Y|}\int_Y \boldsymbol{\sigma}\,\mathrm{d}Y,
\end{equation}
which, under (1), is the work conjugate of \(\boldsymbol{E}\) by the Hill--Mandel condition \cite{ref34}; in the discrete cell of Section 2.4 it is also the reaction at the control degrees of freedom that carry \(\boldsymbol{E}\), and both measures are used in Section 4. Along a macroscopic strain path \(\boldsymbol{E}(s)\), \(s \in [0,1]\), the cell is solved incrementally. Plasticity and interface damage make the response path dependent, so the cell carries a history. Vectors and matrices are written in bold; macroscopic quantities in capitals; Voigt notation with engineering shear strain is used for the in-plane components \((E_{11}, E_{22}, 2E_{12})\) and \((\Sigma_{11}, \Sigma_{22}, \Sigma_{12})\).

\hypertarget{constitutive-laws}{%
\subsection{Constitutive laws}\label{constitutive-laws}}

The stiff phase is linear elastic and isotropic,

\begin{equation}
\boldsymbol{\sigma} = \mathbb{C}_s : \boldsymbol{\varepsilon} \quad \text{in } \Omega_s.
\end{equation}
The ductile phase is elastoplastic with von Mises yield and linear isotropic hardening,

\begin{equation}
\begin{gathered}
\boldsymbol{\sigma} = \mathbb{C}_d : (\boldsymbol{\varepsilon} - \boldsymbol{\varepsilon}^p), \\
f(\boldsymbol{\sigma}, p) = \lVert \operatorname{dev}\boldsymbol{\sigma} \rVert - \sqrt{\tfrac{2}{3}}\,(\sigma_y + H\,p) \le 0, \\
\dot{\boldsymbol{\varepsilon}}^p = \dot{\gamma}\,\frac{\partial f}{\partial \boldsymbol{\sigma}}, \\
\dot{p} = \sqrt{\tfrac{2}{3}}\,\dot{\gamma},
\end{gathered}
\end{equation}
where \(\sigma_y\) is the initial yield stress, \(H\) the plastic modulus and \(p\) the accumulated plastic strain. In plane strain the out-of-plane plastic strain is not zero, so the return mapping of Section 3.1 carries the full deviatoric state and the out-of-plane stress (Appendix A.2).

The interface traction \(\boldsymbol{t}\) depends on the displacement jump \(\boldsymbol{\delta} = \boldsymbol{u}^+ - \boldsymbol{u}^-\) across \(\Gamma_i\), resolved into normal and tangential components \(\delta_n\) and \(\delta_t\), with \(\delta_n > 0\) denoting opening. We use the bilinear mixed-mode law of Alfano and Crisfield \cite{ref35} in the form implemented by the commercial solver \cite{ref36}, so that the comparison in Section 4.7 uses the same law in both codes. Its six constants are the strengths \(\sigma_{\max}\) and \(\tau_{\max}\), the complete-decohesion separations \(\delta_n^c\) and \(\delta_t^c\), the common onset-to-decohesion ratio \(\alpha\), and the tangential weight \(\beta\). The initial stiffnesses \(K_n\) and \(K_t\) follow from the pure-mode branches and need not be equal.

Damage is described by an irreversible scalar \(D \in [0,1]\), governed by the largest effective jump \(\lambda_{\max}\) reached during loading. It begins at \(\lambda_{\max} = \alpha\) and is complete at \(\lambda_{\max} = 1\). Only the opening part of \(\delta_n\) contributes to this effective jump. Damage reduces the tangential stiffness and the tensile normal stiffness; the compressive normal stiffness remains undamaged and provides finite resistance to closure. Unloading is secant to the origin. Appendix A.1 gives the full law, (A.1) to (A.4), including \(G_{Ic} = \sigma_{\max}\delta_n^c/2\) and \(G_{IIc} = \tau_{\max}\delta_t^c/2\).

The finite initial stiffness makes the undamaged interface compliant, reducing the elastic moduli relative to a perfectly bonded cell (Section 4.1). Interface characterisation must therefore identify stiffness, strength and fracture energy in each mode, quantities also relevant to printed multi-material interfaces \cite{ref1,ref2,ref3}.

\hypertarget{periodic-discrete-cell-in-eigenstrain-form}{%
\subsection{Periodic discrete cell in eigenstrain form}\label{periodic-discrete-cell-in-eigenstrain-form}}

The cell is discretised by finite elements. Master-slave elimination imposes periodicity of \(\tilde{\boldsymbol{u}}\) on opposite faces, leaving \(n\) free degrees of freedom. In the commercial solver, constraint equations relate opposite nodes to three control nodes carrying the components of \(\boldsymbol{E}\).

The phases share no nodes across \(\Gamma_i\). Each interface node is duplicated into a coincident pair \((a^+, a^-)\), with jump \(\boldsymbol{\delta}_a = \boldsymbol{N}_a\tilde{\boldsymbol{u}}\). The operator \(\boldsymbol{N}_a\) takes the displacement difference and rotates it to the local normal and tangent. The affine part of (1) contributes no jump because the nodes coincide. Tractions produce equal and opposite nodal forces \(A_a\boldsymbol{t}_a\), where \(A_a\) is the tributary length of the pair.

The key to the condensation is the way the two nonlinearities are written. The elastic stiffness \(\boldsymbol{K}_e\) is assembled with the elastic moduli of both phases and with the undamaged interface springs \(A_a\,\mathrm{diag}(K_n, K_t)\) at every pair, and is constant. The cohesive nonlinearity is then carried by the damage force

\begin{equation}
\boldsymbol{\lambda}_a = A_a\left[\mathrm{diag}(K_n, K_t)\,\boldsymbol{\delta}_a - \boldsymbol{t}(\boldsymbol{\delta}_a, D_a)\right],
\end{equation}
which is the difference between the undamaged spring force and the cohesive traction and vanishes identically on every pair with \(D_a = 0\), in compression included. The plastic strain enters as an eigenstrain through the nodal force \(\boldsymbol{f}_p\) it produces. Equilibrium of the discrete cell at any instant reads

\begin{equation}
\begin{gathered}
\boldsymbol{K}_e\,\tilde{\boldsymbol{u}} = \boldsymbol{f}_E\,\boldsymbol{E} + \boldsymbol{f}_p(\boldsymbol{\varepsilon}^p) + \boldsymbol{N}^{\mathsf T}\boldsymbol{\lambda}, \\
\boldsymbol{f}_p(\boldsymbol{\varepsilon}^p) = \sum_{e \in \Omega_p}\int_{\Omega_e} \boldsymbol{B}_e^{\mathsf T}\,\mathbb{C}_d\,\boldsymbol{\varepsilon}^p_e\,\mathrm{d}\Omega,
\end{gathered}
\end{equation}
where \(\boldsymbol{f}_E\) is the \(n \times 3\) matrix of nodal forces produced by unit macroscopic strains through the affine part of (1), \(\boldsymbol{N}\) stacks the pair operators \(\boldsymbol{N}_a\), \(\boldsymbol{\lambda}\) stacks the damage forces and \(\boldsymbol{B}_e\) is the strain-displacement matrix of element \(e\). The set \(\Omega_p \subset \Omega_d\) is the process zone of the matrix: the elements allowed to yield. It is chosen a priori as the elements of the ductile phase within a distance \(b\) of \(\Gamma_i\) and is verified a posteriori in Section 2.6. Outside \(\Omega_p\) the matrix is treated as elastic, and the verification certifies that this is what the full model would have found.

Equation (6) is the equilibrium of the full finite element cell written so that every nonlinear term has its support on a small set of degrees of freedom: the nodes of the interface pairs and the nodes of the elements of \(\Omega_p\). The reference scheme, used by the commercial solver and by the code in verification mode, advances \(\boldsymbol{E}(s)\) in increments and solves (6) together with the cohesive law and the return mapping of (4) by Newton iteration on all \(n\) degrees of freedom.

\hypertarget{exact-condensation-and-influence-matrices}{%
\subsection{Exact condensation and influence matrices}\label{exact-condensation-and-influence-matrices}}

Partition the degrees of freedom into the process-zone set \(z\), of size \(m\), which contains the nodes of the interface pairs and of the elements of \(\Omega_p\), and its complement \(r\). Since \(\boldsymbol{f}_p\) and \(\boldsymbol{N}^{\mathsf T}\boldsymbol{\lambda}\) have support in \(z\), the \(r\) rows of (6) are linear in \(\tilde{\boldsymbol{u}}_r\) and can be eliminated exactly:

\begin{equation}
\begin{gathered}
\boldsymbol{S}\,\tilde{\boldsymbol{u}}_z = \boldsymbol{s}_E\,\boldsymbol{E} + \boldsymbol{f}_{p,z}(\boldsymbol{\varepsilon}^p) + \boldsymbol{N}_z^{\mathsf T}\boldsymbol{\lambda}, \\
\boldsymbol{S} = \boldsymbol{K}_{zz} - \boldsymbol{K}_{zr}\boldsymbol{K}_{rr}^{-1}\boldsymbol{K}_{rz}, \\
\boldsymbol{s}_E = \boldsymbol{f}_{E,z} - \boldsymbol{K}_{zr}\boldsymbol{K}_{rr}^{-1}\boldsymbol{f}_{E,r}.
\end{gathered}
\end{equation}
\(\boldsymbol{S}\) is the Schur complement of \(\boldsymbol{K}_{rr}\) in \(\boldsymbol{K}_e\): symmetric, positive definite, dense and of size \(m \times m\). With the kernel \(\boldsymbol{G} = \boldsymbol{S}^{-1}\) and the elastic influence matrix \(\boldsymbol{G}_E = \boldsymbol{G}\boldsymbol{s}_E\) of size \(m \times 3\),

\begin{equation}
\tilde{\boldsymbol{u}}_z = \boldsymbol{G}_E\,\boldsymbol{E} + \boldsymbol{G}\left[\boldsymbol{f}_{p,z}(\boldsymbol{\varepsilon}^p) + \boldsymbol{N}_z^{\mathsf T}\boldsymbol{\lambda}\right],
\end{equation}
and the interface jumps and the process-zone strains follow explicitly from \(\tilde{\boldsymbol{u}}_z\) alone,

\begin{equation}
\begin{gathered}
\boldsymbol{\delta} = \boldsymbol{N}_z\,\tilde{\boldsymbol{u}}_z, \\
\boldsymbol{\varepsilon}_e = \boldsymbol{B}_e\,\tilde{\boldsymbol{u}}_z + \boldsymbol{E}, \\
\boldsymbol{\sigma}_e = \mathbb{C}_d : (\boldsymbol{\varepsilon}_e - \boldsymbol{\varepsilon}^p_e) \quad (e \in \Omega_p).
\end{gathered}
\end{equation}
Equations (7) to (9) are algebraic identities under the elastic-complement condition of Section 2.6. For the same mesh, constitutive laws and increments, coupling (8) to the cohesive law and return mapping gives the same discrete equilibrium problem as (6). Cohesive softening can admit several solutions. Exact elimination preserves these solution branches; corresponding Newton iterates coincide in exact arithmetic when both schemes use the same starting iterate and step length. Their line searches can nevertheless select different step lengths (Section 3.3 and Appendix B.2).

Equation (8) is a discrete Green-operator representation of the cell problem. The finite element model and its periodic constraints define the operator, so no reference medium is required and the chosen discrete geometry is preserved. From a structural perspective, this is static condensation \cite{ref27} with the retained set selected by the support of the nonlinearity.

\hypertarget{macroscopic-stress-and-consistent-macroscopic-tangent}{%
\subsection{Macroscopic stress and consistent macroscopic tangent}\label{macroscopic-stress-and-consistent-macroscopic-tangent}}

The average stress (2) is linear in the loads of (6). It can therefore be written as

\begin{equation}
\boldsymbol{\Sigma} = \boldsymbol{C}_{\mathrm{hom}}\,\boldsymbol{E} + \frac{1}{|Y|}\left[\boldsymbol{Q}_\lambda\,\boldsymbol{\lambda} + \boldsymbol{Q}_p\,\boldsymbol{\varepsilon}^p\right],
\end{equation}
where \(\boldsymbol{C}_{\mathrm{hom}}\) is the elastic homogenised stiffness of the cell with undamaged interface, obtained from three elastic solutions, \(\boldsymbol{Q}_\lambda\) is the \(3 \times 2n_i\) matrix whose column \(k\) is \(|Y|\) times the average stress produced by a unit damage force at interface degree of freedom \(k\), and \(\boldsymbol{Q}_p\) is the analogous matrix for a unit plastic strain component at an integration point of \(\Omega_p\), which includes the direct term \(-\mathbb{C}_d : \boldsymbol{\varepsilon}^p\) of the eigenstrain in the element stress. Both influence matrices are by-products of the condensation: forming \(\boldsymbol{K}_{rr}^{-1}\boldsymbol{K}_{rz}\) in (7) already requires one solution of the \(r\) problem per process-zone degree of freedom, and the corresponding average stresses are obtained from those recovered fields, without an additional elastic factorisation. Equation (10) is exact for the discrete cell. It is the counterpart of the influence functions of transformation field analysis \cite{ref15}, with the inelastic variables of every interface pair and every integration point of \(\Omega_p\) kept individually.

At a converged increment the macroscopic tangent is obtained by differentiating (10) along the solution of the reduced problem,

\begin{equation}
\boldsymbol{C}_t = \frac{\mathrm{d}\boldsymbol{\Sigma}}{\mathrm{d}\boldsymbol{E}} = \boldsymbol{C}_{\mathrm{hom}} + \frac{1}{|Y|}\left[\boldsymbol{Q}_\lambda\,\frac{\mathrm{d}\boldsymbol{\lambda}}{\mathrm{d}\boldsymbol{E}} + \boldsymbol{Q}_p\,\frac{\mathrm{d}\boldsymbol{\varepsilon}^p}{\mathrm{d}\boldsymbol{E}}\right],
\end{equation}
where the sensitivities of the damage forces and of the plastic strains are obtained from the converged Jacobian of the reduced problem by one solve with three right-hand sides (Section 3.3 and Appendix B.2), using the cohesive tangent \(\partial\boldsymbol{t}/\partial\boldsymbol{\delta}\) of (A.2) to (A.4) and the consistent elastoplastic tangent of the return mapping. The tangent is consistent with the incremental algorithm, which is what a macroscopic Newton scheme needs; its accuracy is checked against finite differences in Appendix C.1.

\hypertarget{a-posteriori-verification-and-growth-of-the-process-zone}{%
\subsection{A posteriori verification and growth of the process zone}\label{a-posteriori-verification-and-growth-of-the-process-zone}}

The set \(\Omega_p\) is chosen before the path is known, so the method must certify that the choice was adequate. At every converged increment the fluctuation outside the zone is recovered by one back-substitution with the stored factorisation of \(\boldsymbol{K}_{rr}\),

\begin{equation}
\tilde{\boldsymbol{u}}_r = \boldsymbol{K}_{rr}^{-1}\left(\boldsymbol{f}_{E,r}\,\boldsymbol{E} - \boldsymbol{K}_{rz}\,\tilde{\boldsymbol{u}}_z\right),
\end{equation}
and the yield function (4) is evaluated with the elastic stress at every integration point of the ductile phase outside \(\Omega_p\). If \(f < 0\) at all of them, the full model, whose solution coincides with the reduced one up to this increment by (7) to (9), would have found those points elastic as well, and the reduced solution is the full solution at this increment too. The argument is inductive: as long as the check passes at every increment, the reduced response is exact along the whole path. The word certificate is used in this sense throughout: it states whether the reduced and the full solutions coincide; unlike the error estimators of certified reduced-basis methods, it does not bound the error when they do not. If \(f > 0\) at some point, the process zone no longer contains all the plasticity, the reduced result is exact up to the previous increment only, the zone is enlarged (a larger distance \(b\)), the condensation is redone and the path is recomputed. The check costs one back-substitution per increment and is reported with every result of Section 4. The same check applied to the elastic solution \(\tilde{\boldsymbol{u}}_z = \boldsymbol{G}_E\boldsymbol{E}\) tells whether the first point of the cell to yield lies inside the zone, which is the condition for the effective yield surface of Section 3.4 to be exact rather than an upper bound.

A computational advantage requires a sufficiently small retained set and active set. This is most likely when plasticity remains near the interface or other stress raisers. As plasticity spreads through the matrix, the retained set approaches all matrix degrees of freedom. The method then reduces to condensation of the stiff phase alone: the algebraic equivalence remains, but a time saving is no longer assured. Section 4 reports \(m/n\) for each case.

\hypertarget{relation-to-transformation-field-analysis-and-to-reduced-bases}{%
\subsection{Relation to transformation field analysis and to reduced bases}\label{relation-to-transformation-field-analysis-and-to-reduced-bases}}

Transformation field analysis \cite{ref15,ref16} writes the cell solution as the elastic response plus the response to the inelastic strain through elastic influence functions, exactly as (8) and (10) do, and then approximates the inelastic strain field as piecewise uniform over subdomains. Nonuniform transformation field analysis \cite{ref17,ref18,ref19} keeps the influence-function structure and approximates the inelastic field on plastic modes; projection-based reduced bases \cite{ref4,ref5,ref6} use reduced spaces for the cell response. In the snapshot-trained formulations considered in Section 4.10, the unknowns are modal amplitudes and accuracy depends on the retained space, training paths and reduced evolution equations \cite{ref20}. Section 4.10 compares those implementations on the same cell; it does not represent every TFA or certified reduced-basis formulation.

Process-zone condensation keeps the influence-function structure but does not approximate the inelastic field: every integration point of \(\Omega_p\) keeps its own plastic strain and every interface pair its own damage. The reduced system is small because the linear part has been eliminated, not because the inelastic part has been projected. The price is that \(m\) grows with the size of the process zone, whereas the number of modes of a reduced basis does not; the benefits are exactness relative to the discretisation, no training stage, and a check of the elastic-complement assumption at every increment. Reduced homogenisation with cohesive interfaces \cite{ref21,ref22,ref23} and enhanced TFA \cite{ref24,ref25} represent the interface and bulk inelastic fields on modes over subinterfaces and subdomains. With an elastic bulk, one subinterface per cohesive element and matching interpolation modes and shape functions, enhanced TFA reaches the same exactness on the interface \cite{ref24,ref25}; with an inelastic bulk its variables at the nodes of the support mesh differ from those of the finite element model, and its authors point to adaptive refinement where the inelastic processes concentrate as future work. The condensation extends the exactness to a plastic bulk inside the retained zone, and the certificate identifies when the prescribed zone must be expanded. The two approaches are complementary: a reduced basis could be applied inside the process zone if \(m\) became too large. That combination is not pursued here.

\hypertarget{numerical-implementation}{%
\section{Numerical implementation}\label{numerical-implementation}}

\hypertarget{discretization-of-the-cell}{%
\subsection{Discretization of the cell}\label{discretization-of-the-cell}}

The cell is meshed with a periodic butterfly (O-grid) pattern of four-node bilinear quadrilaterals conforming to the circular interface, with uniform layers of prescribed thickness in a band of width \(b\) around the interface, so that the process zone is a union of whole layers (Fig. 1). The interface is represented by node-to-node springs between the duplicated pairs, each with the tributary length of its node, which is the Newton--Cotes integration of a linear interface element. Elements use full \(2\times 2\) Gauss integration without volumetric correction. Plasticity is integrated by the backward-Euler radial return of the plane-strain von Mises model with the consistent elastoplastic tangent \cite{ref37}; only isotropic hardening is implemented. Since \(\operatorname{tr}\boldsymbol{\varepsilon}^p = 0\), the direct term of the eigenstrain in the element stress and in \(\boldsymbol{Q}_p\) reduces to \(-2\mu\,\boldsymbol{\varepsilon}^p\) with \(\mu\) the shear modulus of the matrix. The code is written in Python with sparse and dense linear algebra from NumPy and SciPy and runs the full-mesh reference and the reduced cell from the same mesh, laws and increment sequence.

\hypertarget{setup-of-the-condensation}{%
\subsection{Setup of the condensation}\label{setup-of-the-condensation}}

The condensation is set up once per cell, mesh and process zone (Appendix B.1). The factorisation of \(\boldsymbol{K}_{rr}\) is kept for the a posteriori check (12), and the operators \(\boldsymbol{G}\), \(\boldsymbol{G}_E\), \(\boldsymbol{Q}_\lambda\), \(\boldsymbol{Q}_p\) and \(\boldsymbol{C}_{\mathrm{hom}}\) are shared by every integration point of a macroscopic model that uses the same cell, mesh and retained zone. The setup cost is reported for each configuration in Section 4.

\hypertarget{reduced-incremental-problem}{%
\subsection{Reduced incremental problem}\label{reduced-incremental-problem}}

The reduced unknown is \(\tilde{\boldsymbol{u}}_z\). At increment \(k\), with \(\boldsymbol{E}_k\) given and the history \((\boldsymbol{\varepsilon}^p_{k-1}, p_{k-1}, \lambda_{\max,k-1})\) known, the residual of (8) is

\begin{equation}
\boldsymbol{r}(\tilde{\boldsymbol{u}}_z) = \tilde{\boldsymbol{u}}_z - \boldsymbol{G}_E\,\boldsymbol{E}_k - \boldsymbol{G}\left[\boldsymbol{f}_{p,z}\big(\boldsymbol{\varepsilon}^p_k(\tilde{\boldsymbol{u}}_z)\big) + \boldsymbol{N}_z^{\mathsf T}\boldsymbol{\lambda}\big(\boldsymbol{N}_z\tilde{\boldsymbol{u}}_z, D_k\big)\right] = \boldsymbol{0},
\end{equation}
Here \(\boldsymbol{\varepsilon}^p_k\) follows from the return mapping using (9), and \(D_k = \allowbreak D(\max(\lambda_{\max,k-1},\allowbreak\lambda_k))\) follows from (A.3). The implementation evaluates the equivalent force residual \(\boldsymbol{S}\boldsymbol{r}\) for convergence and line search, then applies \(\boldsymbol{G}\) to form the Newton correction.

Only damaged pairs and yielding elements contribute to the nonlinear force derivative. Newton correction therefore requires a dense solve on the \(k \le m\) active degrees of freedom, followed by products with \(\boldsymbol{G}\). The respective costs are \(O(k^3)\) and \(O(mk)\), whereas the full-cell reference factorises its sparse \(n \times n\) tangent at each iteration. Here \(k\) denotes active-set size, distinct from the increment index.

Elimination of the linear complement gives corresponding Newton corrections in exact arithmetic. Equality of subsequent iterates also requires matching step lengths; the line-search distinction is discussed in Appendix B.2. At convergence, (11) uses the same Jacobian to recover the macroscopic tangent. Appendix B.2 gives the solver details.

\hypertarget{effective-onset-surfaces}{%
\subsection{Effective onset surfaces}\label{effective-onset-surfaces}}

For a direction \(\boldsymbol{e}\) in macroscopic strain space and a non-negative amplitude \(a\), the elastic solution of the virgin cell is \(\tilde{\boldsymbol{u}}_z = a\,\boldsymbol{G}_E\boldsymbol{e}\). The deviatoric stress norm and the effective jump scale linearly with \(a\). The amplitudes at which plasticity first occurs in \(\Omega_p\) and damage first occurs on \(\Gamma_i\) are therefore available in closed form,

\begin{equation}
\begin{gathered}
a_y(\boldsymbol{e}) = \frac{\sqrt{2/3}\,\sigma_y}{\max_{g \in \Omega_p}\lVert\operatorname{dev}\boldsymbol{\sigma}_g(\boldsymbol{e})\rVert}, \\
a_d(\boldsymbol{e}) = \frac{\alpha}{\max_{a \in \Gamma_i}\lambda_a(\boldsymbol{e})},
\end{gathered}
\end{equation}
where the maxima run over the integration points of the zone and over the interface pairs for the unit-amplitude solution. The set of \(a_y\boldsymbol{e}\) and \(a_d\boldsymbol{e}\) over directions \(\boldsymbol{e}\) are the initial effective yield surface and the initial debonding surface in strain space; they map to stress space through \(\boldsymbol{\Sigma} = a\,\boldsymbol{C}_{\mathrm{hom}}\boldsymbol{e}\), which is exact up to the first nonlinear event. The implementation obtains the amplitudes by bracketing and bisection on the reduced elastic solution, which is the general machinery needed for surfaces after a pre-strain, and the closed form (14) is used as a check. The a posteriori check of Section 2.6 at \(a_y\) tells, direction by direction, whether the first yielding point lies in the zone; where it does not, \(a_y\) from the reduced cell is an upper bound of the true onset amplitude. As a complement, the amplitude at which plasticity first occurs in the nonlinear response, with the interface damage that precedes it, is obtained by bisection on the reduced Newton solution from the virgin state.

\hypertarget{reduced-feuxb2}{%
\subsection{Reduced FE²}\label{reduced-feuxb2}}

The macroscopic model uses plane-strain four-node quadrilaterals with \(2\times 2\) integration. Each macroscopic integration point stores its own cell history: the retained fluctuation \(\tilde{\boldsymbol{u}}_z\), macroscopic strain \(\boldsymbol{E}\), matrix plastic variables in the zone, and interface damage history. At each macroscopic Newton iterate, the cell receives \(\boldsymbol{E}\) and returns \(\boldsymbol{\Sigma}\) and \(\boldsymbol{C}_t\) from (10) and (11). These quantities supply the macroscopic residual and tangent. Failed iterations trigger step cutting. The condensation operators are shared by all points using the same cell and zone.

A virgin point uses an elastic shortcut if \(\boldsymbol{G}_E\boldsymbol{E}\) neither causes yielding in \(\Omega_p\) nor initiates interface damage. It returns \(\boldsymbol{C}_{\mathrm{hom}}\boldsymbol{E}\) and \(\boldsymbol{C}_{\mathrm{hom}}\) without an inner Newton solve. Both reduced and full-cell drivers use this shortcut. The full-cell driver otherwise solves (6) on all \(n\) degrees of freedom.

After convergence, the check (12) is evaluated at every macroscopic integration point. It uses the converged cell state at points with history and the elastic solution at virgin points. A failing thin-zone point is promoted to a wider zone, whose operators have also been prepared once. Transfer reconstructs the fluctuation using (8) and (12), restricts it to the wider set, copies plastic variables element by element and preserves interface damage. The two adaptation modes below determine how equilibrium is restored. Thus, each point can retain the thin zone until its loading history requires a wider one.

Two adaptation modes are implemented. In increment-repetition mode, all points return to the previous converged state, the flagged points receive that state mapped to the wide zone, and the increment is solved again, with at most three repetitions. In transfer mode, the flagged points also receive the previous certified history in the wide zone, but the macroscopic Newton iteration resumes from the current displacement. Cached cell solutions are reused only when the driver, committed history and strain are unchanged; other cell problems are solved again as needed. Both modes reproduce the full-cell structural response within the reported tolerances. Transfer reduces repeated work but generally changes the macroscopic iteration counts.

Tables C.2 and 5 report increment-repetition mode. The principal timing comparisons in Sections 4.5 and 4.6 use transfer mode, the default; Section 4.5 also reports repetition on the same machine. Appendix B.3 gives the algorithm.

\hypertarget{reference-solvers}{%
\subsection{Reference solvers}\label{reference-solvers}}

Three reference calculations serve different purposes. The first is the full-cell Newton solver in the same code, using the same mesh, laws, increments and tolerance. It tests the discrete equivalence of Section 2.4.

The second is an independent Ansys Mechanical APDL 2025 R1 cell model with the same geometry and constitutive constants. It uses the solver's own elements, cohesive and hardening laws, periodic constraints and three strain-control nodes, with small displacements and no viscous cohesive regularisation (Appendix A.3). Macroscopic stress is obtained from control-node reactions; its volume average is an internal consistency check.

The third resolves the microstructure of the perforated plate in Section 4.5, for two hole sizes. The quarter plate is tiled with cells. Inclusions lying entirely outside the hole with at least \(0.05L\) clearance retain their interfaces; inclusions cut by or nearly tangent to the hole are replaced by matrix. The constituent laws, constants and load increments match those of the cell model. The edge reaction is compared with FE².

The commercial solver also exported the assembled elastic stiffness and periodic and interface node pairs. These exports check the elastic data; they were not used to construct the condensed operator in this paper. Condensation directly from a commercial model is left to a follow-up.

\hypertarget{verification-against-a-full-field-reference}{%
\section{Verification against a full-field reference}\label{verification-against-a-full-field-reference}}

\hypertarget{test-cell-and-parameters}{%
\subsection{Test cell and parameters}\label{test-cell-and-parameters}}

The test cell is a square of side \(L = 28.3\) µm with a circular inclusion of radius \(R = 8.5\) µm (area fraction 0.283) in plane strain (Fig. 1). The constants are those of a measured pair, E-glass fibre in a DGEBA epoxy (jER 828 cured with triethylenetetramine at 100 °C), for which the elastic constants of both phases, the inelastic response of the resin and the strength of the interface in normal and shear loading were measured on the same system by the cruciform-specimen method \cite{ref28,ref29}, and the interfacial fracture energy by single-fibre fragmentation on E-glass in the same resin family \cite{ref30,ref31}. The fibre diameter of 17 µm in \cite{ref28} fixes \(R\), and \(L\) follows from the area fraction. The inclusion is 16 times stiffer than the matrix, so the matrix carries the inelastic deformation, and the interface strength of 57 MPa is close to the yield stress of the matrix, so decohesion and plasticity interact, which is the situation targeted by the method. The method itself does not depend on the values; they fix the regime, the strain range and the gain. All lengths are in µm and stresses in MPa, so that 1 MPa µm = 1 J/m² and forces per unit thickness are in µN/µm. The constants are listed in Table 1.

\begin{table*}[htbp]
\centering
\footnotesize
\caption{Geometry, material and interface constants of the test cell (E-glass/epoxy pair of [28--31]).}
\setlength{\tabcolsep}{5pt}
\begin{tabular}{>{\raggedright\arraybackslash}p{\dimexpr 0.275\linewidth-2\tabcolsep\relax}>{\raggedright\arraybackslash}p{\dimexpr 0.130\linewidth-2\tabcolsep\relax}>{\raggedright\arraybackslash}p{\dimexpr 0.253\linewidth-2\tabcolsep\relax}>{\raggedright\arraybackslash}p{\dimexpr 0.323\linewidth-2\tabcolsep\relax}}
\toprule
Quantity & Symbol & Value & Source \\
\midrule
Cell side, inclusion radius & $L$, $R$ & 28.33 µm, 8.5 µm (area fraction 0.283) & fibre diameter 17 µm \cite{ref28} \\
Inclusion: Young's modulus, Poisson's ratio & $E_s$, $\nu_s$ & 72 000 MPa, 0.22 & E-glass \cite{ref28} \\
Matrix: Young's modulus, Poisson's ratio & $E_d$, $\nu_d$ & 4550 MPa, 0.39 & neat resin, 100 °C cure \cite{ref28} \\
Matrix: yield stress, plastic modulus & $\sigma_y$, $H$ & 45 MPa, 1780 MPa ($E_t$ = 1279 MPa) & bilinear fit of the measured Ramberg--Osgood law \cite{ref28}, see text \\
Interface: strengths & $\sigma_{\max}$, $\tau_{\max}$ & 57 MPa, 68 MPa & cruciform test, quadratic criterion, $\tau_{\max}/\sigma_{\max} = 1.2$ \cite{ref28,ref29} \\
Interface: fracture energies & $G_{Ic}$, $G_{IIc}$ & 72 J/m\textsuperscript{2}, 120 J/m\textsuperscript{2} & fragmentation, E-glass/epoxy \cite{ref30} and E-glass/Epon 828 \cite{ref31} \\
Interface: decohesion separations, $2G_{Ic}/\sigma_{\max}$, $2G_{IIc}/\tau_{\max}$ & $\delta_n^c$, $\delta_t^c$ & 2.526 µm, 3.529 µm &  \\
Interface: onset ratio, mode weight & $\alpha$, $\beta$ & 0.01, 1.0 &  \\
Interface: initial stiffnesses, from (A.1) & $K_n$, $K_t$ & $2.26\times 10^3$, $1.93\times 10^3$ MPa/µm &  \\
\bottomrule
\end{tabular}
\end{table*}

The resin of \cite{ref28} follows the Ramberg--Osgood law

\begin{equation*}
\varepsilon = \sigma/E + a(\sigma/E)(\sigma/\sigma_0)^{n-1},
\end{equation*}

with \(a = 0.02\), \(n = 4\) and \(\sigma_0 = 20\) MPa; the bilinear law of Section 2.2 was fitted to it by least squares between 45 and 80 MPa, which gives \(\sigma_y = 45\) MPa and \(H = 1780\) MPa with a maximum deviation of 5 MPa. The interface strengths are the values of the quadratic criterion of \cite{ref28} for the 100 °C cure; the single-fibre pull-out test on the same system gave 31 to 40 MPa in normal and 57 to 60 MPa in shear at failure under combined stress, consistent with that envelope. The fracture energies of glass/epoxy interfaces measured by fragmentation and pull-out range from about 15 to 220 J/m² with the sizing and the test \cite{ref30,ref31}; the two values adopted are the friction-corrected arrest energy of E-glass in a stiff epoxy \cite{ref30} for mode I and the debonding toughness of E-glass in Epon 828 with friction and residual stress accounted for \cite{ref31} for mode II. Neither test separates the modes, so the assignment is a modelling choice; it fixes the softening branch, which within the strain range below has little influence, and not the onset.

The mesh has 4480 bilinear elements, 4673 nodes and 128 interface pairs; after the periodic elimination, \(n = 9214\). The process zone of width \(b = 0.1L = 2.83\) µm contains 896 matrix elements, which together with the interface pairs gives \(m = 2304\) and \(m/n = 0.250\). The characteristic length of the cohesive zone, \(E_d G_{Ic}/\sigma_{\max}^2 \approx 100\) µm, exceeds the perimeter of the inclusion (53 µm) and is about 240 times the spacing of the interface pairs (0.42 µm), so the cohesive process zone is resolved; it also explains why damage, once started, spreads over most of the interface, as in the equal biaxial path of Section 4.7. The condensation of Section 3.2 took 0.9 s with the solver and threads described below (between 0.9 and 1.2 s across the runs reported here, which is the run-to-run spread on this machine; 1.1 s at one thread). Unless stated otherwise, the times of this paper were measured on one desktop processor (Intel i5-13500, six performance cores) with the same six threads for the sparse direct solver of the reference (Intel MKL PARDISO) and for the dense BLAS of the reduced cell, on an otherwise idle machine; the reference solver of the implementation defaults to SuperLU, which is single-threaded and 7 to 14 times slower per factorisation on these matrices, and the ratios obtained with it are quoted where they help to show how much of a measured gain is the solver rather than the method. The elastic homogenised stiffness of the cell with undamaged interface is, in MPa and Voigt order \((11, 22, 12)\),

\begin{equation}
\boldsymbol{C}_{\mathrm{hom}} = \begin{bmatrix} 10010.5 & 5230.8 & 0 \\ 5230.8 & 10010.5 & 0 \\ 0 & 0 & 2173.2 \end{bmatrix},
\end{equation}
with normal-shear coupling entries of the order of \(10^{-11}\) MPa, as the square symmetry of the cell requires. The reduced and the full cell agree on \(\boldsymbol{C}_{\mathrm{hom}}\) to \(3.6\times 10^{-15}\) relative to its largest entry. With the interface pairs tied (the limit \(K_n, K_t \to \infty\)), \(C_{11}\) rises from 10010.5 to 13166.9 MPa: the adopted initial compliance of the interface lowers the axial modulus of this cell by 24 \%, an effect that a perfect-bond model would miss. The strain range of the tests below is 1 \%. The cell contains no matrix cracking, whereas the transverse failure of glass/epoxy laminae, which matrix cracking governs, may occur at strains of the same order or lower. Up to the onset of interface damage the response is therefore a prediction for this cell; beyond it, the paths test the reduction and do not predict transverse failure. Within 1 \% the plasticity of this matrix stays confined near the interface in most loading directions; Section 4.3 shows where and when it leaves the interface band.

Four modelling choices make the test cell a benchmark rather than a model of a structural lamina, and their expected effect on the method is stated here. First, the area fraction of 0.283 is lower than the fibre volume fractions of 0.5 to 0.65 of structural laminae; at higher fractions the ligaments between fibres narrow, plasticity leaves a thin interface band earlier and the gain of the reduction decreases, as Sections 4.3 and 4.8 indicate. Second, the cross-section is analysed in plane strain rather than in the generalized plane strain usual for transverse loading of unidirectional laminae; the latter adds one macroscopic degree of freedom, the uniform axial strain, and leaves the condensation unchanged. Third, thermal residual stresses from the cure are not included; they would enter (6) as one further linear load vector, again without changing the condensation, but they would shift the onset of debonding. Fourth, the fibres are in a square array; random arrangements require larger cells, which raise \(m/n\) (Section 4.8).

\hypertarget{reduced-versus-full-cell-along-macroscopic-strain-paths}{%
\subsection{Reduced versus full cell along macroscopic strain paths}\label{reduced-versus-full-cell-along-macroscopic-strain-paths}}

Nine strain paths were run in both modes with 40 equal increments and the same Newton tolerance: uniaxial strain in \(x\) and in \(y\), pure shear, equal biaxial, three tension-shear combinations, the combination \((1,1,1)\), each proportional with final norm \(\lVert\boldsymbol{E}\rVert = 0.010\), and one non-proportional path, \(E_{11}\) to 0.006 followed by \(2E_{12}\) to 0.008 at fixed \(E_{11}\). Table 2 gives, for each path, the onset of damage and of plasticity, the maximum relative error \(\lVert\boldsymbol{\Sigma}_{\mathrm{red}} - \boldsymbol{\Sigma}_{\mathrm{full}}\rVert/\lVert\boldsymbol{\Sigma}_{\mathrm{full}}\rVert\) over the increments before and after the onset of damage, the maximum difference between the two macroscopic tangents relative to \(\max|\boldsymbol{C}_t|\), the extent of plasticity at the end of the path in the full mesh, the first increment at which the a posteriori check (12) fails, and the wall-clock time of the Newton iterations in each mode.

\begin{table*}[htbp]
\centering
\scriptsize
\caption{Reduced versus full cell, nine macroscopic strain paths, process zone $b = 2.83$ µm, $m/n = 0.250$, 40 increments. Errors are maxima over increments; $s^*$ is the path parameter at damage onset (first pair reaching $\lambda = \alpha$) and $s_p$ at plasticity onset. Times are the Newton iterations of each mode, six threads on both sides, and exclude the one-time condensation (0.9 s) and the a posteriori check (0.065 to 0.068 s per path).}
\setlength{\tabcolsep}{3pt}
\resizebox{\linewidth}{!}{%
\begin{tabular}{lccccccccccc}
\toprule
Path & $s^*$ & $\lVert\boldsymbol{E}\rVert$ at $s^*$ & $s_p$ & Error before damage & Error after damage & Diff. in $\boldsymbol{C}_t$ & Plastic points at end / reach [µm] & Check fails at $s$ & Full [s] & Reduced [s] & Ratio \\
\midrule
Uniaxial $x$ & 0.493 & 0.0049 & 0.750 & 1.0e-14 & 1.6e-14 & 2.2e-14 & 424 / 1.13 & — & 3.16 & 0.81 & 3.9 \\
Uniaxial $y$ & 0.493 & 0.0049 & 0.750 & 9.4e-15 & 1.1e-14 & 2.0e-14 & 424 / 1.13 & — & 2.33 & 0.81 & 2.9 \\
Pure shear & — & — & 0.975 & 2.5e-03 & — & 2.6e-02 & 176 / 11.0 & 0.975 & 1.58 & 0.35 & 4.6 \\
Biaxial (1, 1, 0) & 0.519 & 0.0052 & — & 1.1e-14 & 6.6e-12 & 4.5e-14 & 0 / 0 & — & 2.11 & 0.93 & 2.3 \\
Tension-shear (1, 0, 1) & 0.660 & 0.0066 & 0.825 & 1.5e-14 & 5.0e-14 & 2.0e-14 & 262 / 0.89 & — & 2.08 & 0.61 & 3.4 \\
Tension-shear (1, 0, 0.5) & 0.545 & 0.0055 & 0.750 & 1.2e-14 & 1.2e-13 & 2.0e-14 & 378 / 1.52 & — & 2.29 & 0.74 & 3.1 \\
Tension-shear (0.5, 0, 1) & 0.923 & 0.0092 & 1.000 & 2.1e-14 & 1.8e-14 & 8.9e-15 & 72 / 0.32 & — & 1.68 & 0.40 & 4.2 \\
Combined (1, 1, 1) & 0.550 & 0.0055 & — & 1.4e-14 & 1.7e-14 & 3.8e-14 & 0 / 0 & — & 2.23 & 0.75 & 3.0 \\
Non-proportional, $x$ then shear & 0.408 & 0.0049 & 0.875 & 1.5e-14 & 2.6e-12 & 1.9e-14 & 150 / 0.72 & — & 2.33 & 0.69 & 3.4 \\
\bottomrule
\end{tabular}
}
\end{table*}

The first question is whether the thin zone preserves the full-cell response along different loading histories. In eight of the nine paths, the check passes at every increment. The reduced and full cells agree to \(2.1\times 10^{-14}\) before damage and \(6.6\times 10^{-12}\) after it; their macroscopic tangents agree to \(4.5\times 10^{-14}\) (Figs. 2 and 3). Supplementary Figs. S1 to S7 give the remaining stress curves. These discrepancies are consistent with round-off and the Newton tolerance. The accumulated Newton iterations coincide on the eight certified paths, ranging from 51 to 103 per path, with no step cuts. In equal biaxial loading and \((1,1,1)\), interface damage develops without matrix yielding within 1 \%, so these paths test the damage-force contribution alone.

\begin{figure*}[tbp]
\centering
\includegraphics[width=\textwidth]{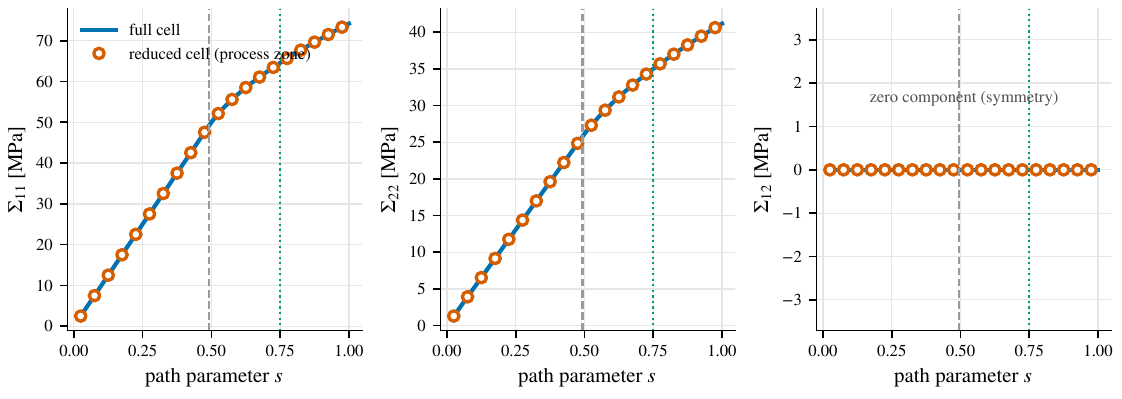}
\caption{Macroscopic stress components against the path parameter \(s\) in uniaxial strain \(x\), full cell (lines) and reduced cell (markers); damage and plasticity onsets marked.}
\end{figure*}

\begin{figure}[tbp]
\centering
\includegraphics[width=0.8\linewidth]{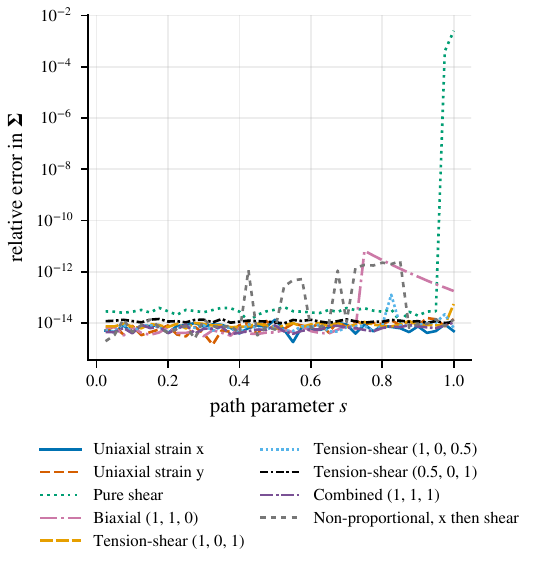}
\caption{Relative error in \(\boldsymbol{\Sigma}\) between reduced and full cell per increment, nine paths, logarithmic scale.}
\end{figure}

Pure shear tests the failure of the thin-zone assumption. The interface remains undamaged, but matrix yielding begins in the ligaments between inclusions, outside the zone. The check fails at \(s = 0.975\) (\(\lVert\boldsymbol{E}\rVert = 0.0098\)). The reduced model then keeps some yielded elements elastic, and the final stress and tangent errors reach \(2.5\times 10^{-3}\) and \(2.6\times 10^{-2}\), respectively (Fig. 4). Before the failing increment, the stress error stays at \(5.0\times 10^{-14}\). The check therefore detects the loss of admissibility without using the reference solution. The Newton counts differ in this path, 46 against 40, because the predictor line searches use different residual norms (Appendix B.2).

\begin{figure*}[tbp]
\centering
\includegraphics[width=\textwidth]{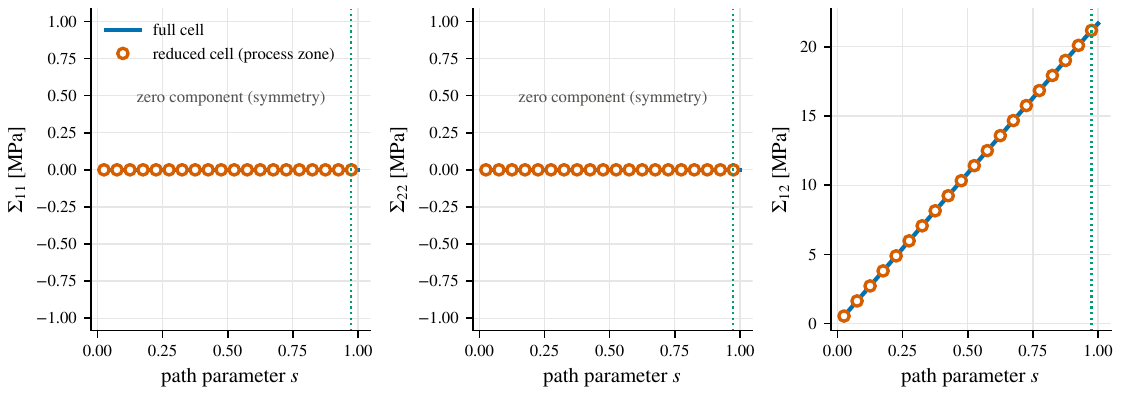}
\caption{Same as Fig. 2 for pure shear, showing the divergence after the a posteriori check fails at \(s = 0.975\).}
\end{figure*}

The second question is the computational cost of this agreement. Over the nine paths, the full-cell calculation takes 19.8 s and the reduced calculation 6.1 s, giving an aggregate ratio of 3.3, excluding checks and the one-time condensation of 0.9 s. With one thread in both solvers, the corresponding times are 27.3 s and 8.9 s, a ratio of 3.1. Per-path ratios range from 2.3 to 4.6 and are highest for small active sets. The nominal-zone pure-shear timing is included as a diagnostic, although that path is uncertified at its final increments. Thus, the aggregate ratio is not a measurement for nine fully certified paths.

The reference solver materially affects the ratio. With single-threaded SuperLU, the nine paths give a ratio of 10 on this machine, 92 s against 8.9 s: one SuperLU factorisation costs 0.08 s compared with 0.011 s for PARDISO. The smaller PARDISO-based gain is therefore the relevant comparison for an efficient sparse reference.

The instrumented profile in Table B.1 identifies the remaining costs. Sparse factorisation and back-substitution account for 9.6 s of the full-cell total of 21.3 s; tangent assembly takes 4.8 s and macroscopic tangent recovery 3.7 s. The reduced calculation takes 6.2 s excluding the check, distributed mainly across active-block solves, tangent recovery and zone evaluation. These measured totals give a ratio of 3.4. Removing all 8.4 s attributed to full-cell tangent assembly and recovery would reduce this ratio to 2.1. This subtraction is an optimistic estimate of the saving from further reference optimisation, not a measurement of an implemented solver. It suggests a cell-level gain of about two to three for this configuration; the measured uninstrumented ratio remains 3.3. Sections 4.5 and 4.6 examine whether a gain persists when the cells are coupled in FE².

\hypertarget{growth-of-the-process-zone}{%
\subsection{Growth of the process zone}\label{growth-of-the-process-zone}}

Two cases in which the a priori zone is too small were repeated with zones of increasing width until the check passes along the whole path (Table 3, Fig. 5 and Fig. 6). The zone of \(0.5L = 14.17\) µm covers the whole matrix, so the condensation degenerates into that of the inclusion alone, still exact.

\begin{figure}[tbp]
\centering
\includegraphics[width=0.8\linewidth]{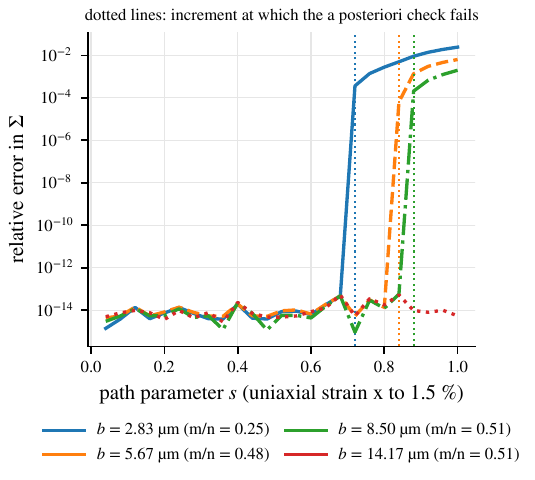}
\caption{Growth of the process zone, uniaxial strain \(x\) to 1.5 \%: error per increment for \(b\) = 2.83, 5.67, 8.50 and 14.17 µm, with the increment at which the check fails.}
\end{figure}

\begin{figure}[tbp]
\centering
\includegraphics[width=0.8\linewidth]{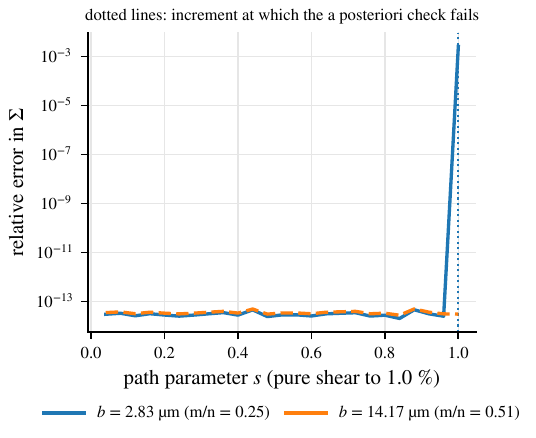}
\caption{Growth of the process zone, pure shear to 1 \%: \(b\) = 2.83 and 14.17 µm.}
\end{figure}

\begin{table*}[htbp]
\centering
\scriptsize
\caption{Growth of the process zone. Uniaxial $x$ to $\lVert\boldsymbol{E}\rVert = 0.015$ and pure shear to $\lVert\boldsymbol{E}\rVert = 0.010$, 25 increments each, with the solver and threads of Table 2. In the full mesh the plasticity reaches the far corner of the matrix in both cases (11.0 µm from the interface); the full Newton took 1.69 s and 0.98 s.}
\setlength{\tabcolsep}{3pt}
\resizebox{\linewidth}{!}{%
\begin{tabular}{lccccccccc}
\toprule
Case & $b$ [µm] & Elements in zone & $m$ & $m/n$ & Condensation [s] & Check fails at $s$ & Max. error while check passes & Max. error on path & Reduced [s] \\
\midrule
Uniaxial $x$, 0.015 & 2.83 & 896 & 2304 & 0.250 & 1.2 & 0.720 & 4.8e-14 & 2.4e-02 & 2.24 \\
 & 5.67 & 1936 & 4380 & 0.475 & 2.0 & 0.840 & 4.7e-14 & 6.3e-03 & 6.33 \\
 & 8.50 & 2136 & 4700 & 0.510 & 2.1 & 0.880 & 5.8e-14 & 2.0e-03 & 6.69 \\
 & 14.17 & 2176 & 4732 & 0.514 & 2.2 & — & 5.3e-14 & 5.3e-14 & 6.85 \\
Pure shear, 0.010 & 2.83 & 896 & 2304 & 0.250 & 1.2 & 1.000 & 4.6e-14 & 2.5e-03 & 0.21 \\
 & 14.17 & 2176 & 4732 & 0.514 & 1.8 & — & 5.0e-14 & 5.0e-14 & 0.68 \\
\bottomrule
\end{tabular}
}
\end{table*}

The error remains below \(7\times 10^{-14}\) while the check passes and grows only after a failing increment. Increasing the zone restores the full-cell response, as expected from Section 2.6. In uniaxial strain, plasticity begins near the interface at 0.75 \%, reaches the ligaments at 1.1 \% and spreads throughout the matrix by 1.3 \%. Between 1.1 \% and 1.5 \%, only the zone covering the matrix passes the check.

This recovery has a cost. At \(m/n = 0.51\), the uniaxial reduced solve takes 6.85 s against 1.69 s for the full cell. The active block has thousands of degrees of freedom, making a reduced basis within the retained zone a possible extension (Section 2.7). Even the inadequate thin zone takes 2.24 s under the extended uniaxial loading. By contrast, the thin-zone pure-shear run takes 0.21 s against 0.98 s, a ratio of 4.7, but fails the check at the final increment. Its certified, wider-zone counterpart is reported in Table 3. Zone adequacy and time saving must therefore be assessed separately.

\hypertarget{effective-onset-surfaces-and-macroscopic-tangent}{%
\subsection{Effective onset surfaces and macroscopic tangent}\label{effective-onset-surfaces-and-macroscopic-tangent}}

The initial debonding and yield surfaces were swept over 36 directions in each of the planes \(E_{11}\)--\(E_{22}\) (with \(2E_{12} = 0\)) and \(E_{11}\)--\(2E_{12}\) (with \(E_{22} = 0\)), 72 directions in all, by bisection on the reduced elastic solution with a relative tolerance of \(10^{-10}\) (about 38 evaluations per amplitude, each a product \(\boldsymbol{G}_E\boldsymbol{E}\) followed by the zone strains and the pair jumps). Three checks were made (Table 4): against the closed forms (14); against the same bisection on the elastic solution of the full mesh with the yield function evaluated at every integration point of the matrix; and the a posteriori check (12) at the yield amplitude.

\begin{table*}[htbp]
\centering
\footnotesize
\caption{Initial effective surfaces: checks over 72 directions.}
\setlength{\tabcolsep}{5pt}
\begin{tabular}{>{\raggedright\arraybackslash}p{\dimexpr 0.567\linewidth-2\tabcolsep\relax}>{\raggedright\arraybackslash}p{\dimexpr 0.413\linewidth-2\tabcolsep\relax}}
\toprule
Check & Result \\
\midrule
Bisection against closed form (14), max. relative difference & yield $9.0\times 10^{-11}$, debonding $8.3\times 10^{-11}$ \\
Debonding onset, reduced against full mesh, max. relative difference & 0 (all 72 directions) \\
Yield onset, reduced against full mesh, where the first yielding point lies in the zone & 0 (58 of 72 directions) \\
Directions in which the first yielding point lies outside the zone of 2.83 µm & 14 of 72, at 5.5 to 10.7 µm from the interface \\
A posteriori check at the yield amplitude against the full mesh & agrees in all 72 directions, no false positive \\
Time: reduced sweep (144 bisections) / full-mesh check & 0.9 s / 1.9 s \\
\bottomrule
\end{tabular}
\end{table*}

The initial debonding surface is reproduced in all sampled directions because every interface pair is retained. Initial yield onset agrees in 58 directions. In the other 14, the first yielding point lies outside the zone, so the reduced elastic onset is an upper bound. These directions include shear-dominated portions of the \(E_{11}\)--\(2E_{12}\) plane, 70° to 110° and 250° to 290°, and four biaxial directions at 30° and 60° from the axes in the \(E_{11}\)--\(E_{22}\) plane. First yielding occurs in ligaments 5.5 to 10.7 µm from the interface, outside the 2.83 µm band. The check identifies all 14 cases without false positives, indicating where the zone must be widened. Fig. 7 and Fig. 8 show the strain-space surfaces and their elastic stress-space images.

\begin{figure*}[tbp]
\centering
\includegraphics[width=\textwidth]{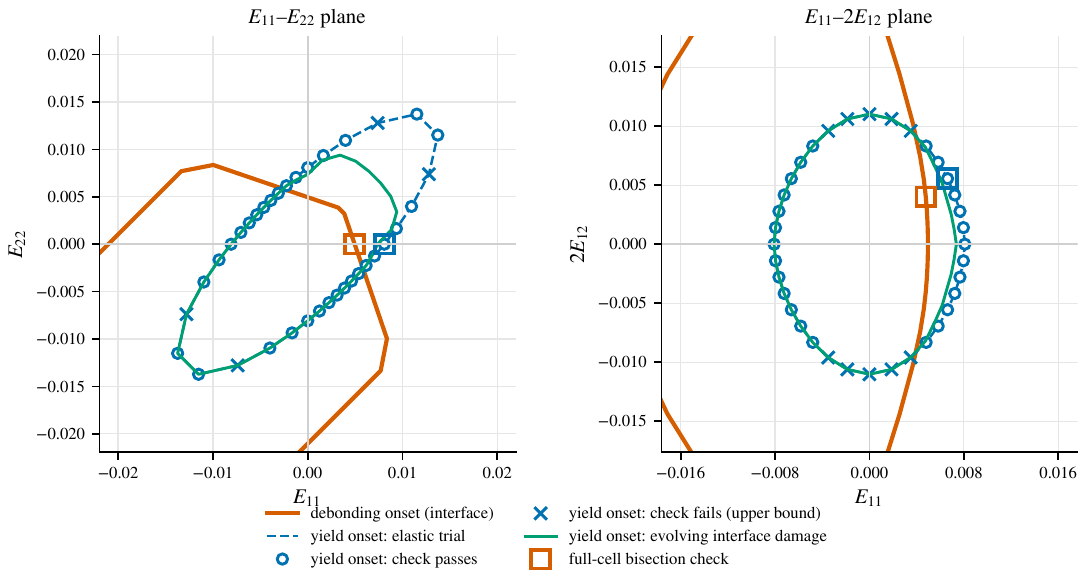}
\caption{Initial debonding and yield surfaces in the planes \(E_{11}\)--\(E_{22}\) and \(E_{11}\)--\(2E_{12}\); reduced cell and full-mesh check; directions in which the first yielding point lies outside the zone marked.}
\end{figure*}

\begin{figure*}[tbp]
\centering
\includegraphics[width=\textwidth]{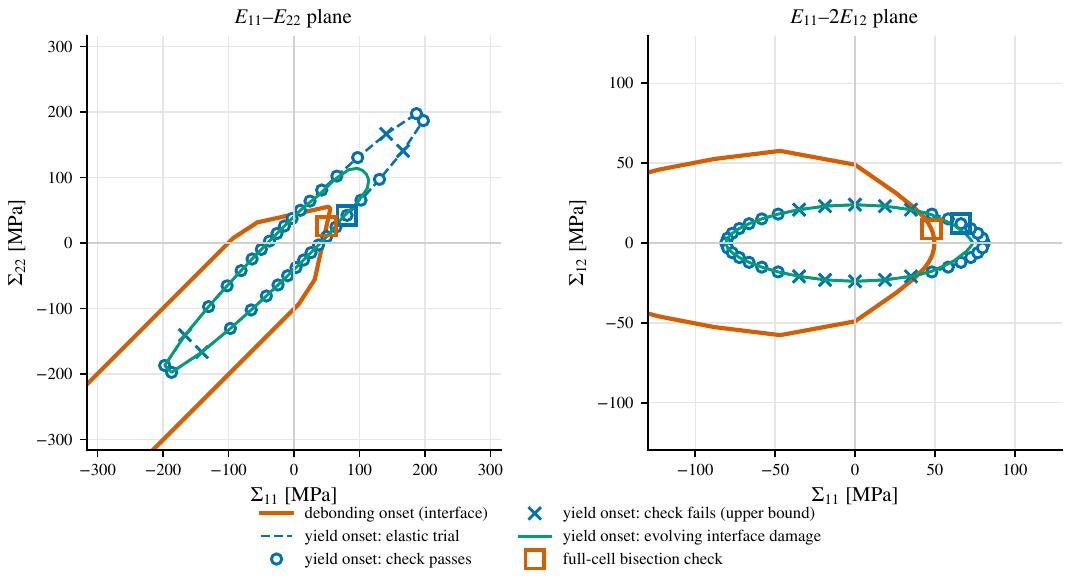}
\caption{The same surfaces in stress space through \(\boldsymbol{\Sigma} = a\,\boldsymbol{C}_{\mathrm{hom}}\boldsymbol{e}\).}
\end{figure*}

These are initial elastic surfaces. Once interface damage develops, stress redistribution can change the subsequent yield onset. In uniaxial strain, damage begins at \(\lVert\boldsymbol{E}\rVert = 0.0050\), and the nonlinear response yields at 0.0074, compared with 0.0081 for the undamaged elastic cell. For the tested tensile directions, interface damage precedes matrix yielding. In compression-dominated directions the order reverses: matrix yield amplitudes are 0.006 to 0.015, while shear-induced interface damage would begin at 0.015 to 0.057.

The consistent tangent (11) agrees with central finite differences of \(\boldsymbol{\Sigma}\) to the cancellation floor of the differences in the elastic, damaging and plastic regimes along three paths (Appendix C.1, Table C.1, Fig. C.1).

\hypertarget{reduced-feuxb2-on-a-plate-with-a-hole}{%
\subsection{Reduced FE² on a plate with a hole}\label{reduced-feuxb2-on-a-plate-with-a-hole}}

The plate provides a structural test of the cell solver and its adaptive zone. With fibres normal to the plane, it represents a transverse section of a unidirectional lamina with a cylindrical hole parallel to the fibres. Its size, 20 cells, makes a microstructure-resolved reference feasible (Table 6). It is a verification example rather than a prediction for a structural component.

The macroscopic model is a quarter of a square plate of side \(20L = 567\) µm, with a central hole of radius \(4L = 113\) µm, in plane strain. Symmetry conditions are imposed on \(x = 0\) and \(y = 0\), the top edge is free, and displacement \(U\) is prescribed at \(x = 567\) µm. Loading reaches 1.70 µm, or 0.30 \% nominal strain, in 16 equal increments. The macroscopic relative force tolerance is \(10^{-9}\) and the inner tolerance is \(10^{-11}\). Each integration point starts with the thin zone \(b = 2.83\) µm. Failed checks trigger transfer to \(b = 8.50\) µm (\(m = 4700\), \(m/n = 0.510\), Table 3). The tests below use increment repetition unless transfer mode is specified.

The thin-zone driver stores 16 771 numbers per point, compared with 33 000 for the wide zone and 81 025 for the full cell. The operators \(\boldsymbol{G}\), \(\boldsymbol{G}_E\), \(\boldsymbol{Q}_\lambda\) and \(\boldsymbol{Q}_p\) are shared by all points using a given zone. The kernel \(\boldsymbol{G}\) occupies 42 MB for the thin zone and 177 MB for the wide zone. Thus, the per-point history storage is about five times smaller for thin-zone points, in addition to the shared operators.

A coarse mesh of 8 elements, with 32 integration points, first checks the two FE² drivers (Appendix C.2, Table C.2, Fig. C.2). Loading is doubled to \(U = 3.40\) µm, or 0.60 \%, so that points farther from the hole reach damage and plasticity. The drivers reproduce the reaction, displacement and cell stresses within the reported macroscopic accuracy.

A run without adaptation shows why the check is needed. At 14 of the 32 points, matrix plasticity starts outside the thin band under shear-dominated loading near the plate diagonal. The check flags the first failing increment, from which the reduced response departs from the full one. Promoting these points to the wide zone and repeating the affected increments, four times in all, restores admissibility throughout loading. The reduced FE² is then about three times slower: 14 points require the wide zone and repeated increments add inner solves. The mechanism is the same as in cell-level shear: yielding in the ligaments requires a retained zone extending far from the interface. Structural cost consequently depends on the cell states visited and on the work spent solving and adapting them.

The structural response was then computed with the reduced driver only on a mesh of 192 elements (768 integration points), graded towards the hole (Table 5, Fig. 9, Fig. 10).

\begin{figure*}[tbp]
\centering
\includegraphics[width=\textwidth]{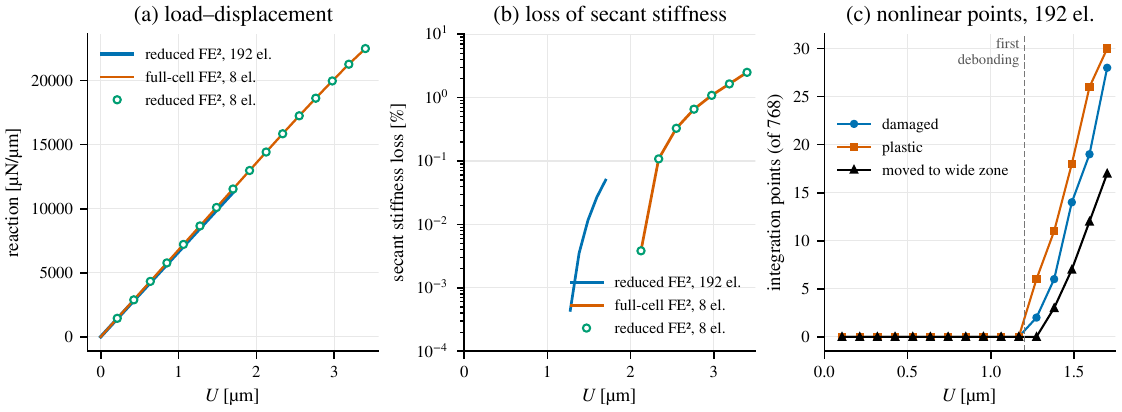}
\caption{Quarter plate with a hole in transverse tension, epoxy pair, increment-repetition mode: (a) reaction against prescribed displacement, reduced and full-cell FE² on the 8-element mesh (to \(U\) = 3.40 µm) and reduced FE² on the 192-element mesh (to 1.70 µm); (b) loss of secant stiffness \(1 - (F/U)/(F/U)_0\), logarithmic scale; (c) integration points of the 192-element mesh that are damaged, plastic and moved to the wide zone, with the increment of first debonding.}
\end{figure*}

\begin{figure*}[tbp]
\centering
\includegraphics[width=\textwidth]{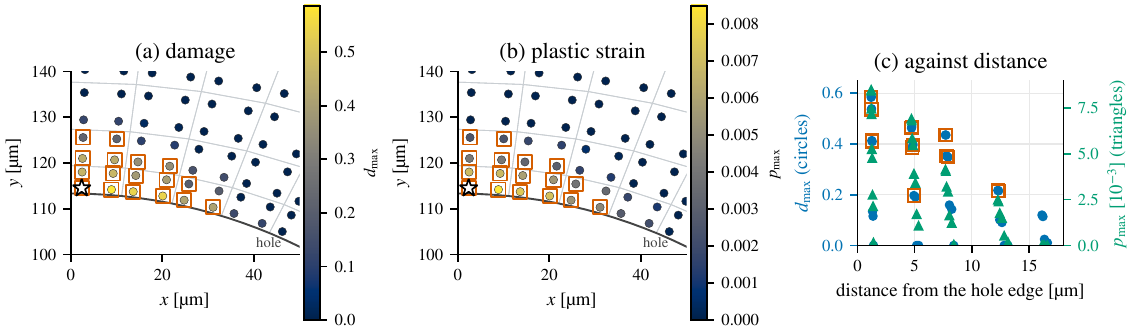}
\caption{State of the 192-element plate at \(U\) = 1.70 µm near the hole edge, increment-repetition mode: (a) largest interface damage and (b) largest accumulated plastic strain per macroscopic integration point, in the region \(0 \le x \le 50\) µm, \(100 \le y \le 140\) µm, with the macroscopic mesh, the point of first debonding (star) and the 17 points moved to the wide zone (open squares); (c) \(d_{\max}\) (circles, left axis) and \(p_{\max}\) (triangles, right axis) against the distance from the hole edge for all nonlinear points, with the points moved to the wide zone in open squares. All other points are undamaged and elastic.}
\end{figure*}

\begin{table*}[htbp]
\centering
\footnotesize
\caption{Reduced FE\textsuperscript{2} on the 192-element macroscopic mesh, plate with a hole, $U$ up to 1.70 µm (0.30 \%), with the increment repeated after each move to the wide zone; times with the solver and threads of Table 2.}
\setlength{\tabcolsep}{5pt}
\begin{tabular}{>{\raggedright\arraybackslash}p{\dimexpr 0.548\linewidth-2\tabcolsep\relax}>{\raggedright\arraybackslash}p{\dimexpr 0.432\linewidth-2\tabcolsep\relax}}
\toprule
Quantity & Value \\
\midrule
First debonding: $U^*$, nominal strain, increment & 1.203 µm, 0.212 \%, increment 12 \\
First debonding: integration point & $(x, y) = (2.38, 114.51)$ µm, 1.20 µm from the hole edge, polar angle 88.8° \\
First plasticity: $U$ & 1.275 µm (0.225 \%, increment 12) \\
At $U = 1.70$ µm: points damaged / plastic & 28 / 30 of 768 (3.6 \% / 3.9 \%) \\
Plastic reach from the hole edge & 13.1 µm (0.12 hole radii) \\
Largest interface damage over all cells & 0.586 \\
Points moved to the wide zone / increments repeated & 17 of 768 / 4 \\
Points uncertified at the end, largest $f$ outside the zones, time of the certificates & 0 of 768, $f_{\max} = -0.32$, 17.6 s \\
Reaction at the end / initial elastic slope & 11 268.5 µN/µm / 6631.9 µN/µm per µm \\
Macroscopic iterations / inner Newton iterations, total & 27 / 872 \\
Wall-clock time, reduced / full-cell FE\textsuperscript{2} & 83.8 s (38.6 s in the last increment) / 34.5 s \\
\bottomrule
\end{tabular}
\end{table*}

Debonding begins at the integration point nearest the hole edge, at 89° from the loading direction and 0.212 \% nominal strain. Plasticity follows in the same increment. This location is consistent with the strain concentration near the hole: the cell-level uniaxial damage and yield thresholds are about 0.50 \% and 0.74 \%, respectively, and the local strain concentration is about 2.5. The structural loading is multiaxial, so these cell thresholds provide an interpretation rather than a direct onset criterion.

The load-displacement curve loses stiffness after \(U^*\) without global softening (Fig. 9a). Secant stiffness falls by 0.05 \% at \(U = 1.70\) µm on the 192-element mesh and by 2.5 \% at \(U = 3.40\) µm on the coarse mesh (Fig. 9b). Thus, a nearly linear global curve coexists with increasing numbers of damaged and plastic points (Fig. 9c). Figure 10 locates these points and the 17 promotions to the wide zone. Plasticity extends only 13 µm from the hole edge, less than half a cell.

For increment repetition, the reduced calculation takes 83.8 s, including 38.6 s in the last increment and 17.6 s for the checks. The reported inner Newton count is 872; this iteration count differs from the number of cell solves, which includes repeated work. Checks at all points cost about 1.1 s per increment and dominate the elastic increments. The full-cell FE² takes 34.5 s, giving a full/reduced ratio of 0.41. It agrees with the reduced calculation to \(2\times 10^{-14}\) in reaction and \(8.5\times 10^{-13}\) in \(\boldsymbol{\Sigma}\), with identical macroscopic iteration counts. The reduced driver performs 708 inner solves against 380, at about 84 ms per solve against 57 ms. Wide-zone points dominate this cost: only 34 of 768 points are nonlinear at the final increment, and 17 of them require the wide zone. The elastic shortcut avoids inner Newton solves at the other 734 points in both drivers.

Transfer mode reduces the repeated work to 530 inner solves against 380 for the full cell. The full and reduced calculations take 34.6 s and 64.5 s, respectively, a ratio of 0.54. The reduced driver uses 33 macroscopic iterations against 27; the reaction and cell-stress differences are \(2.5\times 10^{-11}\) and \(1.3\times 10^{-9}\), respectively. These are close structural responses, but equality of iteration counts is no longer expected. Neither mode accelerates this epoxy example. Section 4.6 tests whether a material pair requiring fewer wide-zone points changes this balance.

Refinement near the hole tests the sensitivity of these structural quantities (Appendix C.3, Table C.3). The 24×18 and 32×24 meshes have first-ring sizes of 1.2 and 0.25 µm, compared with 6.1 µm on the original mesh. Debonding onset shifts by 1.7 \%, first plasticity moves one increment earlier, and the final reaction changes by less than 0.06 \%. The sampled nonlinear region nevertheless grows as points approach the hole edge. This observation raises a scale-separation issue for first-order homogenisation with a 28 µm cell, examined next against the resolved plate. Three meshes alone do not establish convergence or its absence.

The microstructure-resolved Ansys plate provides a different check (Section 3.6, Table 6, Fig. 11). The first plate contains 385 inclusions, 419 679 quadratic bulk elements, 24 640 interface elements and 1 308 418 nodes, giving 2.6 million equations. Four inclusions cut by or nearly tangent to the hole are replaced by matrix, leaving an epoxy ring at the edge. The same 16 increments reach \(U = 1.70\) µm. Solution takes 4.4 min on eight desktop cores; this is a record of computational cost, not a speed comparison with FE², which solves a homogenised problem.

\begin{figure*}[tbp]
\centering
\includegraphics[width=\textwidth]{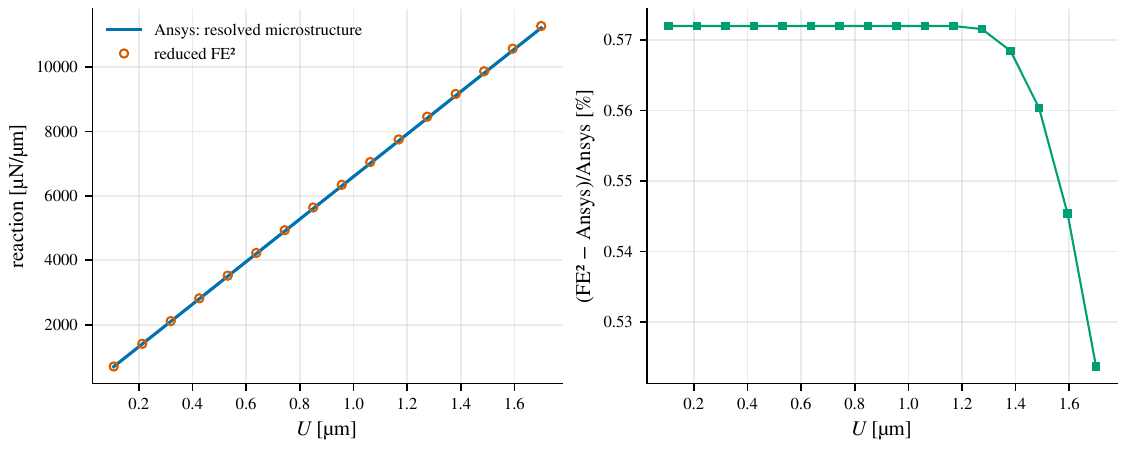}
\caption{Quarter plate with a hole of 4 cells: reaction against prescribed displacement, plate resolved down to the microstructure in Ansys and reduced FE², and relative difference per increment.}
\end{figure*}

A second \(30L \times 30L\) plate has a hole radius of \(10L = 283\) µm, improving the separation of scales. It contains 817 inclusions, 414 203 quadratic bulk elements, 52 288 interface elements and 1 345 402 nodes, or 2.7 million equations. Meshing uses blocks of \(6 \times 6\) cells, limiting each area to 36 holes. This reduces meshing time from the 1 h 46 min of the first plate's single-area procedure to 2 min; solution takes 3.8 min. The matching FE² mesh has 520 elements and 2080 points, with the same element size at the hole edge. Its reduced and full-cell drivers agree to \(2\times 10^{-11}\) in reaction.

\begin{table*}[htbp]
\centering
\footnotesize
\caption{Quarter plate with a hole, nominal strain up to 0.30 \%: plate resolved down to the microstructure in Ansys against the reduced FE\textsuperscript{2}, for a hole of 4 cells ($20L$ plate, $U$ up to 1.70 µm) and a hole of 10 cells ($30L$ plate, $U$ up to 2.55 µm). A dash denotes a plastic-reach value unavailable in the saved export.}
\setlength{\tabcolsep}{5pt}
\begin{tabular}{>{\raggedright\arraybackslash}p{\dimexpr 0.204\linewidth-2\tabcolsep\relax}>{\raggedright\arraybackslash}p{\dimexpr 0.226\linewidth-2\tabcolsep\relax}>{\raggedright\arraybackslash}p{\dimexpr 0.159\linewidth-2\tabcolsep\relax}>{\raggedright\arraybackslash}p{\dimexpr 0.230\linewidth-2\tabcolsep\relax}>{\raggedright\arraybackslash}p{\dimexpr 0.161\linewidth-2\tabcolsep\relax}}
\toprule
Quantity & Hole of 4 cells: Ansys, resolved & Reduced FE\textsuperscript{2} & Hole of 10 cells: Ansys, resolved & Reduced FE\textsuperscript{2} \\
\midrule
Initial slope $F/U$ [µN/µm per µm] & 6594.1 & 6631.9 (+0.57 \%) & 5642.0 & 5676.0 (+0.60 \%) \\
Reaction at the end [µN/µm] & 11 209.8 & 11 268.5 (+0.52 \%) & 14 385.3 & 14 466.4 (+0.56 \%) \\
Max. relative difference in the reaction over the 16 increments &  & 5.7e-3 &  & 6.0e-3 \\
First plasticity in the matrix, nominal strain & 0.28 \% & 0.225 \% & 0.263 \% & 0.225 \% \\
First interface damage, nominal strain & 0.30 \% (3 interface elements) & 0.212 \% & 0.281 \% (5 interface elements) & 0.229 \% \\
Where the interface first damages & pole of the inclusion nearest the top of the hole, 14 µm from the edge ($0.12a$) & integration point 1.2 µm from the edge & poles of the inclusion nearest the top of the hole, 14.6 µm from the edge ($0.05a$) & integration point 1.25 µm from the edge \\
At the end: plastic / damaged & 17 elements / 3 interface elements & 30 / 28 points of 768 & 70 elements / 18 interface elements & 106 / 82 points of 2080 \\
Plastic reach from the hole edge at the end & — & 13 µm & 24 µm & 27 µm \\
\bottomrule
\end{tabular}
\end{table*}

For the smaller hole, reactions agree to 0.57 \% along the path. The resolved plate is more compliant, consistent with its matrix ring and different discretisation. Local onsets differ more substantially. FE² predicts debonding at 0.21 \% and yielding at 0.23 \%, at the integration point nearest the edge. In the resolved plate, the matrix yields at 0.28 \% and three interface elements begin to damage at 0.30 \%, at an inclusion pole 14 µm from the edge. At the final load, FE² has 28 damaged points with \(D\) up to 0.59. The resolved model has three interface elements with reported damage energy below 0.03; damage energy and the dimensionless variable \(D\) are different quantities and are not compared numerically here.

The onset discrepancy cannot be attributed to condensation: the reduced solver reproduces full-cell FE² to \(10^{-12}\) on this mesh in repetition mode. Instead, it exposes a limitation of the homogenised description near the hole. The cell side is a quarter of the hole radius. A macroscopic integration point lies 1.2 µm from the edge, whereas the first resolved interface is 14 µm away. FE² applies the local strain to a periodic cell, although the actual edge neighbourhood has a different microstructure.

For the larger hole, the resolved-to-homogenised onset ratios fall from 1.25 to 1.17 for plasticity and from 1.41 to 1.23 for debonding. The final plastic reach differs by 3 µm, and reactions agree to about 0.6 \%. The first damaged interface remains 14 to 15 µm from the edge, a cell-scale distance, but this decreases from \(0.12a\) to \(0.05a\) relative to the hole radius. Improved scale separation therefore narrows the local onset gap without eliminating it. A boundary treatment accounting for the cell size or a higher-order homogenisation scheme may address this limitation; neither is tested here. The cell solver preserves the response of the homogenisation model and inherits its limits.

\hypertarget{a-second-material-pair-weak-interface-and-soft-matrix}{%
\subsection{A second material pair: weak interface and soft matrix}\label{a-second-material-pair-weak-interface-and-soft-matrix}}

The epoxy plate suggests that the fraction of nonlinear points requiring a wide zone affects structural cost. A second material pair tests a regime in which yielding remains closer to the interface. Single-fibre pull-out tests on aminosilane-sized E-glass in polypropylene with 2 \% maleated polypropylene provide an interfacial shear strength of 26.6 MPa and a critical energy release rate of 9.7 J/m² on the same specimen \cite{ref32,ref33}.

Matrix properties are taken from a polypropylene homopolymer data sheet: \(E = 2200\) MPa, \(\nu = 0.42\), \(\sigma_y = 26.7\) MPa and \(H = 319\) MPa. This is not a measurement of the pull-out specimen's matrix grade. Two further assumptions complete the interface law: \(\tau_{\max}/\sigma_{\max} = 1.2\), giving 22.2 MPa normal strength, and \(G_{Ic} = G_{IIc}\) because only one energy was measured. The resulting separations are \(\delta_n^c = 0.875\) µm and \(\delta_t^c = 0.729\) µm. Geometry, mesh, zones, increments and tolerances match Sections 4.1 to 4.5. Both material pairs use PARDISO for the full cell and six threads in both solvers (Table 7).

\begin{table*}[htbp]
\centering
\footnotesize
\caption{The two measured pairs on the same cell, mesh and plate: onsets, certificate, and full/reduced time ratios with the reference solver and threads of Table 2; the transfer to the wide zone does not repeat the increment.}
\setlength{\tabcolsep}{5pt}
\begin{tabular}{>{\raggedright\arraybackslash}p{\dimexpr 0.437\linewidth-2\tabcolsep\relax}>{\raggedright\arraybackslash}p{\dimexpr 0.274\linewidth-2\tabcolsep\relax}>{\raggedright\arraybackslash}p{\dimexpr 0.269\linewidth-2\tabcolsep\relax}}
\toprule
Quantity & E-glass/epoxy [28--31] & E-glass/polypropylene \cite{ref32,ref33} \\
\midrule
$\sigma_y/\sigma_{\max}$; $\sigma_y/E_m$ & 0.79; 0.99 \% & 1.20; 1.21 \% \\
Uniaxial $x$: damage onset $\lVert\boldsymbol{E}\rVert$; yield onset $s_p$; plastic reach at 1 \% & 0.0049; 0.750; 1.13 µm & 0.0026; 0.825; 1.12 µm \\
Pure shear: plastic reach at 1 \%; certificate with $b = 2.83$ µm & 11.0 µm; fails at $s = 0.975$ & no yielding; passes \\
Certificate over the nine paths & passes on 8 & passes on 9 \\
Max. error while certified & 6.6e-12 & 1.1e-11 \\
Cell, nine paths: full / reduced time, ratio & 19.8 s / 6.1 s, 3.3 & 23.1 s / 9.1 s, 2.5 \\
Plate, 8-element mesh: points in the wide zone; full / reduced, ratio & 14 of 32; 16.6 s / 33.0 s, 0.50 & 9 of 32; 54.8 s / 39.6 s, 1.38 \\
Plate, 192-element mesh: nonlinear points at the end, of which in the wide zone & 34, 17 (50 \%) & 154, 17 (11 \%) \\
Plate, 192-element mesh: damaged / plastic points at the end & 28 / 30 & 154 / 63 \\
Plate, 192-element mesh: inner solves full / reduced & 380 / 530 & 2857 / 3744 \\
Plate, 192-element mesh: full / reduced time, ratio & 34.6 s / 64.5 s, 0.54 & 176.2 s / 116.7 s, 1.51 \\
Plate, 192-element mesh: time per inner solve, full / reduced & 57.1 ms / 76.4 ms & 55.6 ms / 24.3 ms \\
Plate, 192-element mesh: max. error in reaction / displacement / $\boldsymbol{\Sigma}$ & 2.5e-11 / 1.2e-10 / 1.3e-9 & 1.1e-14 / 3.7e-14 / 6.0e-13 \\
\bottomrule
\end{tabular}
\end{table*}

For polypropylene, interface damage starts at about half the epoxy onset strain. Matrix yielding occurs later and remains near the interface along the tested cell paths. The thin-zone check passes on all nine paths, including pure shear, which causes no yielding within 1 \%. The reduced cell takes 9.1 s against 23.1 s for the full cell, a ratio of 2.5, with identical Newton counts and errors below \(1.1\times 10^{-11}\) (Supplementary Table S1).

The structural results show why the gain differs between the pairs. At the final plate increment, polypropylene has 154 nonlinear points out of 768, of which 17 require the wide zone. This is 11 \% of the nonlinear points, compared with 50 \% for epoxy. The mean reduced inner solve costs 24 ms against 56 ms for the full cell; for epoxy it costs 76 ms against 57 ms. Thus, the larger nonlinear region in polypropylene does not itself prevent a gain: most of its cell problems retain the cheaper thin zone.

The polypropylene reduced driver performs 31 \% more inner solves, 3744 against 2857, through transfers and additional macroscopic iterations. It still finishes 1.5 times faster, 117 s against 176 s, and reproduces reaction and cell stress to \(10^{-14}\) and \(6\times 10^{-13}\), respectively. Macroscopic counts are 50 against 42. At one thread in both solvers, the ratio is 1.45, with times of 165 s against 239 s. For epoxy, the full-cell driver is 1.9 times faster.

These two cases identify wide-zone occupancy as a useful diagnostic, rather than a general predictor of speed. The benefit also depends on the active sets, the loading history, certification and transfer work. For polypropylene, the structural ratio of 1.5 is below the cell-path ratio of 2.5; about a quarter of the reduced inner solves are additional solves associated with adaptation and macroscopic iteration. Recording the nonlinear and wide-zone point counts alongside cost helps identify this balance during a structural calculation.

\hypertarget{independent-validation-against-a-commercial-finite-element-reference}{%
\subsection{Independent validation against a commercial finite element reference}\label{independent-validation-against-a-commercial-finite-element-reference}}

The Python code and the commercial solver use different elements, different interface integration and different meshes, so this comparison tests the model and its implementation rather than the identity of Section 2.4. The Ansys model (Section 3.6) has 1667 quadratic plane-strain elements, 96 quadratic interface elements with 192 interface node pairs, element size 0.57 µm within 2.83 µm of the interface and 1.42 µm elsewhere; the periodic constraints and the three control nodes were verified on the mesh (39 node pairs per face pair). Eight paths were run with \(E_{\max} = 0.01\) on the dominant component and 50 substeps, the shear component being the tensorial \(E_{12}\) in this set, and the Python full cell of Section 4.1 (4480 bilinear elements, 128 node-to-node pairs) was run through the same strain history. The comparison uses the full Python cell. It also supports the reduced cell wherever its zone passes the incremental check; the eight loading histories here differ from the nine paths of Section 4.2, so that qualification is essential.

Appendix C.4 gives the cell comparisons (Tables C.4 and C.5, Figs. C.3 and C.4). The bonded elastic stiffnesses agree to \(3.2\times 10^{-4}\) despite the different meshes and element orders. Over the complete nonlinear paths, the global stress discrepancy ranges from \(3.4\times 10^{-4}\) to \(9.3\times 10^{-4}\). The regime-specific values are tabulated separately because plasticity can precede interface damage. The reaction-based and volume-averaged Ansys stresses differ by at most \(2.2\times 10^{-4}\), providing an internal consistency check rather than an uncertainty bound. The global comparisons support consistency of the implementations, including mixed-mode response. They do not establish pointwise equality of interface damage.

\hypertarget{cells-with-several-inclusions}{%
\subsection{Cells with several inclusions}\label{cells-with-several-inclusions}}

The gain of the reduction depends on \(m/n\), and a natural question is how \(m/n\) behaves when the cell contains several inclusions at the same area fraction. Four cells were compared at fixed mesh cost (about 4500 elements, the same element size on the interface): the production cell, cells with 4 and 9 inclusions in a square array built as \(2\times 2\) and \(3\times 3\) tilings of the production block, with \(R = 0.15L\) and \(0.10L\), and a staggered cell with 4 inclusions on a \(64\times 64\) pixel grid, the inclusions being defined by the element centroids and one of them sitting on the periodic boundary (Fig. 12). Each was run in both modes along uniaxial strain, pure shear and tension-shear \((1,0,1)\) to 1 \%, with the nominal zone \(b = 2.83\) µm around all interfaces, with the zone enlarged until the certificate passes, and with the geometrically similar zone \(b = R/3\) (Table 8; every run is listed in Supplementary Table S2).

\begin{figure*}[tbp]
\centering
\includegraphics[width=\textwidth]{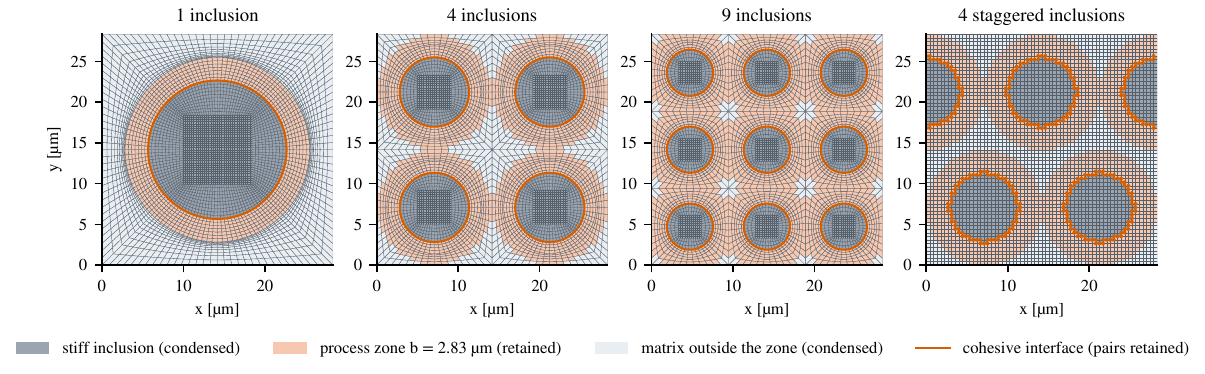}
\caption{Cells with 1, 4 and 9 inclusions in a square array and 4 staggered inclusions on a pixel grid, with the nominal process zone of 2.83 µm.}
\end{figure*}

\begin{table*}[htbp]
\centering
\scriptsize
\caption{Cells with several inclusions at area fraction 0.28, same mesh cost, $\lVert\boldsymbol{E}\rVert$ up to 0.010, 40 increments. $m/n$ for the nominal zone $b = 2.83$ µm and for the similar zone $b = R/3$; error while the certificate passes; time ratio full/reduced in uniaxial strain and in tension-shear with the smallest certified zone of the ladder that starts at $b = 2.83$ µm, with the solver and threads of Table 2.}
\setlength{\tabcolsep}{3pt}
\begin{tabular}{>{\raggedright\arraybackslash}p{\dimexpr 0.167\linewidth-2\tabcolsep\relax}>{\raggedright\arraybackslash}p{\dimexpr 0.084\linewidth-2\tabcolsep\relax}>{\raggedright\arraybackslash}p{\dimexpr 0.117\linewidth-2\tabcolsep\relax}>{\raggedright\arraybackslash}p{\dimexpr 0.102\linewidth-2\tabcolsep\relax}>{\raggedright\arraybackslash}p{\dimexpr 0.089\linewidth-2\tabcolsep\relax}>{\raggedright\arraybackslash}p{\dimexpr 0.144\linewidth-2\tabcolsep\relax}>{\raggedright\arraybackslash}p{\dimexpr 0.126\linewidth-2\tabcolsep\relax}>{\raggedright\arraybackslash}p{\dimexpr 0.151\linewidth-2\tabcolsep\relax}}
\toprule
Cell & $n$ & Interface pairs & $m/n$, $b = 2.83$ µm & $m/n$, $b = R/3$ & Max. error while certified & Ratio, uniaxial $x$ & Ratio, tension-shear \\
\midrule
1 inclusion, $R = 8.5$ µm & 9214 & 128 & 0.250 & 0.250 & 5.5e-14 & 3.9 & 3.5 \\
4 inclusions, square, $R = 4.25$ µm & 9726 & 256 & 0.508 & 0.316 & 1.0e-12 & 1.0 & 1.3 \\
9 inclusions, square, $R = 2.83$ µm & 10096 & 396 & 0.585 & 0.392 & 1.0e-13 & 0.8 & 1.4 \\
4 inclusions, staggered, pixel grid & 8830 & 320 & 0.637 & 0.373 & 4.6e-13 & 1.0 & 0.7 \\
\bottomrule
\end{tabular}
\end{table*}

All four cells preserve the full discrete response while certified: the stress error is at most \(1.0\times 10^{-12}\) and the elastic homogenised stiffness agrees to \(1.9\times 10^{-14}\). Increasing the inclusion count nevertheless increases \(m/n\) in these meshes. At fixed area fraction, total interface length grows as \(\sqrt{N}\), so a fixed-width band covers more matrix. With nine inclusions, the farthest matrix point is only 3.8 µm from an interface, making the 2.83 µm band nearly matrix-wide. Scaling the band as \(b = R/3\) reduces its width, but the number of duplicated interface nodes still increases at fixed interface element size.

The corresponding uniaxial time ratio falls from 3.9 for one inclusion to 0.8 to 1.0 for four and nine; tension-shear ratios range from 0.7 to 1.4 for the multiple-inclusion cells (Fig. 13). With \(b = R/3\), ratios are 1.9 to 3.1 on six certified paths, while three paths fail the check. The active set also grows: nine inclusions produce 2124 yielding points and 180 damaged pairs, entering the costly regime seen in Table 3.

\begin{figure*}[tbp]
\centering
\includegraphics[width=\textwidth]{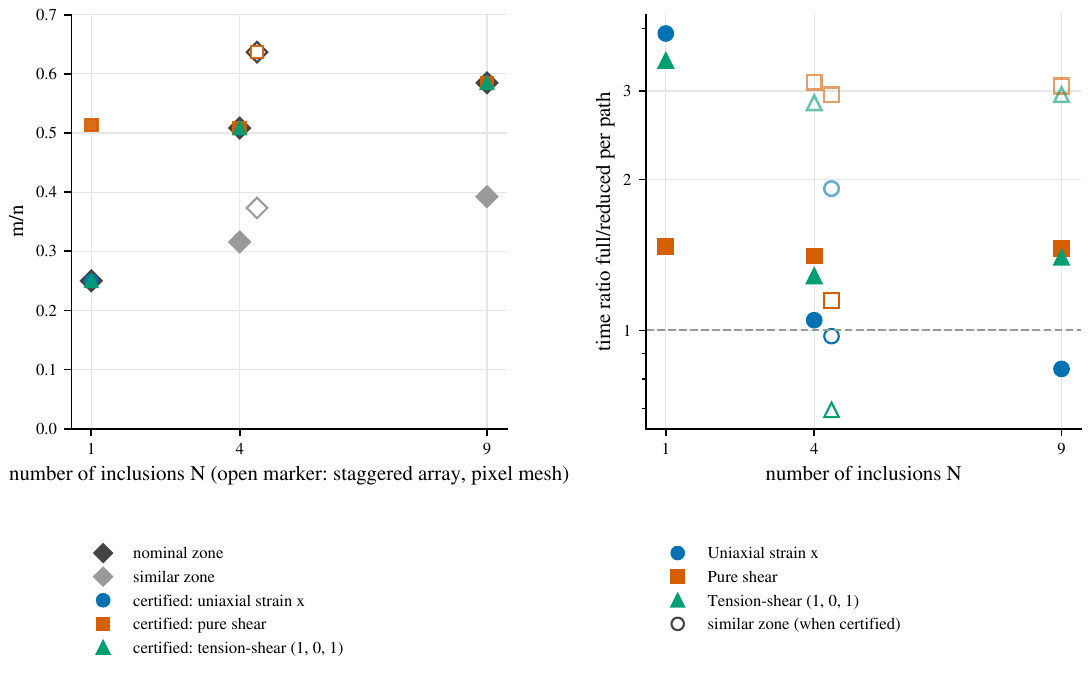}
\caption{\(m/n\) and time ratio full/reduced (PARDISO reference, six threads on both sides) against the number of inclusions for the nominal and the similar zone.}
\end{figure*}

Thus, adding inclusions does not itself improve reduction. Both the retained fraction and the active set matter, and their dependence on interface geometry must be measured for the chosen mesh. Shared condensation operators still reduce repeated setup in FE², but sharing does not guarantee a cheaper cell solve. The three-dimensional test next examines how interface discretisation changes this balance.

\hypertarget{three-dimensional-cell}{%
\subsection{Three-dimensional cell}\label{three-dimensional-cell}}

The three-dimensional test asks whether discrete equivalence and a computational advantage persist beyond plane strain. The cell is cubic, with side \(L = 20.88\) µm and a spherical inclusion of radius 8.5 µm. It uses the constituent properties of Section 4.1, with the fibre replaced geometrically by a particle. A \(16^3\) trilinear-hexahedral voxel grid assigns phases by element centroid; adjusting the radius gives a grid volume fraction of 0.295, compared with the target of 0.283.

The interface follows faces between different phases. Its duplicated nodes form 842 pairs, each using the nominal sphere normal and a tributary area equal to one quarter of its adjacent interface faces. The voxel interface has 1.56 times the sphere area, so it is not a quantitative replacement for a conforming interface mesh. It does allow reduced and full solvers to be compared on exactly the same discrete geometry. Matrix plasticity is three-dimensional von Mises plasticity, and the cohesive measure uses both tangential jump components.

The reference uses PARDISO at six threads. One \(16^3\) tangent factorisation takes 0.14 s, compared with 6.3 s for single-threaded SuperLU. Condensation uses dense Cholesky for \(\boldsymbol{K}_{rr}\) because repeated sparse back-substitutions become expensive at this size, 490 s against 6 s. Three paths reach \(\lVert\boldsymbol{E}\rVert = 0.010\) in 20 increments: uniaxial \(x\), pure shear \(2E_{xy}\) and tension-shear \((1,0,0,0,0,1)\). The nominal zone is \(b = 0.1L\); widths 0.15, 0.2, 0.3 and 0.5 \(L\) are tested when it fails (Tables 9 and 10, Fig. 14). Supplementary Table S3 lists the full ladders.

\begin{figure}[tbp]
\centering
\includegraphics[width=\linewidth]{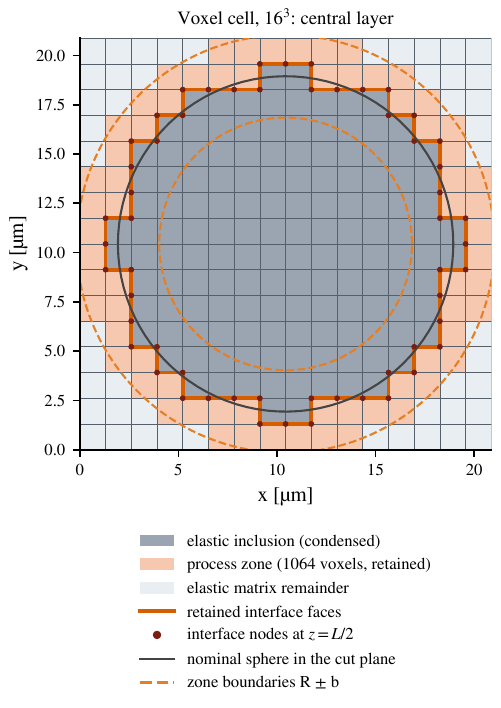}
\caption{Three-dimensional voxel cell, \(16^3\) voxels: cut through the central voxel layer showing the phases, the interface faces and the process zone of \(0.1L\).}
\end{figure}

\begin{table*}[htbp]
\centering
\scriptsize
\caption{Three-dimensional voxel cell, $16^3$ voxels, $n = 14\,811$: process zone against width, and the two finer grids (condensation at six threads; one reference factorisation each: PARDISO at six threads / SuperLU at one thread). The $24^3$ entry is a size and factorisation probe; no condensation time or nonlinear path results are recorded for that grid. A dash denotes a value unavailable in the saved results.}
\setlength{\tabcolsep}{3pt}
\begin{tabular}{lcccccc}
\toprule
Grid & $b/L$ & Voxels in zone & $m$ & $m/n$ & Condensation [s] & Reference factorization [s] \\
\midrule
$16^3$ & 0.10 & 1064 & 8469 & 0.572 & 8.5 & 0.14 / 6.3 \\
$16^3$ & 0.15 & 1664 & 10 017 & 0.676 & 8.7 &  \\
$16^3$ & 0.20 & 2152 & 11 157 & 0.753 & 9.5 &  \\
$16^3$ & 0.30 & 2736 & 12 162 & 0.821 & 10.3 &  \\
$16^3$ & 0.50 & 2888 & 12 324 & 0.832 & 10.8 &  \\
$20^3$ ($n = 27\,747$) & 0.10 & 2144 & 14 535 & 0.524 & 47.5 & 0.50 / 32.5 \\
$24^3$ ($n = 46\,875$) & 0.10 & 3656 & 22 707 & 0.484 & — & 1.5 / — \\
\bottomrule
\end{tabular}
\end{table*}

\begin{table*}[htbp]
\centering
\scriptsize
\caption{Three-dimensional cell, three paths, 20 increments, grids $16^3$ and $20^3$: certificate and error with the nominal zone $b = 0.1L$, and the smallest zone of the ladder for which the certificate passes along the whole path. Times are the Newton iterations of the path, reference (PARDISO, six threads) against reduced (six threads). Iterations are identical in the certified runs (57, 35, 53 on $16^3$; 61, 35, 48 on $20^3$).}
\setlength{\tabcolsep}{3pt}
\resizebox{\linewidth}{!}{%
\begin{tabular}{lcccccccccccc}
\toprule
Grid & Path & $b = 0.1L$: check fails at $s$ & Error while certified & Error at end of path & Reference [s] & Reduced [s] & Ratio & Certified zone $b/L$ & $m/n$ & Error, certified zone & Reduced [s] & Ratio \\
\midrule
$16^3$ & Uniaxial $x$ & 0.700 & 1.3e-12 & 9.5e-04 & 19.9 & 19.6 & 1.0 & 0.30 & 0.821 & 1.3e-12 & 32.9 & 0.6 \\
$16^3$ & Pure shear $2E_{xy}$ & 0.750 & 2.9e-13 & 3.2e-02 & 14.2 & 3.4 & 4.2 & 0.50 & 0.832 & 4.1e-13 & 12.3 & 1.2 \\
$16^3$ & Tension-shear $(1,0,0,0,0,1)$ & 0.800 & 2.4e-13 & 1.4e-03 & 18.6 & 12.8 & 1.5 & 0.30 & 0.821 & 2.4e-13 & 22.2 & 0.8 \\
$20^3$ & Uniaxial $x$ & 0.650 & 4.1e-13 & 4.8e-04 & 64.2 & 62.3 & 1.0 & 0.30 & 0.802 & 4.1e-13 & 95.8 & 0.7 \\
$20^3$ & Pure shear $2E_{xy}$ & 0.750 & 3.0e-13 & 2.7e-02 & 43.0 & 8.1 & 5.3 & 0.50 & 0.817 & 3.0e-13 & 34.0 & 1.3 \\
$20^3$ & Tension-shear $(1,0,0,0,0,1)$ & 0.750 & 3.3e-13 & 8.2e-04 & 52.2 & 32.7 & 1.6 & 0.30 & 0.802 & 3.5e-13 & 54.1 & 1.0 \\
\bottomrule
\end{tabular}
}
\end{table*}

Both three-dimensional grids preserve the discrete equivalence while certified. The reduced elastic homogenised stiffness agrees with the full-cell value to \(10^{-14}\); stress and tangent differences are at most \(1.3\times 10^{-12}\) and \(4\times 10^{-13}\), with identical Newton counts in the certified runs. On the \(16^3\) grid, the nominal zone fails on all three paths between \(s = 0.70\) and \(0.80\). The \(20^3\) failures are listed separately in Table 10. On the coarse voxel mesh, plasticity extends 0.28 to 0.40 \(L\) from the interface by 1 \% strain. The stepped interface concentrates stress at corners, while the nominal band is only 1.6 voxels thick.

The first change from two dimensions is the retained fraction. On \(16^3\), \(m/n\) rises from 0.25 in the two-dimensional test to 0.57 for the nominal shell and about 0.82 for an adequate shell. The geometric shell and ring occupy similar cell fractions, 0.21 and 0.19, but their meshes distribute retained nodes differently. Duplicated interface degrees of freedom account for 0.34 of \(n\) in the voxel cell, compared with 0.056 in two dimensions. Surface and volume nodes scale as \(h^{-2}\) and \(h^{-3}\), respectively, so the relative interface contribution decreases with refinement. The nominal retained fractions fall to 0.52 and 0.48 on the finer grids; this does not establish that the entire retained fraction scales linearly with \(h\).

The second change is cost. For the smallest certified zones, full/reduced ratios range from 0.6 to 1.3 per path, with aggregate ratios of 0.8 and 0.9 on the two grids. Some individual paths are faster, but neither grid gives an aggregate gain. The nominal-zone ratios of 1.0 to 1.6, and 4.2 and 5.3 in shear, refer to paths that fail the check. They cannot establish the cost of an admissible solution. Earlier comparisons with single-threaded SuperLU gave ratios of 4 to 18 for certified zones; those larger gains depend on the slower reference solver. With PARDISO, a large dense active block can cost as much as sparse solution of the full-cell tangent.

Storage is a separate limit. The dense kernel contains \(m^2\) numbers: 1.2 GB for a certified \(16^3\) zone and 4.1 GB for \(m = 22\,707\) in the \(24^3\) size probe. Larger cells may therefore require a reduced basis within the zone or a different operator representation. These results establish equivalence on the tested voxel meshes, not performance for production-quality three-dimensional discretisations.

\hypertarget{comparison-with-trained-reduced-order-models}{%
\subsection{Comparison with trained reduced-order models}\label{comparison-with-trained-reduced-order-models}}

This comparison examines the accuracy and cost of two snapshot-trained reductions on the same cell, mesh, constitutive laws and nine loading paths as Table 2. POD--Galerkin reduces the periodic fluctuation. NTFA \cite{ref17,ref18} reduces matrix plasticity, with either (a) an elastic interface or (b) interface damage retained pointwise. No hyper-reduction is used; the comparison therefore tests the reduced representations and their evolution laws. Appendix D defines the training sets A and B and the three truncation thresholds.

Table 11 reports the worst stress errors on training and untrained paths, the non-proportional path, and the accumulated Newton time. Fig. 15 follows uniaxial \(x\) and the non-proportional path. These calculations use one thread; the timed full-cell reference uses SuperLU. The condensation and trained models are compared within that campaign, whose timed operations differ from Table 2 as stated in the caption. These snapshot-trained implementations do not represent the analytical-mode methods of \cite{ref23,ref24,ref25}, with which no numerical comparison was performed.

\begin{figure*}[tbp]
\centering
\includegraphics[width=\textwidth]{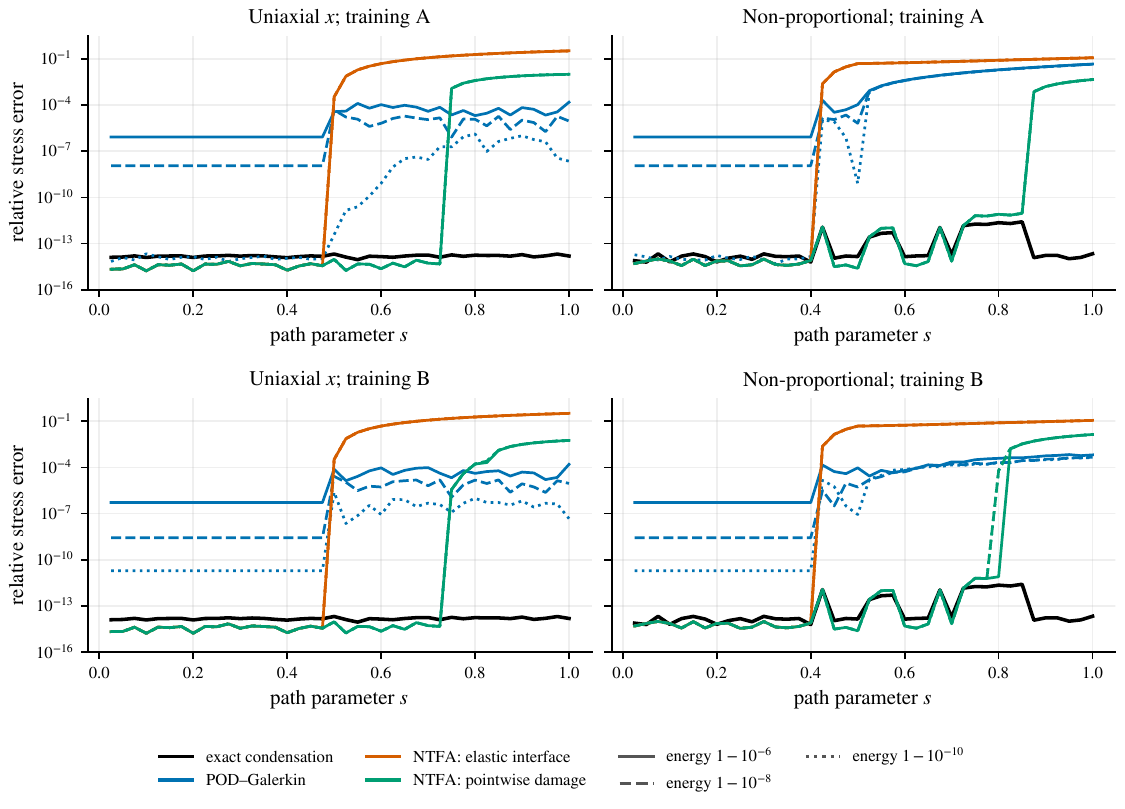}
\caption{Trained reduced models against the exact condensation: relative error in \(\boldsymbol{\Sigma}\) per increment on uniaxial strain \(x\) and on the non-proportional path, POD--Galerkin (training sets A and B), NTFA (a) and (b), and condensation.}
\end{figure*}

\begin{table*}[htbp]
\centering
\scriptsize
\caption{Trained reduced models against the exact condensation on the nine paths of Table 2: worst relative error in $\boldsymbol{\Sigma}$ over the increments, and Newton time summed over the nine paths. The offline cost is the full-cell snapshot generation (27 s for A, 60 s for B) against 1.1 s of condensation. The timed full-cell and condensation runs exclude macroscopic tangent recovery, and the condensation timings exclude the outside-zone check; Table 2 timings include macroscopic tangent recovery and report certification costs separately. Errors of the condensation are those of Table 2.}
\setlength{\tabcolsep}{3pt}
\begin{tabular}{>{\raggedright\arraybackslash}p{\dimexpr 0.151\linewidth-2\tabcolsep\relax}>{\raggedright\arraybackslash}p{\dimexpr 0.109\linewidth-2\tabcolsep\relax}>{\raggedright\arraybackslash}p{\dimexpr 0.102\linewidth-2\tabcolsep\relax}>{\raggedright\arraybackslash}p{\dimexpr 0.198\linewidth-2\tabcolsep\relax}>{\raggedright\arraybackslash}p{\dimexpr 0.124\linewidth-2\tabcolsep\relax}>{\raggedright\arraybackslash}p{\dimexpr 0.176\linewidth-2\tabcolsep\relax}>{\raggedright\arraybackslash}p{\dimexpr 0.120\linewidth-2\tabcolsep\relax}}
\toprule
Model & Training & Size & Worst error, trained paths & Worst error, untrained paths & Non-proportional path & Time, 9 paths [s] \\
\midrule
Full cell &  & $n = 9214$ &  &  &  & 87.4 (SuperLU, one thread) \\
Exact condensation, zone 2.83 µm & none & $m = 2304$ & 6.6e-12 on the 8 certified paths; 2.5e-3 on pure shear, flagged at $s = 0.975$ & same & 2.6e-12 & 5.3 \\
POD--Galerkin & A & $k$ = 8 / 15 / 26 & 8.3e-4 / 4.7e-5 / 1.3e-6 & 4.6e-2 / 4.5e-2 / 4.4e-2 & 4.6e-2 / 4.5e-2 / 4.4e-2 & 1.5 to 2.1 \\
POD--Galerkin & B & $k$ = 16 / 35 / 65 & 8.0e-4 / 3.1e-5 / 2.5e-6 & 1.0e-3 / 6.0e-4 / 5.1e-4 & 6.9e-4 / 4.6e-4 / 4.4e-4 & 1.7 / 2.6 / 3.9 \\
NTFA (a), elastic interface & A / B & $M$ = 13 / 32 & 0.52 / 0.52 & 0.36 / 0.27 & 0.12 / 0.11 & 0.02 \\
NTFA (b), hybrid, pointwise damage & A / B & $M$ = 13 / 32 & 1.3e-2 / 2.9e-2 & 6.7e-3 / 1.8e-2 & 4.5e-3 / 1.4e-2 & 1.5 \\
\bottomrule
\end{tabular}
\end{table*}

Three observations follow. First, agreement deteriorates after damage onset in the trained models, with the amount depending on the representation. NTFA with an elastic interface has errors of 11 to 52 \%. Retaining damage pointwise reduces them to about \(10^{-2}\), but adding plastic modes does not improve every result: one reported error rises from \(2.7\times 10^{-3}\) with one mode to \(2.9\times 10^{-2}\) with 32. This behaviour concerns the tested mode-averaged evolution law and should not be generalised to all NTFA formulations. For POD--Galerkin, enlarging the basis trained on set A does not remove the error on untrained paths. Set B improves those errors to about \(5\times 10^{-4}\) and the non-proportional error to about \(4\times 10^{-4}\), compared with \(3\times 10^{-12}\) for certified condensation.

Second, a training path can favour the trained model. POD--Galerkin reproduces its pure-shear training path to \(10^{-12}\), whereas the nominal-zone condensation reaches \(2.5\times 10^{-3}\) at the last increment. The condensation flags its loss of admissibility and can recover agreement by enlarging the zone. The tested trained implementations have no corresponding check for a new path.

Third, the trained models are cheaper per path in this comparison. POD--Galerkin and hybrid NTFA are cheaper than condensation by factors of 1.4 to 4, and elastic-interface NTFA by about 250. Their offline snapshot generation costs 25 to 55 times the condensation setup. For POD--Galerkin set B, the additional offline cost, 58.8 s of snapshots and 0.1 s of decomposition less the 1.1 s condensation setup, is recovered after \(N \approx 57.8\,\mathrm{s}/[(5.3 - t_{\mathrm{POD}})/9]\) paths of this length, where \(t_{\mathrm{POD}}\) is the nine-path time in Table 11: about 150, 190 and 370 paths for \(k\) = 16, 35 and 65. This estimate uses the one-thread timings of Table 11, which exclude macroscopic tangent recovery and the outside-zone check. Table 2 separately gives the PARDISO-based comparison with the full cell. The evidence therefore supports reduction without training and with a zone-admissibility check, but not a per-path cost advantage over the trained models.

\hypertarget{scope-of-the-verification}{%
\subsection{Scope of the verification}\label{scope-of-the-verification}}

The structural tests use two-dimensional plane strain, small strains and rate-independent laws with isotropic hardening. Each structural integration point uses a square cell with one circular inclusion; Section 4.8 separately tests multiple-inclusion cells. The three-dimensional test uses a voxel cell with one spherical inclusion. The condensation and influence matrices require a linear elastic complement and retained nonlinear variables, rather than these particular geometries. Their computational value nevertheless depends on the retained and active dimensions and on reference-solver cost.

For the tested epoxy cell, most paths up to about 1 \% strain keep plasticity near the interface, but shear can require a larger zone earlier. The missing matrix-cracking mechanism and the assumptions in Section 4.1 limit physical interpretation beyond damage onset. The cell is a reduction benchmark, not a complete model of transverse lamina failure.

Independent Ansys comparisons test the constitutive implementation with different discretisations (Table C.5). The resolved plates also expose differences between periodic first-order homogenisation and an explicit edge microstructure (Table 6). Condensation preserves the full-cell FE² response, so it cannot remove these modelling differences. This separates the verified algebraic reduction from the physical limits of the homogenisation model.

\hypertarget{conclusions}{%
\section{Conclusions}\label{conclusions}}

Process-zone condensation eliminates the linear elastic remainder of a periodic composite cell while preserving every cohesive interface variable and every plastic internal variable in the retained matrix zone. The cohesive law enters as a damage force on damaged pairs and plasticity as an eigenstrain; macroscopic stress and the consistent tangent follow from influence matrices. The reduction is exact relative to the full discretisation while the excluded matrix remains elastic throughout the loading history, and no training is involved. An incremental yield check verifies this condition using one back-substitution and identifies when the zone must grow.

For the E-glass/epoxy cell, eight of nine two-dimensional strain paths agree with the full cell to 1e-11 in stress and tangent, with identical Newton iterations. The certificate flags the pure-shear path when plasticity develops outside the nominal zone. Three paths on each of two three-dimensional voxel meshes require larger zones to recover agreement. The initial debonding surface agrees in all 72 sampled directions; the initial yield surface agrees in 58, and the certificate identifies the remaining 14. The consistent tangent agrees with central finite differences in the elastic, damaging and plastic regimes. These tests support conditional equivalence to the full discretisation.

Computational benefit varies with the retained and active dimensions. With equal thread counts and a PARDISO reference, the aggregate two-dimensional path calculation is 3.3 times faster, excluding setup and certification, with those additional costs reported separately. The adequate three-dimensional zones show no aggregate gain. On the 192-element plate, adaptation reproduces the full-cell structural response within solver tolerances. The full/reduced time ratio is 1.51 for glass/polypropylene and 0.54 for glass/epoxy, with points transferred to the wide zone without repeating the increment. The larger share of nonlinear points requiring the wide zone in the epoxy case, 50 \% against 11 \%, helps explain this difference alongside the active-set and adaptation costs. The check identifies points requiring promotion; the driver records their share at each increment. The certificate establishes adequacy, while the measured costs establish efficiency.

The independent Ansys cell and the Python full cell agree to below 1e-3 in global stress. Reduced FE² and the Ansys plates with resolved microstructure agree to approximately 0.6 \% in reaction. Local damage and plasticity onsets differ, exposing limitations of first-order homogenisation near the hole despite close agreement in the global response. The trained POD--Galerkin and NTFA models are cheaper per path in the tested comparison, with errors that depend on the training and representation. No numerical comparison with enhanced TFA \cite{ref24,ref25} was made; the published studies report efficiency gains for coarse subdivisions, but those results do not establish a direct cost ranking against the present implementation. Process-zone condensation offers an alternative without training when preserving the discrete response and checking zone adequacy are priorities.

The structural tests are two-dimensional, the three-dimensional tests use voxel cells, and most comparisons use strains up to 1 \%, with an extended uniaxial test to 1.5 \%. Under extended loading, plasticity spreads through the matrix, requiring a larger retained zone. Extended loading and multiple inclusions show how growth of the retained zone can remove the computational advantage. Frictional contact at a debonded interface and cohesive cracks through the matrix are localised nonlinearities that can join the retained set; these extensions and finite strains are left to a second paper.

\section*{CRediT authorship contribution statement}
\textbf{Leonilson Toffoli Pedrini}: Methodology, Software, Writing -- original draft. \textbf{Maycoln Conci}: Conceptualization, Validation, Writing -- review and editing. \textbf{Rafael Marin Ferro}: Validation (reference analyses), Resources, Writing -- review and editing.

\section*{Declaration of competing interest}
The authors declare that they have no known competing financial interests or personal relationships that could have appeared to influence the work reported in this paper.

\section*{Funding}
This research did not receive any specific grant from funding agencies in the public, commercial, or not-for-profit sectors.

\section*{Data availability}
The reduced-cell code and the data supporting this study are openly available on Zenodo at \url{https://doi.org/10.5281/zenodo.23512947} \cite{ref38}. The deposit comprises the Python implementation and its tests, the Ansys Mechanical APDL inputs and macros of the reference analyses, the saved numerical results, the timing logs and the scripts that redraw each figure, with code under the MIT licence and data under CC BY 4.0.

\section*{Declaration of generative AI and AI-assisted technologies in the manuscript preparation process}
During the preparation of this work, the authors used OpenAI Codex and Anthropic Claude to assist with manuscript preparation and code development. The authors reviewed and edited the outputs and take full responsibility for the content of the published article.

\section*{Acknowledgements}
The authors acknowledge the CAPES Journals Portal for providing access to the scientific literature.

\appendix
\makeatletter
\@addtoreset{figure}{section}
\@addtoreset{table}{section}
\makeatother
\renewcommand{\thesubsection}{\Alph{section}.\arabic{subsection}}
\hypertarget{constitutive-laws-and-correspondence-with-the-commercial-solver}{%
\section{Constitutive laws and correspondence with the commercial solver}\label{constitutive-laws-and-correspondence-with-the-commercial-solver}}

\hypertarget{cohesive-law}{%
\subsection{Cohesive law}\label{cohesive-law}}

The bilinear mixed-mode law of Section 2.2 is written as follows. The initial stiffnesses follow from the pure-mode branches,

\begin{equation}
\begin{gathered}
\delta_n^0 = \alpha\,\delta_n^c, \\
\delta_t^0 = \alpha\,\delta_t^c, \\
K_n = \frac{\sigma_{\max}}{\alpha\,\delta_n^c}, \\
K_t = \frac{\tau_{\max}}{\alpha\,\delta_t^c},
\end{gathered}
\end{equation}
and are in general different in the two modes. The mixed-mode state is measured by the dimensionless effective jump

\begin{equation}
\lambda = \sqrt{\left(\frac{\langle \delta_n \rangle_+}{\delta_n^c}\right)^2 + \left(\frac{\beta\,\delta_t}{\delta_t^c}\right)^2},
\end{equation}
where \(\langle\cdot\rangle_+\) is the positive part, so that closing of the interface does not contribute to damage. The scalar damage variable \(D \in [0,1]\) is a function of the largest value \(\lambda_{\max}\) reached in the history,

\begin{equation}
\begin{gathered}
D = 0 \ \ (\lambda_{\max} \le \alpha), \\
D = \frac{\lambda_{\max} - \alpha}{\lambda_{\max}\,(1-\alpha)} \ \ (\alpha < \lambda_{\max} < 1), \\
D = 1 \ \ (\lambda_{\max} \ge 1),
\end{gathered}
\end{equation}
and is therefore irreversible. The tractions are

\begin{equation}
\begin{gathered}
t_n = (1-D)\,K_n\,\delta_n \ \ (\delta_n \ge 0), \\
t_n = K_n\,\delta_n \ \ (\delta_n < 0), \\
t_t = (1-D)\,K_t\,\delta_t.
\end{gathered}
\end{equation}
In compression the normal stiffness is the undamaged one, which provides finite normal resistance to closure without a separate contact formulation. In pure mode I the law reduces to the bilinear branch \(t_n = \sigma_{\max}(\delta_n^c - \delta_n)/(\delta_n^c - \delta_n^0)\) for \(\delta_n^0 \le \delta_n \le \delta_n^c\), and in pure mode II to its tangential analogue; the fracture energies are \(G_{Ic} = \sigma_{\max}\delta_n^c/2\) and \(G_{IIc} = \tau_{\max}\delta_t^c/2\). Unloading from a damaged state is secant to the origin, so the interface stores no residual separation.

\hypertarget{return-mapping-in-plane-strain}{%
\subsection{Return mapping in plane strain}\label{return-mapping-in-plane-strain}}

The out-of-plane strain is zero and the plastic strain is deviatoric, so the plastic strain is stored by its in-plane components \((\varepsilon^p_{11}, \varepsilon^p_{22}, 2\varepsilon^p_{12})\) with \(\varepsilon^p_{33} = -(\varepsilon^p_{11} + \varepsilon^p_{22})\), and the stress has four components, including the out-of-plane one. With the elastic trial stress \(\boldsymbol{\sigma}^{\mathrm{tr}} = \mathbb{C}_d : (\boldsymbol{\varepsilon}_k - \boldsymbol{\varepsilon}^p_{k-1})\), its deviator \(\boldsymbol{s}^{\mathrm{tr}}\) and \(f^{\mathrm{tr}} = \lVert\boldsymbol{s}^{\mathrm{tr}}\rVert - \sqrt{2/3}\,(\sigma_y + H p_{k-1})\), the backward-Euler radial return for linear isotropic hardening is available in closed form \cite{ref37},

\begin{equation}
\begin{gathered}
\Delta\gamma = \frac{\langle f^{\mathrm{tr}} \rangle_+}{2\mu + \tfrac{2}{3}H}, \\
\boldsymbol{n} = \frac{\boldsymbol{s}^{\mathrm{tr}}}{\lVert\boldsymbol{s}^{\mathrm{tr}}\rVert}, \\
\boldsymbol{\sigma} = \boldsymbol{\sigma}^{\mathrm{tr}} - 2\mu\,\Delta\gamma\,\boldsymbol{n}, \\
\boldsymbol{\varepsilon}^p_k = \boldsymbol{\varepsilon}^p_{k-1} + \Delta\gamma\,\boldsymbol{n}, \\
p_k = p_{k-1} + \sqrt{\tfrac{2}{3}}\,\Delta\gamma,
\end{gathered}
\end{equation}
For a plastic trial state, \(f^{\mathrm{tr}} > 0\), the consistent elastoplastic tangent is

\begin{equation}
\begin{gathered}
\mathbb{C}^{ep} = \kappa\,\boldsymbol{1}\otimes\boldsymbol{1} + 2\mu\,\theta\,\mathbb{I}_{\mathrm{dev}} - 2\mu\,\bar{\theta}\,\boldsymbol{n}\otimes\boldsymbol{n}, \\
\theta = 1 - \frac{2\mu\,\Delta\gamma}{\lVert\boldsymbol{s}^{\mathrm{tr}}\rVert}, \\
\bar{\theta} = \frac{1}{1 + H/(3\mu)} - (1 - \theta),
\end{gathered}
\end{equation}
Here \(\kappa\) and \(\mu\) are the matrix bulk and shear moduli, \(\boldsymbol{1}\) is the second-order identity and \(\mathbb{I}_{\mathrm{dev}}\) is the three-dimensional deviatoric projector. The tensor norm includes \(s_{33}\) and twice the squared shear component. The tangent is restricted to in-plane stress and engineering-strain components, so the plastic shear update stored by the code is \(2\Delta\gamma n_{12}\). For \(f^{\mathrm{tr}} \le 0\), the elastic branch gives \(\Delta\gamma = 0\) and \(\mathbb{C}^{ep} = \mathbb{C}_d\) directly; no direction \(\boldsymbol{n}\) is needed when the trial deviator vanishes.

\hypertarget{correspondence-with-the-commercial-solver}{%
\subsection{Correspondence with the commercial solver}\label{correspondence-with-the-commercial-solver}}

The Ansys model of Section 3.6 uses eight-node quadratic plane-strain elements (PLANE183) in both phases, six-node interface elements (INTER203) generated by splitting the mesh along \(\Gamma_i\), the bilinear cohesive-zone material option TB,CZM with the BILI law of Section 2.2 and its constants \((\sigma_{\max}, \delta_n^c, \tau_{\max}, \delta_t^c, \alpha, \beta)\), bilinear isotropic hardening (TB,BISO), periodic constraint equations between opposite boundary nodes with the macroscopic strain imposed through three control nodes, small displacements, automatic time stepping with a minimum of 50 substeps per path and no viscous regularisation of the cohesive law. The bilinear stress-strain table of the solver is parameterised by the tangent modulus \(E_t = EH/(E+H)\) rather than by \(H\); the two descriptions are the same law. The cohesive law (A.1) to (A.4) is the BILI option of TB,CZM \cite{ref36} with the same six constants, so the two codes integrate the same law with different interface discretisations: quadratic six-node interface elements in Ansys and node-to-node springs with the tributary length of each node in the Python code (Section 3.1).

\hypertarget{algorithms-and-cost}{%
\section{Algorithms and cost}\label{algorithms-and-cost}}

\hypertarget{steps-of-the-condensation-setup}{%
\subsection{Steps of the condensation setup}\label{steps-of-the-condensation-setup}}

The setup is done once per cell, mesh and process zone and consists of: a sparse factorisation of \(\boldsymbol{K}_{rr}\); \(m\) back-substitutions to form \(\boldsymbol{K}_{rr}^{-1}\boldsymbol{K}_{rz}\) and three more for \(\boldsymbol{K}_{rr}^{-1}\boldsymbol{f}_{E,r}\); the dense Schur complement \(\boldsymbol{S}\) and its inverse \(\boldsymbol{G}\); the influence matrix \(\boldsymbol{G}_E\); the matrices \(\boldsymbol{Q}_\lambda\) and \(\boldsymbol{Q}_p\), read off from the average stress of the same \(m\) solutions; and \(\boldsymbol{C}_{\mathrm{hom}}\) from three elastic solutions of the full cell. The factorisation of \(\boldsymbol{K}_{rr}\) is kept for the a posteriori check (12). The kernel \(\boldsymbol{G}\) holds \(m^2\) numbers and sets the memory of the method (Sections 4.5 and 4.9). In three dimensions \(\boldsymbol{K}_{rr}\) is factorised by dense Cholesky instead, because the fill-in of the sparse factor makes the \(m\) back-substitutions prohibitive (Section 4.9).

\hypertarget{reduced-newton-iteration}{%
\subsection{Reduced Newton iteration}\label{reduced-newton-iteration}}

The Jacobian of (13), \(\boldsymbol{I} - \boldsymbol{G}\,\partial(\boldsymbol{f}_{p,z} + \boldsymbol{N}_z^{\mathsf T}\boldsymbol{\lambda})/\partial\tilde{\boldsymbol{u}}_z\), differs from the identity only in the columns of the active degrees of freedom, and the Newton correction is obtained from a dense solve on the active block alone, followed by matrix-vector products with \(\boldsymbol{G}\). A bisection line search and step cutting are used in both schemes with the same relative tolerance on the residual, \(10^{-11}\) of \(\lVert\boldsymbol{f}_E\boldsymbol{E}\rVert\). At the predictor step the line search compares residual norms that include the \(r\) part in the full scheme and not in the reduced one, which explains the different iteration counts of the pure-shear path in Section 4.2. A pair enters the active set once \(D > 0\) and remains in it whatever the sign of \(\delta_n\); under closure the normal component of its damage force vanishes, because \(t_n = K_n\delta_n\) by (A.4), and only the tangential component remains. The macroscopic tangent (11) is obtained at convergence from the same Jacobian by one dense solve with the three columns of \(\boldsymbol{G}_E\) plus the explicit dependence of the trial strain on \(\boldsymbol{E}\) as right-hand sides, followed by the chain rule through the cohesive tangent of (A.2) to (A.4) and the consistent elastoplastic tangent (A.6). No arc-length control was used in the reported cell tests, which required no step cuts. These calculations do not assess continuation through snap-back.

\hypertarget{one-increment-of-the-reduced-feuxb2}{%
\subsection{One increment of the reduced FE²}\label{one-increment-of-the-reduced-feuxb2}}

One increment of the reduced FE², from the converged state at \(U_{k-1}\) to \(U_k\), proceeds as follows.

\begin{enumerate}
\def\labelenumi{\arabic{enumi}.}
\item
  Save the previous converged displacement, cell histories and zone assignments. At each macroscopic Newton iterate, compute \(\boldsymbol{E}_g\) at every point. A virgin point satisfying the elastic-shortcut conditions returns \(\boldsymbol{C}_{\mathrm{hom}}\boldsymbol{E}_g\) and \(\boldsymbol{C}_{\mathrm{hom}}\). Otherwise solve (13) from the committed history in the current zone and recover (10) and (11). Assemble the macroscopic residual and tangent and update displacement. Failed inner or macroscopic iterations trigger step cutting.
\item
  After macroscopic convergence, commit the trial states provisionally and evaluate (12) at every point. Points with history use their converged state; virgin points use the elastic solution.
\item
  If all points pass, accept the increment. Otherwise map the previous certified histories of failing thin-zone points to the wide zone using (8) and (12). Copy matrix plastic variables element by element and retain interface damage. A virgin point remains virgin in the wider driver.
\item
  In repetition mode, restore the previous displacement and histories, retain the promotions and solve the increment again, with at most three repetitions. In transfer mode, retain the current displacement and restore the previous certified histories, including wide-zone histories mapped in any earlier transfer round of this increment. Resume macroscopic Newton. Reuse a cached cell solution only if its driver, history object and strain are unchanged; solve all other cell problems again as needed. Recheck after convergence, with at most three transfer rounds.
\item
  Record any points still failing after adaptation or the round limit. An uncertified increment is not an exact full-cell result; equivalence is established only through the preceding certified increment. No such remaining point occurred in the reported adaptive FE² cases.
\end{enumerate}

\hypertarget{cost-profile-of-the-cell}{%
\subsection{Cost profile of the cell}\label{cost-profile-of-the-cell}}

The profile quoted in Section 4.2 wraps each stage of the two schemes with exclusive timers, so that a stage called inside another is not counted twice, on the nine paths of Table 2 with the solver and threads of Section 4.1; the timers add about 1 s to the reference. Table B.1 gives the breakdown. Sparse factorisation and back-substitution take 9.6 s of the 21.3 s of the reference (45 \%); tangent assembly over the whole mesh and the macroscopic tangent take another 8.4 s, which a reference that stored the matrices of the elastic elements could partly avoid. In the reduced cell the dense solves on the active block, the macroscopic tangent and the evaluation of the laws in the zone take comparable shares, and the check outside the zone, which is the a posteriori certificate of Section 2.6, takes 0.5 s.

\begin{table*}[htbp]
\centering
\footnotesize
\caption{Cost profile of the nine paths of Table 2: exclusive time per stage summed over the paths, in seconds, six threads on both sides; the one-time condensation (1.2 s) is excluded.}
\setlength{\tabcolsep}{5pt}
\begin{tabular}{>{\raggedright\arraybackslash}p{\dimexpr 0.468\linewidth-2\tabcolsep\relax}>{\raggedright\arraybackslash}p{\dimexpr 0.237\linewidth-2\tabcolsep\relax}>{\raggedright\arraybackslash}p{\dimexpr 0.274\linewidth-2\tabcolsep\relax}}
\toprule
Stage & Full cell & Reduced cell \\
\midrule
Sparse factorization & 7.08 & 0.00 \\
Tangent assembly & 4.78 & 0.00 \\
Macroscopic tangent & 3.66 & 1.68 \\
Sparse back-substitution & 2.50 & 0.00 \\
Dense active-set solve & 0.00 & 1.78 \\
Zone evaluation & 0.00 & 1.23 \\
Internal force & 1.06 & 0.00 \\
Return mapping & 1.01 & 0.43 \\
Mesh evaluation & 0.53 & 0.00 \\
Outside-zone check & 0.00 & 0.51 \\
other (Newton loop and path driver) & 0.15 & 0.49 \\
Active-set tangent & 0.00 & 0.48 \\
Strains & 0.47 & 0.08 \\
Cohesive law & 0.08 & 0.07 \\
Total & 21.31 & 6.76 \\
\bottomrule
\end{tabular}
\end{table*}

\hypertarget{complementary-verification}{%
\section{Complementary verification}\label{complementary-verification}}

\hypertarget{consistent-tangent-against-finite-differences}{%
\subsection{Consistent tangent against finite differences}\label{consistent-tangent-against-finite-differences}}

The consistent tangent (11) was compared with central finite differences of \(\boldsymbol{\Sigma}(\boldsymbol{E}_k \pm h\boldsymbol{e}_j)\), \(h = 10^{-7}\), restarted from the converged history at \(\boldsymbol{E}_{k-1}\), at 16 increments of each of three paths (Table C.1, Fig. C.1).

\begin{figure}[tbp]
\centering
\includegraphics[width=0.8\linewidth]{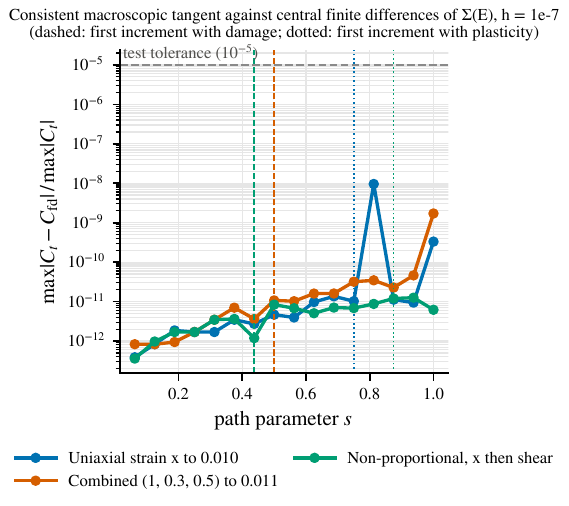}
\caption{Consistent macroscopic tangent against central finite differences along three paths, relative difference per increment with the regimes marked.}
\end{figure}

\begin{table*}[htbp]
\centering
\footnotesize
\caption{Consistent macroscopic tangent (11) against central finite differences, maximum relative difference per regime.}
\setlength{\tabcolsep}{5pt}
\begin{tabular}{>{\raggedright\arraybackslash}p{\dimexpr 0.321\linewidth-2\tabcolsep\relax}>{\raggedright\arraybackslash}p{\dimexpr 0.148\linewidth-2\tabcolsep\relax}>{\raggedright\arraybackslash}p{\dimexpr 0.256\linewidth-2\tabcolsep\relax}>{\raggedright\arraybackslash}p{\dimexpr 0.256\linewidth-2\tabcolsep\relax}}
\toprule
Path & Elastic & Damage, no plasticity & Damage and plasticity \\
\midrule
Uniaxial $x$ to 0.010 & 3.5e-12 & 1.4e-11 & 9.6e-09 \\
Combined (1, 0.3, 0.5) to 0.011 & 7.0e-12 & 3.5e-11 & 1.7e-09 \\
Non-proportional, $x$ then shear & 3.6e-12 & 8.7e-12 & 1.3e-11 \\
\bottomrule
\end{tabular}
\end{table*}

Differences are of order \(10^{-11}\) in the damage regime and \(10^{-9}\) with plasticity. The larger values occur near yield transitions, where central differences sample different one-sided responses. With an inner Newton tolerance of \(10^{-11}\) and the stated perturbation, the results support consistency of the implemented tangent. A perturbation-size study would be needed to isolate a finite-difference cancellation floor.

\hypertarget{equivalence-of-the-feuxb2-drivers-on-a-coarse-mesh}{%
\subsection{Equivalence of the FE² drivers on a coarse mesh}\label{equivalence-of-the-feuxb2-drivers-on-a-coarse-mesh}}

The equivalence of the two drivers was tested on a coarse macroscopic mesh of 8 elements (32 integration points), with the same macroscopic code and the load doubled to \(U = 3.40\) µm (0.60 \%) so that this coarse mesh, whose points lie farther from the hole, reaches damage and plasticity, comparing increment by increment the reaction, the displacement field, \(\boldsymbol{\Sigma}\) at every integration point, the iteration counts and the time (Table C.2, Fig. C.2).

\begin{figure}[tbp]
\centering
\includegraphics[width=\linewidth]{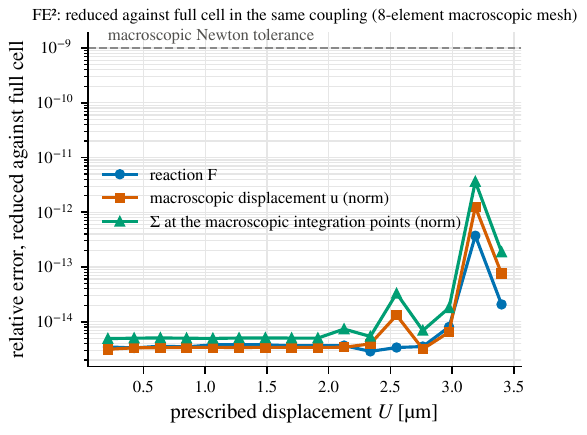}
\caption{Reduced against full-cell FE² on the 8-element macroscopic mesh: relative error in reaction, displacement and \(\boldsymbol{\Sigma}\) per increment, increment-repetition mode; the peak at \(U\) = 2.98 µm is the increment in which the two drivers take different numbers of macroscopic iterations (Table C.2).}
\end{figure}

\begin{table*}[htbp]
\centering
\footnotesize
\caption{Reduced against full-cell FE\textsuperscript{2} on the 8-element macroscopic mesh, 16 increments to $U = 3.40$ µm, with the increment repeated after each move to the wide zone (Section 3.5); times with the solver and threads of Table 2.}
\setlength{\tabcolsep}{5pt}
\begin{tabular}{>{\raggedright\arraybackslash}p{\dimexpr 0.526\linewidth-2\tabcolsep\relax}>{\raggedright\arraybackslash}p{\dimexpr 0.454\linewidth-2\tabcolsep\relax}}
\toprule
Quantity & Value \\
\midrule
Max. relative error in the reaction, increments 1 to 13 / increment 14 / increments 15 and 16 & 2.3e-14 / 2.5e-10 / 1.3e-13 \\
Max. relative error in the macroscopic displacement (norm), same increments & 6.5e-14 / 4.3e-10 / 8.8e-13 \\
Max. relative error in $\boldsymbol{\Sigma}$ at the integration points (norm), same increments & 1.5e-13 / 8.4e-10 / 1.7e-12 \\
Macroscopic iterations per increment, full / reduced & 1 (increments 1 to 9), 3, 4, 3, 4, 3 / 4, 3, 4 (increment 14 differs) \\
Points damaged / plastic at the end & 15 / 21 of 32 \\
Points moved to the wide zone / increments repeated & 14 of 32 / 4 \\
Points uncertified at the end, time of the certificates & 0 of 32, 0.8 s \\
Wall-clock time, full / reduced & 16.1 s / 47.6 s (ratio 0.34) \\
Inner Newton solves, full / reduced (including the repeated increments) & 257 / 467 \\
Mean time per inner Newton solve, full / reduced & 60 ms / 100 ms (ratio 0.60) \\
Condensation of the thin and of the wide zone, once, shared by all points & 1.2 s and 2.2 s \\
\bottomrule
\end{tabular}
\end{table*}

In repetition mode, the two drivers agree closely at 15 of 16 increments and use identical substeps. At increment 14, the reduced driver needs four macroscopic iterations against three. Round-off near the \(10^{-9}\) convergence threshold changes the stopping iteration; reaction and cell-stress differences reach \(2.5\times 10^{-10}\) and \(8.4\times 10^{-10}\). They fall back to about \(10^{-13}\) at the next increment. Damage and plasticity onsets, promoted points and final damage coincide. An earlier laptop run with SuperLU gave identical iteration counts throughout, illustrating sensitivity of the stopping test to arithmetic.

Default transfer mode follows a different macroscopic sequence. On this mesh with the Table 2 solver, reduced and full drivers use 39 and 34 macroscopic iterations. Their largest differences over the 16 increments, \(2.5\times 10^{-10}\) in reaction and \(8.4\times 10^{-10}\) in cell stress, again occur at increment 14 and coincide with the repetition-mode values.

The repetition-mode timing is 47.6 s reduced against 16.1 s full. Four repeated increments and the extra macroscopic iteration add 210 cell solves, 467 against 257. The wide zone is required at 14 of 32 points, and mean time per inner solve is 100 ms reduced against 60 ms full. Both repeated work and expensive wide-zone solves therefore contribute to the slowdown. These measured costs support the mechanism discussed in Section 4.5.

\hypertarget{refinement-of-the-macroscopic-mesh}{%
\subsection{Refinement of the macroscopic mesh}\label{refinement-of-the-macroscopic-mesh}}

Both FE² drivers were run in transfer mode on the refined macroscopic meshes. Debonding onset changes from \(U^* = 1.203\) µm to 1.186 and 1.183 µm. First plasticity moves one increment earlier to 1.169 µm, while the final reaction changes by less than 0.06 \%. On both refined meshes plasticity therefore precedes debonding, the order found in the two resolved Ansys plates and in the reduced FE² of the hole of 10 cells (Table 6), whereas on the 16×12 mesh debonding comes first. On the finest mesh, reduced and full results agree to \(2\times 10^{-12}\) in reaction and \(4.4\times 10^{-10}\) in cell stress.

The local plastic reach changes more substantially, from 13.1 to 15.0 and 17.7 µm. Nonlinear point counts rise from 34 of 768 to 140 of 1728 and 368 of 3072. The nearest point moves to within 0.03 µm of the hole edge, raising the scale-separation issue discussed in Section 4.5. These three meshes show sensitivity of the local response; they do not establish non-convergence. Full/reduced time ratios fall from 0.54 to 0.39 and 0.32, while wide-zone counts rise from 17 of 34 nonlinear points to 74 of 140 and 215 of 368. The timing trend accompanies increased wide-zone occupancy and additional work; the data do not isolate a single cause.

\begin{table*}[htbp]
\centering
\scriptsize
\caption{Refinement of the macroscopic mesh, plate with a hole, epoxy pair, $U$ up to 1.70 µm, transfer mode (Section 3.5), reduced and full-cell FE\textsuperscript{2} with the solver and threads of Table 2. The first ring is the radial size of the elements at the top of the hole; the onset point is the integration point at which debonding starts.}
\setlength{\tabcolsep}{3pt}
\resizebox{\linewidth}{!}{%
\begin{tabular}{lcccccccccc}
\toprule
Mesh & Elements / points & First ring [µm] & $U^*$ [µm] & Onset point from the edge [µm] & First plasticity, $U$ [µm] & Reaction at the end [µN/µm] & Nonlinear / wide-zone points & Plastic reach [µm] & Full / reduced [s], ratio & Max. error, reaction / $\boldsymbol{\Sigma}$ \\
\midrule
16$\times$12 & 192 / 768 & 6.1 & 1.203 & 1.20 & 1.275 & 11 268.5 & 34 / 17 & 13.1 & 34.6 / 64.5, 0.54 & 2.5e-11 / 1.3e-09 \\
24$\times$18 & 432 / 1728 & 1.2 & 1.186 & 0.22 & 1.169 & 11 263.1 & 140 / 74 & 15.0 & 131.3 / 338.4, 0.39 & 1.5e-11 / 1.2e-09 \\
32$\times$24 & 768 / 3072 & 0.25 & 1.183 & 0.03 & 1.169 & 11 261.7 & 368 / 215 & 17.7 & 364.1 / 1128.4, 0.32 & 2.0e-12 / 4.4e-10 \\
\bottomrule
\end{tabular}
}
\end{table*}

\hypertarget{ansys-against-python-on-the-cell}{%
\subsection{Ansys against Python on the cell}\label{ansys-against-python-on-the-cell}}

The elastic check is reported first (Table C.4). The elastic Ansys cases were run with the interface bonded by coupling, so the comparable Python quantity is \(\boldsymbol{C}_{\mathrm{hom}}\) with the pairs tied; the undamaged-spring \(\boldsymbol{C}_{\mathrm{hom}}\) of (15) is listed for reference.

\begin{table*}[htbp]
\centering
\footnotesize
\caption{Elastic homogenised stiffness, MPa, Voigt order (11, 22, 12) with engineering shear; matrix indices 1, 2 and 3 refer to these three components. Ansys: perfectly bonded interface, from control-node reactions and from the volume average. Python: interface pairs tied, and with undamaged springs. The deviation is $\lvert C^{\mathrm{Ansys}}_{ij} - C^{\mathrm{Python}}_{ij}\rvert/\max_{kl}\lvert C^{\mathrm{Ansys}}_{kl}\rvert$, with the Ansys reaction values and the largest entry $C_{11} = 13164.29$ MPa as the common denominator.}
\setlength{\tabcolsep}{5pt}
\begin{tabular}{>{\raggedright\arraybackslash}p{\dimexpr 0.111\linewidth-2\tabcolsep\relax}>{\raggedright\arraybackslash}p{\dimexpr 0.141\linewidth-2\tabcolsep\relax}>{\raggedright\arraybackslash}p{\dimexpr 0.161\linewidth-2\tabcolsep\relax}>{\raggedright\arraybackslash}p{\dimexpr 0.122\linewidth-2\tabcolsep\relax}>{\raggedright\arraybackslash}p{\dimexpr 0.270\linewidth-2\tabcolsep\relax}>{\raggedright\arraybackslash}p{\dimexpr 0.176\linewidth-2\tabcolsep\relax}}
\toprule
 & Ansys, reactions & Ansys, volume average & Python, tied & Deviation, Ansys reactions vs Python tied, relative to $\max\lvert C^{\mathrm{Ansys}}_{ij}\rvert$ & Python, undamaged springs \\
\midrule
$C_{11}$ & 13164.29 & 13165.54 & 13166.89 & 2.0e-04 & 10010.54 \\
$C_{22}$ & 13164.29 & 13165.91 & 13166.89 & 2.0e-04 & 10010.54 \\
$C_{33}$ & 2403.97 & 2403.12 & 2405.06 & 8.2e-05 & 2173.20 \\
$C_{12}$ & 7310.37 & 7309.01 & 7306.18 & 3.2e-04 & 5230.84 \\
$C_{13}$, $C_{23}$ & 0.00 & -0.02 & 0.00 & 3e-08 & 0.00 \\
\bottomrule
\end{tabular}
\end{table*}

The two codes agree on the bonded elastic stiffness to \(3.2\times 10^{-4}\) of the largest entry, and to \(5.7\times 10^{-4}\) of each entry's own value (for \(C_{12}\); \(4.5\times 10^{-4}\) for \(C_{33}\)), despite using 1667 quadratic and 4480 bilinear elements, respectively; the Ansys matrix is symmetric to \(10^{-15}\). The nonlinear paths are compared in Table C.5 by the relative error \(\max_s|\Sigma_{\mathrm{Ansys}} - \Sigma_{\mathrm{Python}}|/\max_s|\Sigma_{\mathrm{Ansys}}|\) per component and globally, over the whole path and over two subsets of increments: those in which the Python interface is still undamaged, and those in which no Python integration point has yielded, whether or not damage has started. The last column is the difference between the two Ansys stress measures, reactions against volume average, which provides an internal consistency check on the reference stress extraction.

\begin{table*}[htbp]
\centering
\scriptsize
\caption{Ansys against Python full cell, eight paths, $E_{\max} = 0.01$, 50 substeps. $s_d$ and $s_p$: first increment with damage and with plasticity in the Python cell. The two subset columns use the same global normaliser; in paths 3 and 8 the interface never damages, so the first subset is the whole path and the second contains only elastic increments. The three tension-shear paths are given by the ratio $E_{11} : E_{12}$ with tensorial $E_{12}$. Figures: Fig. C.3 and Fig. C.4.}
\setlength{\tabcolsep}{3pt}
\resizebox{\linewidth}{!}{%
\begin{tabular}{lcccccccccc}
\toprule
\# & Path & $\Sigma_{11}$ & $\Sigma_{22}$ & $\Sigma_{12}$ & Global & Increments without damage & Increments without plasticity & $s_d$ & $s_p$ & Ansys reactions vs volume \\
\midrule
1 & Uniaxial strain $x$ & 3.42e-04 & 1.81e-04 & — & 3.42e-04 & 5.4e-05 & 1.02e-04 & 0.50 & 0.74 & 6.2e-05 \\
2 & Uniaxial strain $y$ & 1.87e-04 & 3.61e-04 & — & 3.61e-04 & 4.4e-05 & 1.01e-04 & 0.50 & 0.74 & 6.3e-05 \\
3 & Shear ($E_{12}$ = 0.01) & — & — & 9.32e-04 & 9.32e-04 & 9.3e-04 (whole path) & 3.28e-04 & — & 0.50 & 2.2e-04 \\
4 & Equal biaxial & 8.00e-04 & 8.30e-04 & — & 8.30e-04 & 4.1e-05 & 4.12e-04 & 0.38 & 0.72 & 6.8e-05 \\
5 & Tension-shear 1 : 0.5 & 3.39e-04 & 3.72e-04 & 1.36e-03 & 3.39e-04 & 6.9e-05 & 7.9e-05 & 0.48 & 0.60 & 6.5e-05 \\
6 & Tension-shear 1 : 1 & 5.84e-04 & 4.32e-04 & 1.39e-03 & 6.11e-04 & 1.4e-04 & 1.41e-04 & 0.42 & 0.44 & 1.0e-04 \\
7 & Tension-shear 0.5 : 1 & 3.20e-04 & 5.88e-04 & 9.35e-04 & 7.02e-04 & 5.7e-04 & 2.39e-04 & 0.70 & 0.48 & 1.6e-04 \\
8 & Compression $x$ with shear, -1 : 1 & 2.11e-04 & 1.67e-04 & 1.09e-03 & 3.43e-04 & 3.4e-04 (whole path) & 9.6e-05 & — & 0.44 & 1.1e-04 \\
\bottomrule
\end{tabular}
}
\end{table*}

\begin{figure*}[tbp]
\centering
\includegraphics[width=\textwidth]{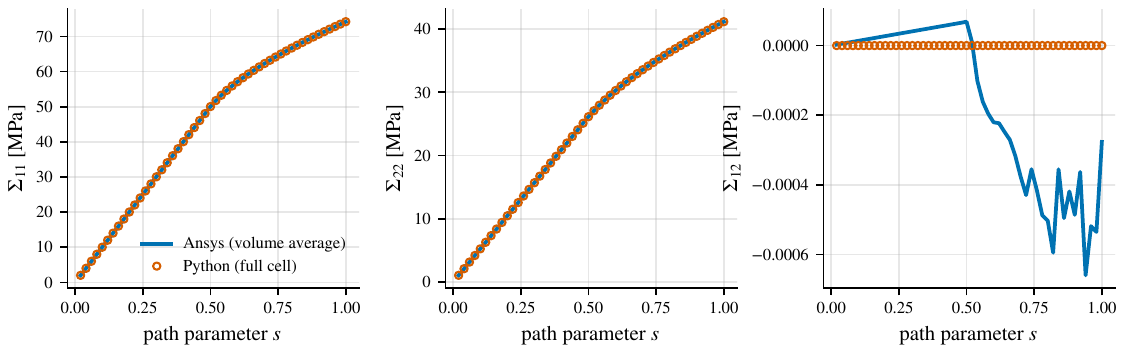}
\caption{Ansys against Python, uniaxial strain \(x\): the three macroscopic stress components against \(s\).}
\end{figure*}

\begin{figure*}[tbp]
\centering
\includegraphics[width=\textwidth]{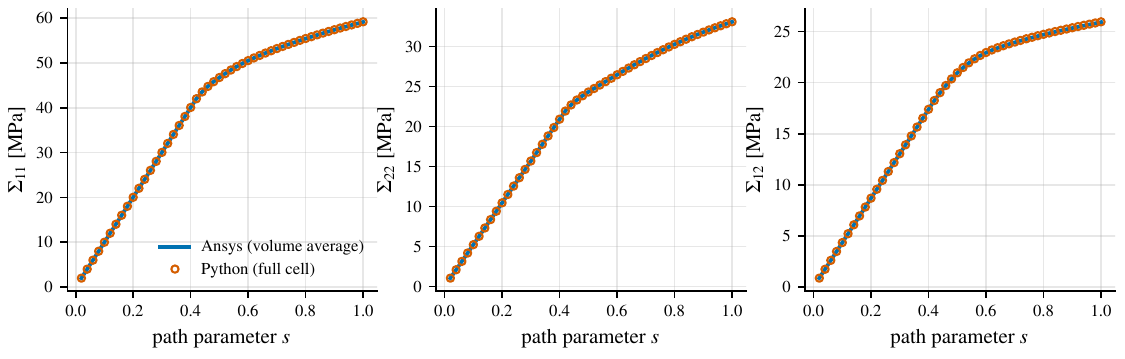}
\caption{Ansys against Python, tension-shear 1 : 1: the path in which damage and plasticity start closest together (\(s_d = 0.42\), \(s_p = 0.44\), Table C.5).}
\end{figure*}

The global stress discrepancies over the eight paths range from \(3.4\times 10^{-4}\) to \(9.3\times 10^{-4}\). The two subset columns of Table C.5 overlap and are not regimes: an increment can be undamaged and plastic, as in shear, or damaged and still elastic. For equal biaxial loading, the interface is damaged from \(s = 0.38\), and over the increments before plasticity begins at \(s = 0.72\) the discrepancy stays below \(4.2\times 10^{-4}\).

The global agreement supports consistency of the implemented laws across different discretisations. It does not establish equality of damage at corresponding pairs or quantify all reference uncertainty. Differences between the two Ansys stress measures, up to \(2.2\times 10^{-4}\), check stress extraction internally. Smaller stress components have larger relative discrepancies when normalised by their own maxima: \(\Sigma_{12}\) reaches \(1.4\times 10^{-3}\) in the tension-shear paths, although the global discrepancy remains smaller.

\hypertarget{trained-reduced-order-models}{%
\section{Trained reduced-order models}\label{trained-reduced-order-models}}

The two trained models in Section 4.10 evaluate the constitutive laws at all integration points, without hyper-reduction. POD--Galerkin uses the \(k\) leading singular directions of periodic-displacement snapshots. The full residual (6) is projected onto that basis, with all integration points and interface pairs evaluated in Newton iteration.

NTFA \cite{ref17,ref18} uses \(M\) plastic modes from matrix plastic-strain snapshots, orthonormal in the matrix average, with their elastic influence computed once on the full cell. The reduced evolution law \(\phi = \lVert\boldsymbol{\tau}\rVert - \sqrt{2/3}\,(\sigma_y + H\bar{q})\) is integrated implicitly. Because the plastic modes do not represent interface damage, variant (a) keeps the interface elastic and variant (b) retains damage pointwise, reducing only matrix plasticity.

Training set A contains uniaxial \(x\), pure shear and equal biaxial loading, giving 120 snapshots. Set B contains six proportional paths from Table 2, giving 240 snapshots: uniaxial \(x\) and \(y\), pure shear, equal biaxial, tension-shear \((1,0,1)\) and combined \((1,1,1)\). The other two tension-shear paths and the non-proportional path are excluded from both sets. The retained energy fractions are \(1 - 10^{-6}\), \(1 - 10^{-8}\) and \(1 - 10^{-10}\), giving \(k = 8,15,26\) for A and \(16,35,65\) for B. All thresholds retain the full plastic snapshot rank, \(M = 13\) and 32, respectively.

\end{document}


\hypertarget{supplementary-material}{%
\section*{Supplementary material}\label{supplementary-material}}
\addcontentsline{toc}{section}{Supplementary material}

\textbf{Process-zone condensation: a discretization-exact reduced cell model for FE² analysis of composites with fibre--matrix debonding and matrix plasticity}

L. T. Pedrini, M. Conci, R. M. Ferro

This file collects results that repeat, path by path or zone by zone, a pattern shown in the article; no statement of the article depends on them alone. Section, table, figure and equation numbers without the prefix S refer to the article. All values come from the saved results of the campaigns described in the article; the data and code deposit is cited in the Data availability statement.

\hypertarget{s1.-reduced-against-full-cell-on-the-remaining-strain-paths}{%
\subsection*{S1. Reduced against full cell on the remaining strain paths}\label{s1.-reduced-against-full-cell-on-the-remaining-strain-paths}}
\addcontentsline{toc}{subsection}{S1. Reduced against full cell on the remaining strain paths}

Figs. S1 to S7 complete Figs. 2 and 4 for the other seven paths of Table 2: macroscopic stress components against the path parameter \(s\), full cell (lines) and reduced cell with the process zone of 2.83 µm (markers). The dashed line marks the onset of interface damage and the dotted line the onset of plasticity; the errors and times are those of Table 2.

\begin{figure}[htbp]
\centering
\includegraphics[width=\linewidth]{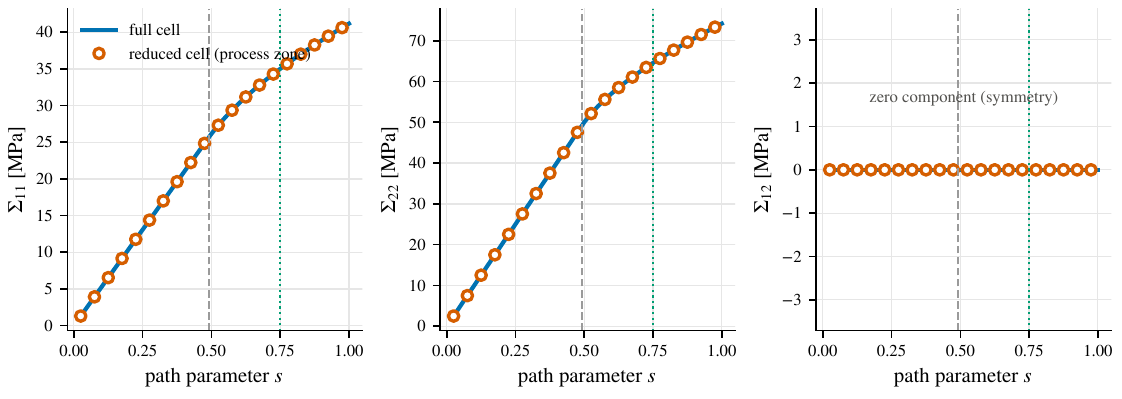}
\caption{Uniaxial \(y\): full cell against reduced cell, epoxy pair.}
\end{figure}

\begin{figure}[htbp]
\centering
\includegraphics[width=\linewidth]{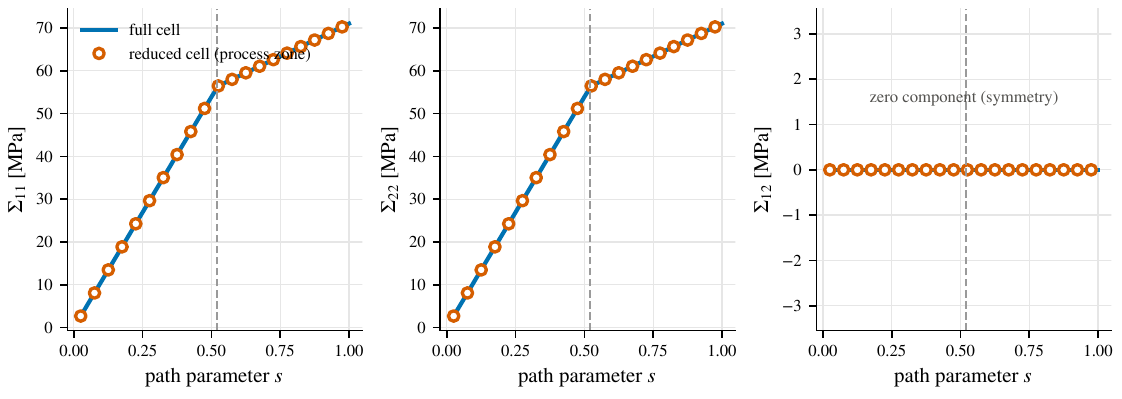}
\caption{Biaxial (1, 1, 0): full cell against reduced cell, epoxy pair.}
\end{figure}

\begin{figure}[htbp]
\centering
\includegraphics[width=\linewidth]{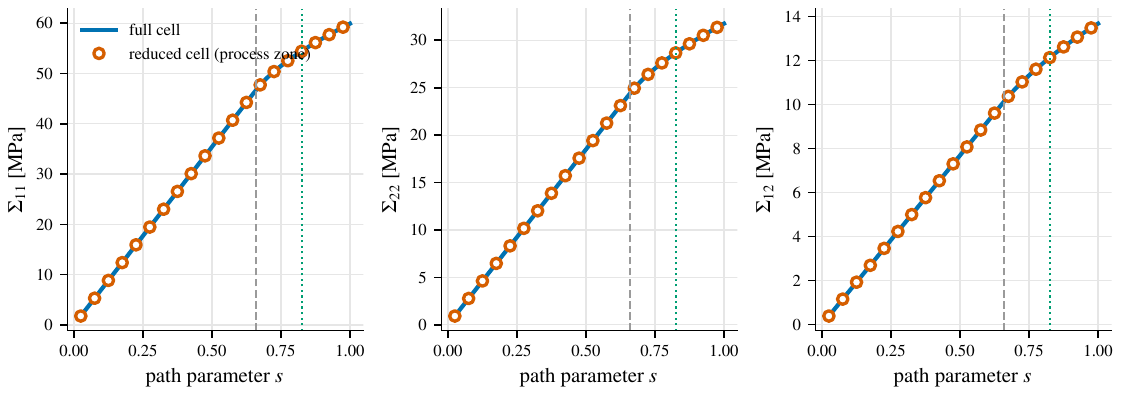}
\caption{Tension-shear (1, 0, 1): full cell against reduced cell, epoxy pair.}
\end{figure}

\begin{figure}[htbp]
\centering
\includegraphics[width=\linewidth]{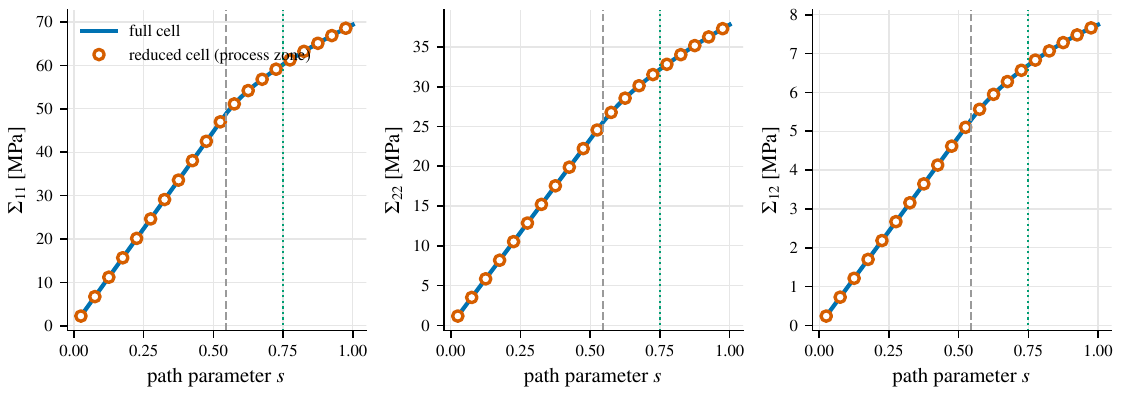}
\caption{Tension-shear (1, 0, 0.5): full cell against reduced cell, epoxy pair.}
\end{figure}

\begin{figure}[htbp]
\centering
\includegraphics[width=\linewidth]{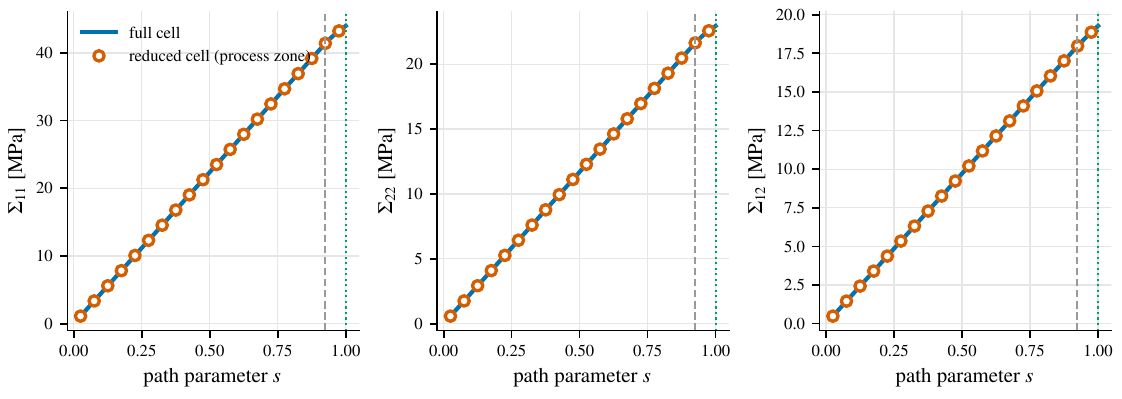}
\caption{Tension-shear (0.5, 0, 1): full cell against reduced cell, epoxy pair.}
\end{figure}

\begin{figure}[htbp]
\centering
\includegraphics[width=\linewidth]{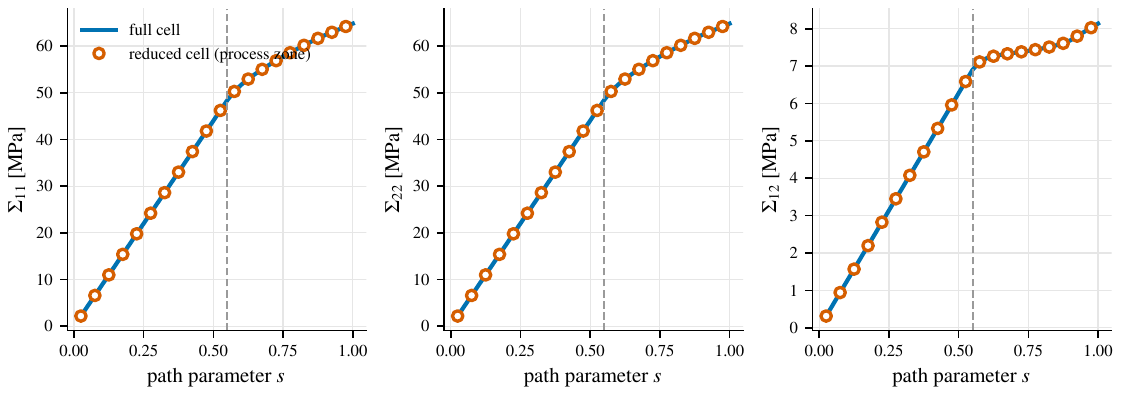}
\caption{Combined (1, 1, 1): full cell against reduced cell, epoxy pair.}
\end{figure}

\begin{figure}[htbp]
\centering
\includegraphics[width=\linewidth]{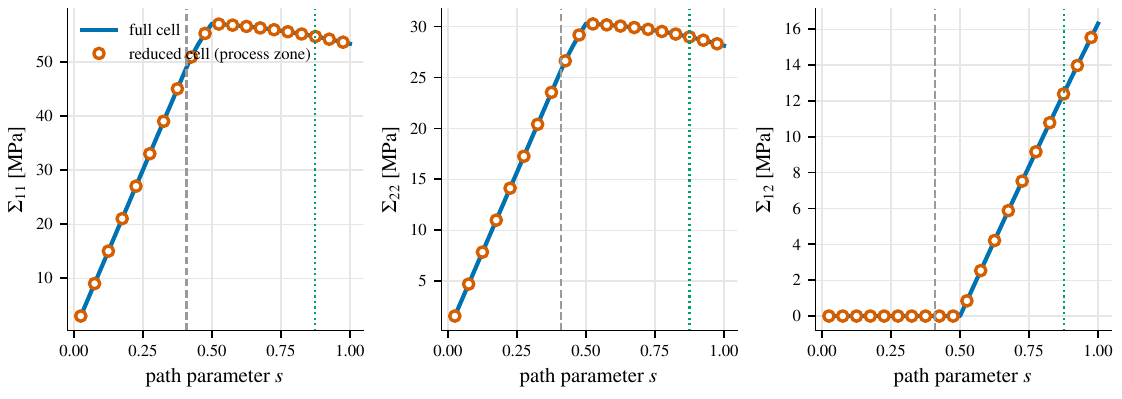}
\caption{Non-proportional, \(x\) then shear: full cell against reduced cell, epoxy pair.}
\end{figure}

\hypertarget{s2.-ansys-against-python-on-the-remaining-paths}{%
\subsection*{S2. Ansys against Python on the remaining paths}\label{s2.-ansys-against-python-on-the-remaining-paths}}
\addcontentsline{toc}{subsection}{S2. Ansys against Python on the remaining paths}

Figs. S8 to S13 complete Figs. C.3 and C.4 for paths 2 to 5, 7 and 8 of Table C.5: macroscopic stress components against \(s\), Ansys model with the microstructure of the cell (line) and Python full cell (markers). The errors are those of Table C.5.

\begin{figure}[htbp]
\centering
\includegraphics[width=\linewidth]{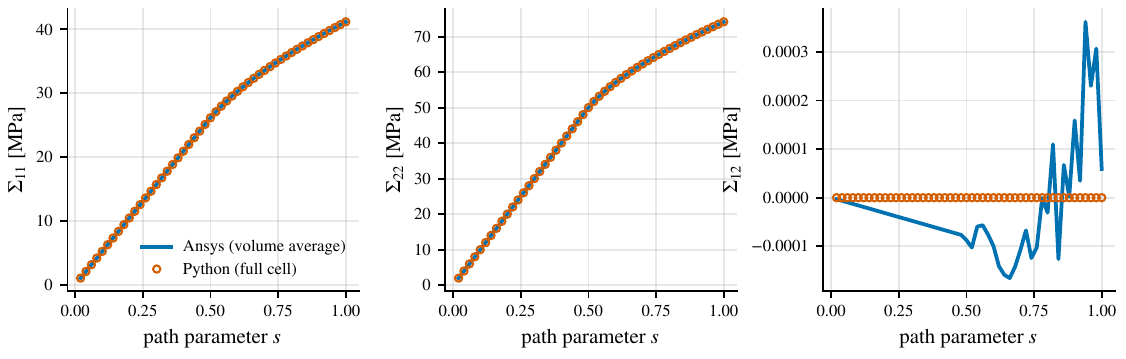}
\caption{Path 2, Uniaxial strain \(y\): Ansys against Python.}
\end{figure}

\begin{figure}[htbp]
\centering
\includegraphics[width=\linewidth]{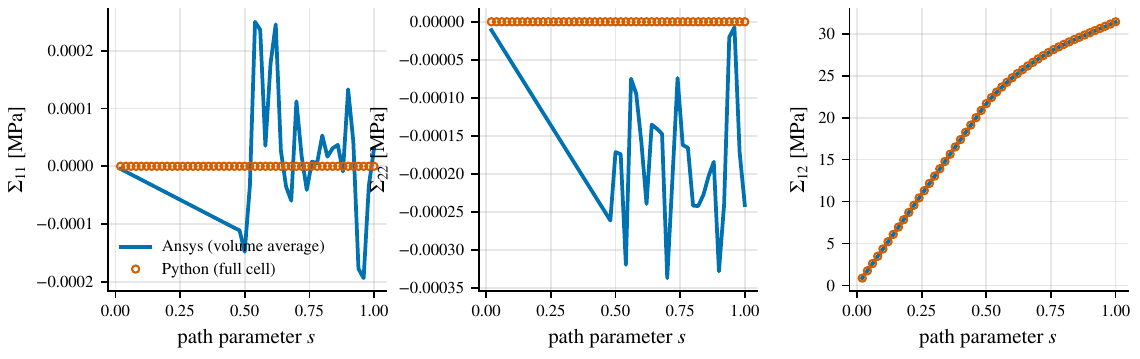}
\caption{Path 3, Shear, \(E_{12} = 0.01\): Ansys against Python.}
\end{figure}

\begin{figure}[htbp]
\centering
\includegraphics[width=\linewidth]{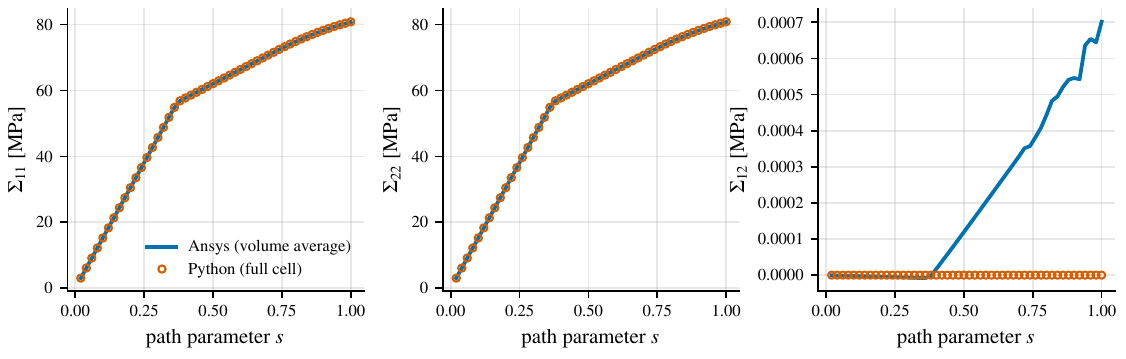}
\caption{Path 4, Equal biaxial: Ansys against Python.}
\end{figure}

\begin{figure}[htbp]
\centering
\includegraphics[width=\linewidth]{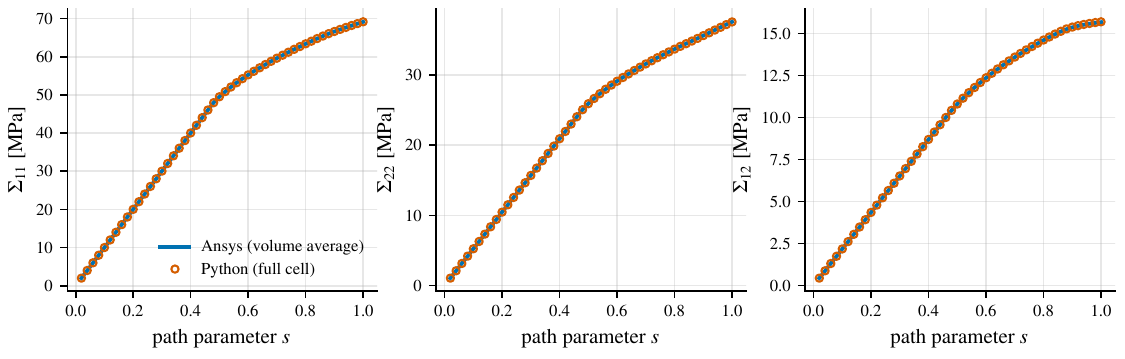}
\caption{Path 5, Tension-shear 1 : 0.5: Ansys against Python.}
\end{figure}

\begin{figure}[htbp]
\centering
\includegraphics[width=\linewidth]{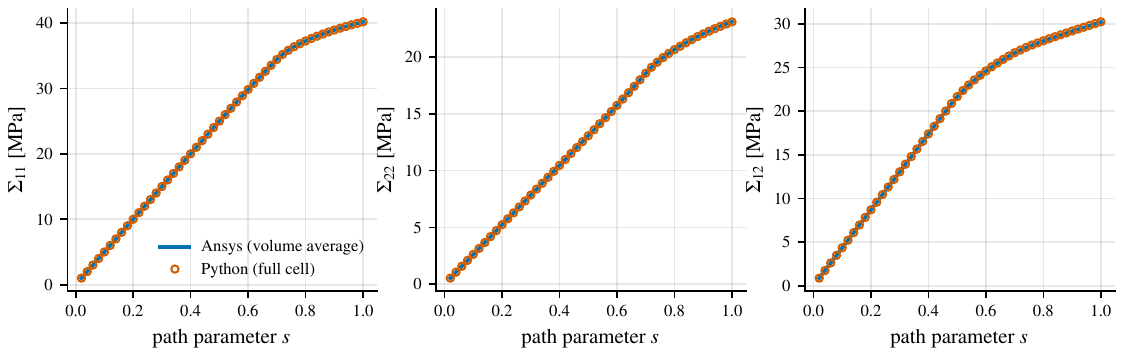}
\caption{Path 7, Tension-shear 0.5 : 1: Ansys against Python.}
\end{figure}

\begin{figure}[htbp]
\centering
\includegraphics[width=\linewidth]{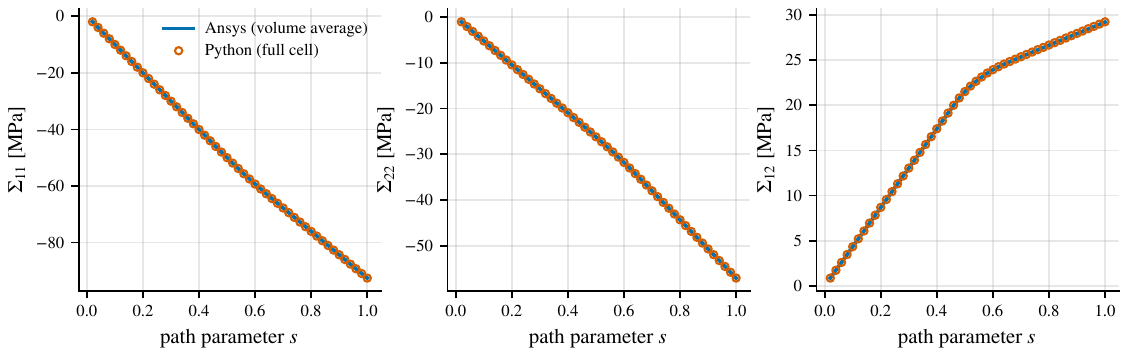}
\caption{Path 8, Compression \(x\) with shear, -1 : 1: Ansys against Python.}
\end{figure}

\hypertarget{s3.-glasspolypropylene-cell-per-path}{%
\subsection*{S3. Glass/polypropylene cell per path}\label{s3.-glasspolypropylene-cell-per-path}}
\addcontentsline{toc}{subsection}{S3. Glass/polypropylene cell per path}

Table S1 gives for the second material pair of Section 4.6 the quantities that Table 2 gives for the epoxy pair, on the same cell, mesh, process zone and increments, with the solver and threads of Table 2. The sums enter Table 7.

\begin{sidewaystable}[htbp]
\centering
\scriptsize
\caption{Glass/polypropylene cell, nine macroscopic strain paths, process zone $b = 2.83$ µm, 40 increments. $s^*$: path parameter at damage onset; $s_p$: first increment with plasticity; errors are maxima over increments; times are the Newton iterations of each mode, six threads on both sides.}
\setlength{\tabcolsep}{3pt}
\resizebox{\linewidth}{!}{%
\begin{tabular}{lccccccccccc}
\toprule
Path & $s^*$ & $s_p$ & Error before damage & Error after damage & Diff. in $\boldsymbol{C}_t$ & Plastic points / reach [µm] & Check & Iterations full / reduced & Full [s] & Reduced [s] & Ratio \\
\midrule
Uniaxial $x$ & 0.258 & 0.825 & 4.4e-14 & 6.4e-13 & 2.0e-13 & 376 / 1.12 & passes & 130 / 130 & 3.45 & 1.14 & 3.0 \\
Uniaxial $y$ & 0.258 & 0.825 & 5.5e-14 & 6.6e-13 & 2.0e-13 & 376 / 1.12 & passes & 130 / 130 & 2.71 & 1.13 & 2.4 \\
Pure shear & — & — & 6.7e-14 & — & 6.5e-14 & 0 / 0.00 & passes & 40 / 40 & 1.51 & 0.34 & 4.4 \\
Biaxial (1, 1, 0) & 0.256 & 0.875 & 3.7e-14 & 1.1e-11 & 3.6e-13 & 976 / 1.30 & passes & 117 / 117 & 2.51 & 1.45 & 1.7 \\
Tension-shear (1, 0, 1) & 0.360 & 0.800 & 4.2e-14 & 9.3e-13 & 1.4e-13 & 222 / 0.72 & passes & 119 / 119 & 2.55 & 0.89 & 2.9 \\
Tension-shear (1, 0, 0.5) & 0.291 & 0.800 & 3.9e-14 & 4.5e-13 & 1.7e-13 & 294 / 1.12 & passes & 129 / 129 & 2.67 & 1.04 & 2.6 \\
Tension-shear (0.5, 0, 1) & 0.532 & 0.850 & 3.9e-14 & 8.4e-13 & 1.2e-13 & 132 / 0.32 & passes & 100 / 100 & 2.30 & 0.70 & 3.3 \\
Combined (1, 1, 1) & 0.291 & 0.925 & 3.6e-14 & 5.4e-13 & 2.9e-13 & 220 / 0.72 & passes & 127 / 127 & 2.65 & 1.31 & 2.0 \\
Non-proportional, $x$ then shear & 0.217 & 0.875 & 4.4e-14 & 2.0e-12 & 1.3e-13 & 66 / 0.32 & passes & 139 / 139 & 2.76 & 1.07 & 2.6 \\
Sum &  &  &  &  &  &  &  &  & 23.1 & 9.1 & 2.5 \\
\bottomrule
\end{tabular}
}
\end{sidewaystable}

\hypertarget{s4.-cells-with-several-inclusions-every-run}{%
\subsection*{S4. Cells with several inclusions: every run}\label{s4.-cells-with-several-inclusions-every-run}}
\addcontentsline{toc}{subsection}{S4. Cells with several inclusions: every run}

Table S2 lists every run behind Table 8 and Fig. 13: for each cell and path, the nominal zone \(b = 2.83\) µm, the larger zones of the ladder where the certificate fails, and the geometrically similar zone \(b = R/3\).

\begin{sidewaystable}[htbp]
\centering
\scriptsize
\caption{Cells with several inclusions, $\lVert\boldsymbol{E}\rVert$ up to 0.010, 40 increments, solver and threads of Table 2. "Error while certified" is the largest stress error over the increments at which the certificate passes; "error over the path" includes the increments after a failure.}
\setlength{\tabcolsep}{3pt}
\resizebox{\linewidth}{!}{%
\begin{tabular}{lcccccccccc}
\toprule
Cell & Path & $b$ [µm] & $m/n$ & Check & Error while certified & Error over the path & Iterations full / reduced & Full [s] & Reduced [s] & Ratio \\
\midrule
1 inclusion & Uniaxial $x$ & 2.83 & 0.250 & passes & 1.3e-14 & 1.3e-14 & 103 / 103 & 3.15 & 0.81 & 3.9 \\
1 inclusion & Pure shear & 2.83 & 0.250 & fails at 0.975 & 4.2e-14 & 2.5e-03 & 46 / 40 & 1.64 & 0.34 & 4.8 \\
1 inclusion & Pure shear & 4.25 & 0.411 & fails at 0.975 & 3.9e-14 & 2.5e-03 & 46 / 40 & 1.64 & 0.72 & 2.3 \\
1 inclusion & Pure shear & 5.67 & 0.475 & fails at 0.975 & 4.2e-14 & 2.5e-03 & 46 / 40 & 1.64 & 0.91 & 1.8 \\
1 inclusion & Pure shear & 8.50 & 0.510 & fails at 0.975 & 4.4e-14 & 2.5e-03 & 46 / 42 & 1.64 & 1.04 & 1.6 \\
1 inclusion & Pure shear & 14.17 & 0.514 & passes & 4.6e-14 & 4.6e-14 & 46 / 46 & 1.64 & 1.12 & 1.5 \\
1 inclusion & Tension-shear (1, 0, 1) & 2.83 & 0.250 & passes & 5.5e-14 & 5.5e-14 & 83 / 83 & 2.07 & 0.60 & 3.5 \\
4, square & Uniaxial $x$ & 1.42 ($R/3$) & 0.316 & fails at 0.950 & 1.0e-12 & 6.8e-04 & 72 / 72 & 2.23 & 1.07 & 2.1 \\
4, square & Uniaxial $x$ & 2.83 & 0.508 & passes & 1.0e-12 & 1.0e-12 & 72 / 72 & 2.23 & 2.13 & 1.0 \\
4, square & Pure shear & 1.42 ($R/3$) & 0.316 & passes & 4.6e-14 & 4.6e-14 & 40 / 40 & 1.46 & 0.47 & 3.1 \\
4, square & Pure shear & 2.83 & 0.508 & passes & 4.2e-14 & 4.2e-14 & 40 / 40 & 1.46 & 1.03 & 1.4 \\
4, square & Tension-shear (1, 0, 1) & 1.42 ($R/3$) & 0.316 & passes & 2.6e-13 & 2.6e-13 & 57 / 57 & 1.68 & 0.59 & 2.8 \\
4, square & Tension-shear (1, 0, 1) & 2.83 & 0.508 & passes & 2.6e-13 & 2.6e-13 & 57 / 57 & 1.68 & 1.31 & 1.3 \\
9, square & Uniaxial $x$ & 0.94 ($R/3$) & 0.392 & fails at 0.900 & 8.2e-14 & 1.9e-03 & 58 / 57 & 2.66 & 1.76 & 1.5 \\
9, square & Uniaxial $x$ & 2.83 & 0.585 & passes & 1.0e-13 & 1.0e-13 & 58 / 58 & 2.66 & 3.17 & 0.8 \\
9, square & Pure shear & 0.94 ($R/3$) & 0.392 & passes & 7.3e-14 & 7.3e-14 & 40 / 40 & 2.00 & 0.65 & 3.1 \\
9, square & Pure shear & 2.83 & 0.585 & passes & 6.0e-14 & 6.0e-14 & 40 / 40 & 2.00 & 1.38 & 1.5 \\
9, square & Tension-shear (1, 0, 1) & 0.94 ($R/3$) & 0.392 & passes & 6.0e-14 & 6.0e-14 & 40 / 40 & 1.92 & 0.65 & 3.0 \\
9, square & Tension-shear (1, 0, 1) & 2.83 & 0.585 & passes & 8.2e-14 & 8.2e-14 & 40 / 40 & 1.92 & 1.37 & 1.4 \\
4, staggered & Uniaxial $x$ & 1.42 ($R/3$) & 0.373 & passes & 1.5e-13 & 1.5e-13 & 75 / 75 & 2.29 & 1.20 & 1.9 \\
4, staggered & Uniaxial $x$ & 2.83 & 0.637 & passes & 1.6e-13 & 1.6e-13 & 75 / 75 & 2.29 & 2.35 & 1.0 \\
4, staggered & Pure shear & 1.42 ($R/3$) & 0.373 & passes & 2.2e-13 & 2.2e-13 & 42 / 42 & 1.48 & 0.50 & 2.9 \\
4, staggered & Pure shear & 2.83 & 0.637 & passes & 2.2e-13 & 2.2e-13 & 42 / 42 & 1.48 & 1.29 & 1.1 \\
4, staggered & Tension-shear (1, 0, 1) & 1.42 ($R/3$) & 0.373 & fails at 0.975 & 4.6e-13 & 3.5e-04 & 53 / 53 & 1.61 & 0.64 & 2.5 \\
4, staggered & Tension-shear (1, 0, 1) & 2.83 & 0.637 & fails at 0.975 & 4.6e-13 & 2.1e-04 & 53 / 53 & 1.61 & 1.60 & 1.0 \\
4, staggered & Tension-shear (1, 0, 1) & 4.25 & 0.776 & passes & 4.6e-13 & 4.6e-13 & 53 / 53 & 1.61 & 2.32 & 0.7 \\
\bottomrule
\end{tabular}
}
\end{sidewaystable}

\hypertarget{s5.-three-dimensional-cell-every-run}{%
\subsection*{S5. Three-dimensional cell: every run}\label{s5.-three-dimensional-cell-every-run}}
\addcontentsline{toc}{subsection}{S5. Three-dimensional cell: every run}

Table S3 lists every run behind Tables 9 and 10: for each grid and path, the zones of the ladder 0.1, 0.15, 0.2, 0.3 and 0.5 \(L\) up to the first certified one.

\begin{sidewaystable}[htbp]
\centering
\scriptsize
\caption{Three-dimensional voxel cell, grids $16^3$ and $20^3$, $\lVert\boldsymbol{E}\rVert$ up to 0.010, 20 increments, reference PARDISO with six threads and reduced cell with six threads.}
\setlength{\tabcolsep}{3pt}
\resizebox{\linewidth}{!}{%
\begin{tabular}{lcccccccccc}
\toprule
Grid & Path & $b/L$ & $m/n$ & Check & Error while certified & Error over the path & Iterations full / reduced & Full [s] & Reduced [s] & Ratio \\
\midrule
$16^3$ & Uniaxial $x$ & 0.10 & 0.572 & fails at 0.700 & 1.3e-12 & 9.5e-04 & 57 / 56 & 19.9 & 19.6 & 1.02 \\
$16^3$ & Uniaxial $x$ & 0.15 & 0.676 & fails at 0.800 & 1.3e-12 & 1.7e-04 & 57 / 57 & 19.9 & 28.1 & 0.71 \\
$16^3$ & Uniaxial $x$ & 0.20 & 0.753 & fails at 1.000 & 1.3e-12 & 7.4e-05 & 57 / 57 & 19.9 & 30.6 & 0.65 \\
$16^3$ & Uniaxial $x$ & 0.30 & 0.821 & passes & 1.3e-12 & 1.3e-12 & 57 / 57 & 19.9 & 32.9 & 0.60 \\
$16^3$ & Pure shear & 0.10 & 0.572 & fails at 0.750 & 2.9e-13 & 3.2e-02 & 35 / 29 & 14.2 & 3.4 & 4.20 \\
$16^3$ & Pure shear & 0.15 & 0.676 & fails at 0.750 & 3.0e-13 & 2.6e-02 & 35 / 29 & 14.2 & 5.7 & 2.51 \\
$16^3$ & Pure shear & 0.20 & 0.753 & fails at 0.750 & 2.8e-13 & 1.7e-02 & 35 / 31 & 14.2 & 8.1 & 1.75 \\
$16^3$ & Pure shear & 0.30 & 0.821 & fails at 0.900 & 2.9e-13 & 1.2e-03 & 35 / 35 & 14.2 & 11.5 & 1.23 \\
$16^3$ & Pure shear & 0.50 & 0.832 & passes & 4.1e-13 & 4.1e-13 & 35 / 35 & 14.2 & 12.3 & 1.15 \\
$16^3$ & Tension-shear (1, 0, 0, 0, 0, 1) & 0.10 & 0.572 & fails at 0.800 & 2.4e-13 & 1.4e-03 & 53 / 53 & 18.6 & 12.8 & 1.46 \\
$16^3$ & Tension-shear (1, 0, 0, 0, 0, 1) & 0.15 & 0.676 & fails at 0.850 & 2.4e-13 & 2.5e-04 & 53 / 53 & 18.6 & 17.5 & 1.06 \\
$16^3$ & Tension-shear (1, 0, 0, 0, 0, 1) & 0.20 & 0.753 & fails at 0.900 & 2.4e-13 & 5.3e-05 & 53 / 53 & 18.6 & 19.7 & 0.94 \\
$16^3$ & Tension-shear (1, 0, 0, 0, 0, 1) & 0.30 & 0.821 & passes & 2.4e-13 & 2.4e-13 & 53 / 53 & 18.6 & 22.2 & 0.84 \\
$20^3$ & Uniaxial $x$ & 0.10 & 0.524 & fails at 0.650 & 4.1e-13 & 4.8e-04 & 61 / 60 & 64.2 & 62.3 & 1.03 \\
$20^3$ & Uniaxial $x$ & 0.15 & 0.633 & fails at 0.950 & 4.2e-13 & 9.2e-05 & 61 / 61 & 64.2 & 79.8 & 0.80 \\
$20^3$ & Uniaxial $x$ & 0.20 & 0.722 & fails at 1.000 & 4.3e-13 & 5.0e-05 & 61 / 61 & 64.2 & 89.0 & 0.72 \\
$20^3$ & Uniaxial $x$ & 0.30 & 0.802 & passes & 4.1e-13 & 4.1e-13 & 61 / 61 & 64.2 & 95.8 & 0.67 \\
$20^3$ & Pure shear & 0.10 & 0.524 & fails at 0.750 & 3.0e-13 & 2.7e-02 & 35 / 25 & 43.0 & 8.1 & 5.32 \\
$20^3$ & Pure shear & 0.15 & 0.633 & fails at 0.750 & 2.9e-13 & 2.3e-02 & 35 / 29 & 43.0 & 14.5 & 2.96 \\
$20^3$ & Pure shear & 0.20 & 0.722 & fails at 0.750 & 2.9e-13 & 1.4e-02 & 35 / 32 & 43.0 & 24.8 & 1.73 \\
$20^3$ & Pure shear & 0.30 & 0.802 & fails at 0.850 & 2.8e-13 & 9.0e-04 & 35 / 35 & 43.0 & 31.2 & 1.38 \\
$20^3$ & Pure shear & 0.50 & 0.817 & passes & 3.0e-13 & 3.0e-13 & 35 / 35 & 43.0 & 34.0 & 1.26 \\
$20^3$ & Tension-shear (1, 0, 0, 0, 0, 1) & 0.10 & 0.524 & fails at 0.750 & 3.3e-13 & 8.2e-04 & 48 / 47 & 52.2 & 32.7 & 1.60 \\
$20^3$ & Tension-shear (1, 0, 0, 0, 0, 1) & 0.15 & 0.633 & fails at 0.900 & 3.3e-13 & 6.3e-05 & 48 / 48 & 52.2 & 43.6 & 1.20 \\
$20^3$ & Tension-shear (1, 0, 0, 0, 0, 1) & 0.20 & 0.722 & fails at 1.000 & 3.7e-13 & 1.8e-06 & 48 / 48 & 52.2 & 49.3 & 1.06 \\
$20^3$ & Tension-shear (1, 0, 0, 0, 0, 1) & 0.30 & 0.802 & passes & 3.5e-13 & 3.5e-13 & 48 / 48 & 52.2 & 54.1 & 0.96 \\
\bottomrule
\end{tabular}
}
\end{sidewaystable}